\documentclass[12pt,letterpaper]{article}
\usepackage{amsfonts}
\usepackage{amsmath}
\usepackage{bm}
\usepackage{geometry}
\usepackage{setspace}
\usepackage{graphicx}
\usepackage{lscape}
\usepackage{verbatim}
\usepackage{natbib}
\usepackage{xcolor}
\usepackage{tabularx}
\definecolor{winered}{rgb}{0.5,0,0}
\usepackage[bookmarks=true, bookmarksnumbered=true, allbordercolors={1 1 1}]{hyperref}
\hypersetup{
  colorlinks   = true, 
  urlcolor     = blue, 
  linkcolor    = blue, 
  citecolor    = winered,
}
\usepackage[open=true, numbered=true]{bookmark}
\usepackage{float}
\usepackage{fancyhdr}
\usepackage[toc,page]{appendix}
\usepackage{scalefnt}
\usepackage{afterpage}
\usepackage[left]{lineno}
\usepackage{caption}
\usepackage{subcaption}

\makeatletter
\def\blfootnote{\xdef\@thefnmark{}\@footnotetext}
\makeatother

\emergencystretch=\maxdimen
\begin{document}

\title{\LARGE{A new algorithm for structural restrictions in Bayesian vector autoregressions}\blfootnote{I would like to thank without implicating Christiane Baumeister, Martin Bruns, Fabio Canova, Filippo Ferroni, Luca Gambetti, Toru Kitagawa, Gary Koop, Michele Lenza, Laura Liu, Emanuel M\"{o}nch, Alberto Musso, Serena Ng, Michele Piffer, Davide Pettenuzzo, Francesco Ravazzolo, Frank Schorfheide, Maximilian Schr\"{o}der, Christian Schumacher, Leif Anders Thorsrud, John Tsoukalas and Harald Uhlig, as well as seminar and conference participants, for useful discussions and comments. Any remaining errors should solely be attributed to the author.
\\
\indent MATLAB code that replicates the Monte Carlo and empirical results of this paper is available on \href{https://sites.google.com/site/dimitriskorobilis/}{https://sites.google.com/site/dimitriskorobilis/}.
\\
\indent Correspondence: Professor of Econometrics, Adam Smith Business School, University of Glasgow, 40 University Avenue, Glasgow, G12 8QQ, UK; email: \href{mailto:dikorobilis@googlemail.com}{dikorobilis@googlemail.com}.
}}
\author{Dimitris Korobilis \\
\emph{University of Glasgow}}
\date{\today}

\maketitle
\begin{abstract}
\noindent A comprehensive methodology for inference in vector autoregressions (VARs) using sign and other structural restrictions is developed. The reduced-form VAR disturbances are driven by a few common factors and structural identification restrictions can be incorporated in their loadings in the form of parametric restrictions. A Gibbs sampler is derived that allows for reduced-form parameters and structural restrictions to be sampled efficiently in one step. A key benefit of the proposed approach is that it allows for treating parameter estimation and structural inference as a joint problem. An additional benefit is that the methodology can scale to large VARs with multiple shocks, and it can be extended to accommodate non-linearities, asymmetries, and numerous other interesting empirical features. The excellent properties of the new algorithm for inference are explored using synthetic data experiments, and by revisiting the role of financial factors in economic fluctuations using identification based on sign restrictions.

\bigskip

\noindent \emph{Keywords:} Gibbs sampling; factor model decomposition; large VAR; sign restrictions

\bigskip \medskip

\noindent \emph{JEL Classification:}\ C11, C13, C15, C22, C52, C53, C61
\end{abstract}
\thispagestyle{empty} 

\newpage
\doublespacing

\setcounter{page}{1}
\section{Introduction}
This paper proposes a new Bayesian Markov chain Monte Carlo (MCMC) algorithm for joint estimation of parameters of reduced-form vector autoregressions (VARs) and associated sign restrictions for structural identification. The main idea is to allow the reduced-form VAR disturbances to have a static factor model structure. By doing so, sign and other restrictions can be incorporated via straightforward parametric prior distributions, and the factors can be interpreted as the structural VAR (SVAR) disturbances. A new, computationally efficient, algorithm is able to jointly sample VAR parameters and identification restrictions. The implication of this feature is that the parameter estimates and the fit of the VAR depend on, and interact with, the identification restrictions the researcher has in mind. Existing reduced-form VAR approaches typically follow a two-step procedure in which an estimate of the VAR covariance matrix is obtained in the first step, and some identification scheme that seems plausible to the researcher is imposed in a second step.\footnote{Consider a VAR covariance matrix estimate $\mathbf{\widehat{\Omega}}$, structural identification simply boils down to finding a matrix $\mathbf{A}$ such that $\mathbf{A}\mathbf{A}^{\prime} = \mathbf{\widehat{\Omega}}$. There are infinite such matrices that satisfy this relationship, therefore, it is required to impose some zero or other restrictions on $\mathbf{A}$. However, these restrictions can never be (statistically) tested since $\mathbf{\widehat{\Omega}}$ is fixed and the data likelihood remains unchanged no matter what the researcher thinks restrictions in $\mathbf{A}$ should be.} In the proposed modeling approach, different identification schemes result to different VAR parameter estimates and general model fit. The benefit of the new approach is that the researcher can treat parameter estimation and identification as a joint problem, which is a very advantageous approach towards inference, due to the fact that either the ``true'' VAR parameters or the ``true'' structural restrictions are never known. By extracting unobserved factors from VAR disturbances, sign and zero structural restrictions become, respectively, inequality and zero parametric restrictions in the associated factor loadings matrix.

The sign restrictions approach to identification has become very popular in applied work compared to traditional identification methods such as exclusion restrictions; see \cite{kilianlutkepohl2017} for a detailed review of this literature. The main feature of popular Bayesian algorithms for inference in sign restrictions, such as \cite{RubioRamirezetal2010}, is that they rely on rejection sampling schemes (also known as \emph{accept/reject algorithms}) in order to search for matrices that satisfy the desired restrictions. If restrictions are tight, as it would be the case in models with many variables and many shocks, rejection sampling results in constantly rejecting draws. In contrast, the Gibbs sampler proposed in this paper allows to sample contemporaneous structural matrices from their conditional posterior and these samples are always accepted. The benefits of the new approach are demonstrated by revisiting the empirical results in \cite{Furlanettoetal2017} by using a single 15-variable VAR with many shocks, instead of many 5-variable VARs for identifying a few shocks at a time (which is what these authors do due to the computational constraints of the \citealp{RubioRamirezetal2010} algorithm they adopt). As a rough indication of the computational efficiency of the new algorithm, I find that obtaining 5,000 uncorrelated samples from the benchmark six-variable VAR of \cite{Furlanettoetal2017} using the new algorithm takes less than five minutes; using the original \cite{RubioRamirezetal2010} algorithm that \cite{Furlanettoetal2017} adopted in order to produce their results, it takes roughly four hours to obtain 2,000 draws that satisfy the same restrictions. Empirically, using the same set of sign restrictions, the two algorithms produce comparable results as measured by shapes and magnitudes of impulse response functions.

The new algorithm for structural inference in VARs shares some similarities, from a computational perspective, with the SVAR approach of \cite{BaumeisterHamilton2015}. These authors estimate a joint model for structural restrictions and parameter estimation. However, the need to derive a reasonably simple algorithm algorithm for inference, means that these authors integrate out autoregressive and variance parameters from the joint posterior of parameters and structural restrictions using natural conjugate priors. The result is a Metropolis-Hastings algorithm that is also of the accept-reject form and cannot scale up easily to very high dimensions.\footnote{To be exact, if a candidate sample is not accepted as a sample from the true posterior, then the immediately previous accepted sample values are used. \cite{RePEc:diw:diwwpp:dp1796} propose a more efficient Dynamic Striated Metropolis Hastings algorithm that builds on importance sampling proposals. While this algorithm is appropriate for high-dimensional models, in the case of the VAR it will still hit a computational bottleneck at much lower dimensions than the Gibbs sampler proposed in this paper.} Most importantly, the adoption of a natural conjugate prior means that it is not possible to extend the \cite{BaumeisterHamilton2015} methods with empirically relevant time series features, such as heteroskedasticity or structural breaks. In contrast, the algorithm proposed here builds on a standard reduced-form Bayesian VAR that is easy to work with and extend to large dimensions, nonlinear or asymmetric shocks, nonlinear parameters (e.g. stochastic volatility), and numerous other interesting features; see \cite{KoopKorobilis2010} for a thorough review of Bayesian tools for inference in reduced-form VARs.

The idea to decompose the VAR disturbances into common factors is also related to numerous other modelling approaches. \cite{Gorodnichenko2005} specified an identical VAR model with reduced-rank decomposition of the disturbance term. The purpose of specifying the VAR that way was to replace standard block diagonal restrictions in VARs \citep{BernankeMihov1992} with a more parsimonious identification scheme that imposes less (possibly unreasonable) zero restrictions. More recently, \cite{MatthesSchwartzman2019} specify a closely related VAR model in order to identify the structural impact of sectoral dynamics on GDP. Their identification is via a factor structure on the residuals that has the additional assumption of allowing for correlation within industries but no correlation across industries.

In a different strand of the VAR literature, \cite{StockWatson2005} specify a more general factor-augmented VAR (FAVAR) and discuss in detail how various identification schemes fit in this setting. They note \citep[Section 3.5]{StockWatson2005} that the sign restrictions identification scheme proposed by \cite{Uhlig2005} also fits the FAVAR framework. An application of this idea can be found in \cite{AhmadiUhlig2016}. From a modeling point of view, the model I propose in this paper can be viewed as a special case of the \cite{AhmadiUhlig2016} FAVAR model. However, the specification I use has completely different implications both algorithmically and in terms of inference. \cite{AhmadiUhlig2016} project a large vector of observable macroeconomic variables into a smaller vector of factors and they model VAR dynamics only on these factors. This means that there is some loss of information (not all macro variables are explained well by the factors) and the statistical fit of the factors determines the contribution of each structural shock on each macroeconomic variable. Additionally, the autoregressive dynamics of the large macro dataset is represented only by the autoregressive dynamics of the smaller vector of factors. This modeling choice means that, inevitably, the FAVAR is unable to capture richer patterns of propagation of structural shocks to observed macroeconomic variables. In contrast, in this paper all observable macroeconomic variables are endogenous in the VAR and the sole role of the factors is to represent structural shocks. Additionally, the algorithm derived here is computationally simpler as it relies on posterior formulas for linear regression models, instead of building on more demanding simulation smoothing techniques, as is the case with the FAVAR (see \cite{AhmadiUhlig2016} and \cite{Bernankeetal2005}).

The next Section introduces the new methodology and associated Gibbs sampler algorithm for inference, and it outlines the key components that help speed up and stabilize (numerically) posterior sampling in high dimensions. In Sections 3 I undertake several important exercises using synthetic datasets, in order to test both the computational features of the new algorithm as well as shed light on how joint inference on parameters and structural restriction is implemented. In Section 4 the algorithm is applied to the issue of measuring the impact of a financial shock to the macroeconomy. Section 5 concludes the paper.

\section{A new methodology for sign restrictions in VARs}
The starting point is the reduced-form vector autoregression
\begin{equation}
\mathbf{y}_{t} = \mathbf{\Phi} \mathbf{x}_{t} + \bm{\varepsilon}_{t}, \label{VAR}
\end{equation}
where $\mathbf{y}_{t}$ is a $\left( n \times 1 \right)$ vector of observed variables, $\mathbf{x}_{t} = \left( 1,\mathbf{y}_{t-1}^{\prime},...,\mathbf{y}_{t-p}^{\prime} \right)^{\prime}$ a $\left( k \times 1 \right)$ vector (with $k=np+1$) containing a constant and $p$ lags of $\mathbf{y}$, $\mathbf{\Phi}$ is an $(n \times k)$ matrix of coefficients, and $\bm{\varepsilon}_{t}$ a $\left( n \times 1 \right)$ vector of disturbances distributed as $N\left( \mathbf{0}_{n \times 1},\mathbf{\Omega} \right)$ with $\mathbf{\Omega}$ an $n \times n$ covariance matrix. The structural VAR (SVAR) form associated with the reduced-form model in \eqref{VAR} is
\begin{equation}
\mathbf{A} \mathbf{y}_{t} = \mathbf{B} \mathbf{x}_{t} + \bm{u}_{t}, \label{SVAR}
\end{equation}
where $\mathbf{B} = \mathbf{A}\mathbf{\Phi}$, $\bm{u}_{t} = \mathbf{A}\bm{\varepsilon}_{t}$ and $cov(\bm{u}_{t}) = \mathbf{D}$, with $\mathbf{D}$ an $n \times n$ diagonal matrix which, sometimes, is normalized to be the identity matrix. The SVAR form can be obtained by means of a decomposition of the reduced-form covariance matrix of the form $\mathbf{A}\mathbf{\Omega}\mathbf{A}^{\prime} = \mathbf{D} $ where both sides of equation \eqref{VAR} are left-multiplied with the $n \times n$ matrix $\mathbf{A}$. This decomposition is unique when $\mathbf{A}$ is a lower triangular matrix (typically with a unit diagonal, unless we do the normalization $\mathbf{D}=\mathbf{I}$), but it has infinite solutions for a full matrix $\mathbf{A}$.

\subsection{VARs driven by a few, common shocks}
I begin by building on fundamental ideas introduced in the factor model literature, as applied to empirical problems in macroeconomics: a few common forces (which in a structural setting we desire to identify as ``primitive shocks''; see \citealp{Ramey2016}) are driving the set of reduced-form shocks in the system of $n$ endogenous variables. In order to materialize this idea, the reduced-form VAR disturbances in equation \eqref{VAR} are decomposed using the following static factor model specification
\begin{equation}
\bm{\varepsilon}_{t} = \mathbf{\Lambda} \mathbf{f}_{t} + \mathbf{v}_{t}, \label{factor_model}
\end{equation}
where $\mathbf{\Lambda}$ is an $n \times r$ matrix of factor loadings, $\mathbf{f}_{t}$ is an $r \times 1$ vector of factors, and $\mathbf{v}_{t}$ is an $n \times 1$ vector of idiosyncratic shocks. While the $n$ shocks in $\bm{\varepsilon}_{t}$ are decomposed into $r + n$ shocks, only the $r$ common shocks in $\mathbf{f}_{t}$ are considered structural and the $n$ shocks in $\mathbf{v}_{t}$ are simply nuisance shocks (e.g. due to measurement or expectations error, incomplete information etc). The assumption here is that $n$ is large and that $r \leq n$, and not necessarily $r \ll n$, as is typically assumed in the factor literature. In line with the \emph{exact factor model} literature, let $\mathbf{v}_{t} \overset{\text{i.i.d}}\sim  N \left(\mathbf{0}_{n \times 1}, \mathbf{\Sigma} \right)$, with $\mathbf{\Sigma}$ an $n \times n$ diagonal matrix. Additionally, let $\mathbf{f}_{t} \sim N \left( \mathbf{0}_{r \times 1}, \mathbf{I}_{r}\right)$, such that the conditional covariance matrix of $\bm{\varepsilon}_{t}$ is now of the form
\begin{equation}
cov \left( \bm{\varepsilon}_{t} \vert \mathbf{\Lambda},\mathbf{\Sigma} \right) = \mathbf{\Omega} = \mathbf{\Lambda}\mathbf{\Lambda}^{\prime} + \mathbf{\Sigma} \label{factor_decomp}.
\end{equation}

This factor model decomposition of $\mathbf{\Omega}$ reveals that, as long as $\mathbf{\Sigma}$ is diagonal, identification via sign restrictions can be achieved by imposing the desired signs on $\mathbf{\Lambda}$. Similarly, zero restrictions simply correspond to setting the respective elements of $\mathbf{\Lambda}$ to zero.\footnote{If desired, several other restrictions can be incorporated in a straightforward way, such as ranking restrictions \citep{https://doi.org/10.3982/QE1277}, restrictions on elasticities, or other restrictions on magnitudes of shocks. For example, if for shock $j$ variable $i$ should react with a larger magnitude than variable $k$, then we get the restriction $\mathbf{\Lambda}_{ij} > \mathbf{\Lambda}_{kj}$ which is fairly simple to incorporate within an MCMC sampling setting. The key early reference for inference in the Bayesian regression model with general inequality constraints is \cite{Geweke1996}.} To see this, consider a reduced-rank SVAR representation of this model, which can be obtained by left-multiplying the reduced-form VAR model given by equations \eqref{VAR} - \eqref{factor_model} with the generalized inverse of $\mathbf{\Lambda}$, as follows:
\begin{eqnarray}
\mathbf{y}_{t} & = & \mathbf{\Phi} \mathbf{x}_{t} + \mathbf{\Lambda} \mathbf{f}_{t} + \mathbf{v}_{t} \\
\left( \mathbf{\Lambda}^{\prime} \mathbf{\Lambda} \right)^{-1}\mathbf{\Lambda}^{\prime} \mathbf{y}_{t} & = & \left( \mathbf{\Lambda}^{\prime} \mathbf{\Lambda} \right)^{-1}\mathbf{\Lambda}^{\prime} \mathbf{\Phi} \mathbf{x}_{t} +  \mathbf{f}_{t} + \left( \mathbf{\Lambda}^{\prime} \mathbf{\Lambda} \right)^{-1}\mathbf{\Lambda}^{\prime} \mathbf{v}_{t} \\
\mathbf{A}_{1} \mathbf{y}_{t} & = & \mathbf{B}_{1} \mathbf{x}_{t} + \mathbf{f}_{t} + \left( \mathbf{\Lambda}^{\prime} \mathbf{\Lambda} \right)^{-1}\mathbf{\Lambda}^{\prime}\mathbf{v}_{t}, \label{RR_SVAR} \Rightarrow \\
\mathbf{f}_{t} & \approx & \mathbf{A}_{1} \mathbf{y}_{t} - \mathbf{B}_{1} \mathbf{x}_{t}. \label{struct_shock}
\end{eqnarray}
In the equation above, the SVAR matrix $\mathbf{A}_{1}$ is equivalent to the generalized inverse $\left( \mathbf{\Lambda}^{\prime} \mathbf{\Lambda} \right)^{-1}\mathbf{\Lambda}^{\prime}$. While $\mathbf{\Lambda}$ is not observed, assume that a consistent estimator of this parameter exists. Given that in the exact factor model formulation the $\mathbf{v}_{t}$ are uncorrelated, the Central Limit Theorem in \cite{Bai2003} suggests that for each $t$ and for $n\rightarrow \infty$ we have $\left( \mathbf{\Lambda}^{\prime} \mathbf{\Lambda} \right)^{-1}\mathbf{\Lambda}^{\prime}\mathbf{v}_{t} \rightarrow 0$ making this term asymptotically negligible. Therefore, it is justified to view $\mathbf{v}_{t}$ as a residual or a noise shock that carries no structural interpretation. At the same time, $f_{t}$ can be interpreted as a projection of the SVAR structural shocks $\mathbf{u}_{t}$ into $\mathbb{R}^{r}$.

Impulse response functions (IRFs) can be obtained via the vector moving average (VMA) representation of the VAR. In the particular case with $p=1$ lags (for notational simplicity) the VMA form becomes
\begin{eqnarray}
\mathbf{y}_{t} & = & \mathbf{\phi}_0 + \mathbf{\Phi}_{1} \mathbf{y}_{t-1} + \bm{\varepsilon}_{t}, \Rightarrow \\ 
\mathbf{y}_{t} & = & \mathbf{\mu} + \sum_{i=0}^{\infty} \mathbf{\Psi}_{i}\bm{\varepsilon}_{t},  \Rightarrow \\
\mathbf{y}_{t} & = & \mathbf{\mu} + \sum_{i=0}^{\infty} \mathbf{\Psi}_{i}\mathbf{\Lambda} \mathbf{f}_{t} + \sum_{i=0}^{\infty} \mathbf{\Psi}_{i}\mathbf{v}_{t}, \label{VMA}
\end{eqnarray}
where $\mathbf{\mu} = \left( I -  \mathbf{\Phi}_{1} L \right)^{-1}\mathbf{\phi}_0$ with $L$ is the lag operator, and $\mathbf{\Psi}_{i} = \sum_{j=1}^{i}  \mathbf{\Psi}_{j-i}\mathbf{\Phi}_{1}$ with $\mathbf{\Psi}_{0} = \mathbf{I}$ \citep[see][Section 2.2.2]{kilianlutkepohl2017}. The impulse response on impact is 
\begin{equation}
\frac{\partial \mathbf{y}_{t}}{\partial \mathbf{u}_{t}} \approx \frac{\partial \mathbf{y}_{t}}{\partial \mathbf{f}_{t}} = \mathbf{\Psi}_{0}\mathbf{\Lambda} = \mathbf{\Lambda}, \label{IRF_cond}
\end{equation}
showing that parametric restrictions in $\mathbf{\Lambda}$ correspond to structural restrictions on impact IRFs. This result is true even when considering the effect of $\mathbf{v}_{t}$, since this term has a diagonal covariance matrix and does not affect contemparenous relationships in the variables $\mathbf{y}_{t}$.

Finally, similar to the algorithm in \cite{BaumeisterHamilton2015}, the proposed specification is efficient only for static sign restrictions. This is because dynamic restrictions are nonlinear and cannot be represented (in a straightforward way) as equivalent parametric inequality restrictions in the linear VAR parameters. Therefore, dynamic restrictions would require to turn to less efficient sampling schemes similar to \cite{Ariasetal2018}. In practice, there is little consensus in economic theory about the signs of structural impulse responses at longer horizons \citep[see for example][]{CanovaPaustian2011}, and for that reason the vast majority of empirical papers impose restrictions only on impact \citep{kilianlutkepohl2017}. Nevertheless, as \cite{Uhlig2017} notes, it can be quite useful to have the option to impose sign restrictions in the longer-run responses of macroeconomic variables to shocks. Within the context of the proposed methodology, this issue can be addressed if equation \eqref{factor_model} is specified as a dynamic factor model instead of a static factor model. However, this more general modeling assumption would require to rely on filtering and smoothing sampling steps that would, in turn, lead to a completely different estimation algorithm compared to the algorithm presented in this paper. As a result, I exclusively focus here on impact (contemporaneous) structural restrictions in small and large VARs. Extending to the case of long-horizon sign restrictions is conceptually feasible, but it is left for future research.

\subsection{Identification} \label{sec:identif}
The previous discussion established that the factor decomposition of the VAR disturbances projects the $n$ shocks into $r$ structural plus $n$ nuisance shocks. Therefore, the first identification issue relates to being able to separate the common component ($\mathbf{\Lambda}\mathbf{f}_{t}$) from the idiosyncratic one. Notice that the original VAR covariance matrix $\mathbf{\Omega}$ has $n(n+1)/2$ free elements, while the right-hand side of the factor decomposition in \eqref{factor_decomp} has $nr + n$ free parameters. Therefore, the first condition is that $n(n+1)/2 \ge nr + n$ or that $r \le (n-1)/2$, which implies that the common component will always be identified even if the factors (structural shocks) are not identified. This condition implies that in a 19-variable VAR a reasonable number of nine factors/shocks can be estimated. Next, restrictions are required for uniquely identifying the factors, which are also structural shocks. Following \cite{anderson1956} an additional $r(r-1)/2$ restrictions are needed in order to deal with the \emph{rotation problem}. This is due to the fact that rotating $\mathbf{\Lambda}$ and $\mathbf{f}_{t}$ using an orthogonal matrix $\mathbf{P}$ leads to an observationally equivalent solution, that is,
\begin{equation}
\mathbf{\Lambda} \mathbf{f}_{t}  =  \mathbf{\Lambda} \mathbf{P} \mathbf{P}^{\prime} \mathbf{f}_{t} = \widetilde{\mathbf{\Lambda}} \widetilde{\mathbf{f}}_{t}, \label{ident_fac}
\end{equation}
where $ \widetilde{\mathbf{\Lambda}} = \mathbf{\Lambda} \mathbf{P}$ and $\widetilde{\mathbf{f}}_{t} = \mathbf{P}^{\prime} \mathbf{f}_{t}$. 

From an estimation perspective, the above equation shows that sampling of unique values of the factors in equation \eqref{factor_model} cannot be achieved without these additional $r(r-1)/2$ restrictions. From a structural perspective, this condition shows that the same number of restrictions is required for identification of the structural factors (shocks) in the VMA form of equation \eqref{VMA}. Consequently, under the assumption that $\mathbf{f}_{t} \sim N(\mathbf{0},\mathbf{I})$, placing restrictions on $\mathbf{\Lambda}$ ensures both unique estimation of factors and identification of the structural model. For instance, \cite{anderson1956} show that setting to zero all elements of $\mathbf{\Lambda}$ above the main diagonal achieves the $r(r-1)/2$ restrictions required for identification of the factors. Additionally, the diagonal elements of $\mathbf{\Lambda}$ can be normalized to be nonegative, such that the sign of the factors is also always identified. However, as shown in detail in \cite{StockWatson2005} numerous other identifying assumptions can be used in structural factor models.

In this paper interest lies in structural identification via sign (and possibly some zero) restrictions. However, many other restrictions can be incorporated in a straightforward way. For example, a researcher may want to impose that the impact response of variable $i$ to shock $j$ is lower in magnitude than the response of variable $k$ to the same shock. Due to the fact that impact restrictions on the IRFs are equivalent to imposing restrictions to $\mathbf{\Lambda}$ (see equation \eqref{IRF_cond}), this magnitude restriction can be represented as $\mathbf{\Lambda}_{ij}<\mathbf{\Lambda}_{kj}$. In the proposed model, a large class of desired structural restrictions can be represented using parametric inequalities that are imposed upon estimation of $\mathbf{\Lambda}$. Following \cite{Geweke1996}, these parametric inequality restrictions can be sampled efficiently using the Gibbs sampler. The next subsection derives such a Gibbs sampler algorithm, with particular focus on ensuring computational efficiency in high-dimensions.

\subsection{Gibbs sampler for sign and zero restrictions in reduced-form VARs}
Posterior sampling in the reduced-form VAR with factor structure in the residuals is straightforward due to the fact that posterior conditional distributions have standard forms. To see this, write the model using a single equation for convenience
\begin{equation}
\mathbf{y}_{t} = \mathbf{\Phi} \mathbf{x}_{t} +  \mathbf{\Lambda} \mathbf{f}_{t} + \mathbf{v}_{t}. \label{VAR_one_eq}
\end{equation}
Assume that all sign and zero restrictions in $\mathbf{\Lambda}$ are collected into a matrix $\mathbf{S}$, with entries $+1$ for positive signs, $-1$ for negative signs, $0$ for zero restrictions, and a missing value for no restriction (this case is denoted in this paper using the symbol \emph{NA}, and in the code using the MATLAB value $NaN$). The priors for the VAR parameters are of the form
\begin{eqnarray}
\bm{\phi}_{i} \equiv vec\left( \mathbf{\Phi}_{i} \right) & \sim & N_{k} \left( \mathbf{0}, \underline{\mathbf{V}}_{i} \right), \label{phi_prior} \\
\mathbf{f}_{t} & \sim & N_{r} \left( \mathbf{0}, \mathbf{I} \right), \\
\mathbf{\Lambda}_{ij}  & \sim & \left\lbrace \begin{array}{ll} N\left(0, \underline{h}_{ij} \right) I( \Lambda_{ij} > 0), & \text{if $S_{ij} = 1$}, \\
N\left(0,\underline{h}_{ij} \right) I( \Lambda_{ij} < 0),  & \text{if $S_{ij} = -1$}, \\
\delta_{0} \left( \mathbf{\Lambda}_{ij}\right),  & \text{if $S_{ij} = 0$}, \\
N\left(0,\underline{h}_{ij} \right),  & \text{otherwise},
\end{array} \right.  \label{lam_prior}\\
\sigma^{2}_{i} & \sim & inv-Gamma \left(\underline{\rho}_{i},\underline{\kappa}_{i} \right),
\end{eqnarray}
for $i=1,...,n$, $j=1,...,r$, where $\mathbf{\Phi}_{i}$ is the $i^{th}$ row of $\mathbf{\Phi}$, $\sigma_{i}^{2}$ is the $i^{th}$ diagonal element of the matrix $\mathbf{\Sigma}$, and $\delta_{0} \left( \mathbf{\Lambda}_{ij}\right)$ is the Dirac delta function for $\mathbf{\Lambda}_{ij}$ at zero (i.e. a point mass function with all mass concentrated at zero).

The joint posterior of the parameters is the distribution $p\left(\bm{\Phi}, \bm{\Lambda},\bm{F},\bm{\Sigma} \vert \bm{y}\right)  \equiv p\left(\lbrace\bm{\phi}_{i}\rbrace_{i=1}^{n},\lbrace \mathbf{\Lambda}_{i} \rbrace_{i=1}^{n},\lbrace \mathbf{f}_{t}\rbrace_{t=1}^{T}, \lbrace \sigma_{i} \rbrace_{i=1}^{n} \vert \mathbf{y} \right)$, which by Bayes theorem is the product of the normal likelihood function implied by equation \eqref{VAR_one_eq} and the prior distributions presented above. This product is a complicated function making sampling from the joint posterior infeasible. However, the conditional posteriors are tractable and trivial to derive in this linear model. Therefore, Bayesian inference in this VAR breaks down to sequentially sampling from the following conditional posterior distributions \medskip \\ 
\textbf{Factor-based structural restrictions (FSR) algorithm}
\begin{enumerate}
\item Sample $\bm{\phi}_{i}$ for $i=1,...,n$ from
\begin{equation}
\bm{\phi}_{i} \vert \mathbf{\Sigma},\mathbf{\Lambda}, \mathbf{f}, \mathbf{y} \sim N_{k} \left( \overline{\mathbf{V}}_{i} \left( \sum_{t=1}^{T} \sigma_{i}^{-2} \mathbf{x}_{t}^{\prime} \widetilde{\mathbf{y}}_{it} \right), \overline{\mathbf{V}}_{i} \right), \label{phi_post}
\end{equation}
where $\widetilde{\mathbf{y}}_{it} = \mathbf{y}_{it} - \mathbf{\Lambda}_{i} \mathbf{f}_{t}$ and $\overline{\mathbf{V}}_{i}^{-1} = \left( \underline{\mathbf{V}}_{i}^{-1} + \sum_{t=1}^{T} \sigma_{i}^{-2} \mathbf{x}_{t}^{\prime} \mathbf{x}_{t} \right)$.
\item Sample $\mathbf{\Lambda}_{i}$ for $i=1,...,n$ from
\begin{equation}
\mathbf{\Lambda}_{i} \vert \mathbf{\Phi},\mathbf{\Sigma}, \mathbf{f}, \mathbf{y}  \sim MTN_{\mathbf{a} < vec \left( \mathbf{\Lambda}\right) < \mathbf{b}} \left( \overline{\mathbf{H}}_{i} \left(\sum_{t=1}^{T}  \sigma_{i}^{-2} \mathbf{f}_{t}^{\prime} \widehat{\mathbf{y}}_{it} \right), \overline{\mathbf{H}}_{i} \right) , \label{lambda_post}
\end{equation}
where $\widehat{\mathbf{y}}_{it} \equiv \bm{\varepsilon}_{it} = \mathbf{y}_{it} - \mathbf{\phi}_{i} \mathbf{x}_{t}$, $\overline{\mathbf{H}}_{i}^{-1} = \left( \underline{\mathbf{H}_{i}}^{-1} + \sum_{t=1}^{T} \sigma_{i}^{-2} \mathbf{f}_{t}^{\prime}\mathbf{f}_{t} \right)$, and $\underline{\mathbf{H}}_{i} = diag\left(\underline{h}_{i1},...,\underline{h}_{ir} \right)$. Here we define $MTN\left( \bullet \right)$ to be the multivariate truncated Normal distribution, and $\mathbf{a},\mathbf{b}$ are the vectors indicating the truncation points, with $ij^{th}$ element:
\begin{equation}
(\mathbf{a}_{ij},\mathbf{b}_{ij}) = \left \lbrace \begin{array}{ll}
(-\infty,0) & \text{if $S_{ij} = -1$}, \\
(0, \infty) & \text{if $S_{ij} = 1$}, \\
(0,0) & \text{if $S_{ij} = 0$}, \\
(- \infty, \infty) & \text{otherwise,} 
\end{array} \right. 
\end{equation}
for $i=1,...,n$, $j=1,...,r$.
\item Sample $\mathbf{f}_{t}$ for $t=1,...,T$ from
\begin{equation}
\mathbf{f}_{t} \vert \mathbf{\Lambda}, \mathbf{\Sigma}, \mathbf{\Phi}, \mathbf{y} \sim N\left(\overline{\mathbf{G}} \left( \mathbf{\Lambda} \mathbf{\Sigma}^{-1} \widehat{\mathbf{y}}_{t} \right), \overline{\mathbf{G}} \right),
\end{equation}
where $\overline{\mathbf{G}}^{-1} = \left( \mathbf{I}_{r} + \mathbf{\Lambda}^{\prime} \mathbf{\Sigma} \mathbf{\Lambda} \right)$.
\item Sample $\sigma_{i}^{2} $ for $i=1,...,n$ from
{\footnotesize
\begin{equation}
\sigma_{i}^{2} \vert \mathbf{\Lambda}, \mathbf{f}, \mathbf{\Phi}, \mathbf{y} \sim inv-Gamma \left( \frac{T}{2} + \underline{\rho}_{i}, \left[ \underline{\kappa}_{i}^{-1} + \sum_{t=1}^{T} \left( \mathbf{y}_{it} - \mathbf{\phi}_{i} \mathbf{x}_{t} - \mathbf{\Lambda}_{i} \mathbf{f}_{t} \right)^{\prime} \left( \mathbf{y}_{it} - \mathbf{\phi}_{i} \mathbf{x}_{t} - \mathbf{\Lambda}_{i}\mathbf{f}_{t} \right) \right]^{-1} \right)
\end{equation}
}
\end{enumerate}

Step 1 is efficient because autoregressive coefficients can be sampled equation-by-equation. Further speed enhancements can be achieved by using the sampling methodology of \cite{Bhattacharya2016} in the case where the prior covariance matrices $\underline{\mathbf{V}}_{i}$ are diagonal. This is the case here, as $\underline{\mathbf{V}}_{i}$ is diagonal and its elements follow the horseshoe hierarchical structure of \cite{Carvalhoetal2010} which has the form
\begin{eqnarray}
\bm{\phi}_{i} \vert  \sigma^{2}_{i} \tau_{i}^{2} \mathbf{\Psi}_{i} & \sim & N_{k} \left( \mathbf{0}, \underline{\mathbf{V}}_{i} \right), \\
\underline{\mathbf{V}}_{i,(jj)} & = & \sigma^{2}_{i} \tau_{i}^{2} \psi_{i,j}^{2}, \\
\psi_{i,j} & \sim & Cauchy^{+} \left(0, 1 \right),  \\
\tau_{i}  & \sim & Cauchy^{+} \left(0, 1 \right).
\end{eqnarray}
This prior belongs to the class of \emph{local-global} shrinkage priors, that is a prior that penalizes the likelihood and shrinks coefficients towards zero. In this prior, $\psi_{i,j}$ is the local shrinkage parameter for each scalar coefficient $\bm{\phi}_{i,j}$ while $\tau_{i}$ is the global shrinkage parameter pertaining to all coefficients in equation $i$. Unlike the popular (in macroeconomic VARs) Minnesota prior that typically requires subjective tuning \citep{KoopKorobilis2010}, the horseshoe prior is tuning-free as the local and global shrinkage parameters have their own distributions and are, thus, updated by information in the data. The horseshoe prior has established posterior consistency properties when used in a variety of high-dimensional settings\footnote{See online Appendix for more details and citations.} making it an appropriate choice for penalized estimation in both smaller and higher dimensional VARs. Sampling from the truncated Normal posteriors in step 2 of the algorithm above can be done using the recent contribution of \cite{Botev2017}.\footnote{In an ideal world one would want to sample the vector $\mathbf{\Lambda}_{i}$ in one step and unconditionally from the factors $\mathbf{f}_{t}$, in order to reduce correlation in the MCMC chain. In practice, I follow \citep{Geweke1996} and sample each element $\mathbf{\Lambda}_{ij}$ conditional on $\mathbf{\Lambda}_{ij}$, as this conditioning allow for the algorithm to be extended more easily (e.g. if structural breaks or time-varying loadings are required). This comes at the cost of thinning the Gibbs chain by a factor of 100, that is, every 100-th sample is stored in order to ensure posterior samples are uncorrelated. In all results in this paper the Gibbs chain runs for 550,000 iterations where the first 50,000 iterations are discarded and from the final 500,000 iterations I store every 100th iteration, leaving a total of 5,000 samples from the parameter posterior for inference.} Finally, sampling of the factors $\mathbf{f}_{t}$ for each $t=1,...,T$ is fairly fast for monthly or quarterly macroeconomic data, as is sampling of the scalar parameters $\sigma_{i}^{2}$ for each $i=1,...,n$. Computational details and further discussion on the excellent properties of the horseshoe prior are provided in the online Appendix.

Given that the horseshoe prior requires no subjective tuning, there are only a handful of prior hyperparameters in the whole VAR that need to be elicited by the researcher. The parameter affecting primarily structural identification, is the choice of the prior variance $\underline{h}_{ij}$, regardless of whether the associated loading parameter $\mathbf{\Lambda}_{ij}$ should be truncated (sign restricted) or not. Due to the fact that the coefficients $\mathbf{\Lambda}$ enter the VAR in a linear way, it turns out that a fairly large value of $\underline{h}_{ij}$ implies a diffuse prior that does not impact the posterior asymptotically. Therefore, for the remainder of this paper, I set $\underline{h}_{ij} = 100$ which is a fairly diffuse choice for the typical scale of variables encountered in macroeconomic VARs. Finally, I set $\rho_{i}=1$ and $\kappa_{i}=0.01$ for all $i$, leading to a prior mean of 0.16 and substantially large prior variance, reflecting the belief that the scale of the idiosyncratic variances $\sigma_{i}^{2}$ should be small and most of variability in the VAR disturbances should come from the common component $\mathbf{\Lambda} \mathbf{f}_{t}$. Model fit in this VAR can be assessed with the Deviance Information Criterion (DIC) of \cite{Spiegelhalteretal2002}, due to its simplicity of implementation. The formula and justification for the use of the DIC is provided also in the online Appendix. What suffices to remember is that the DIC has the same interpretation as any other information criterion, that is, lower values signify better fit. The DIC can be used to test any kind of parametric restrictions that are of interest, whether it pertains to lag length selection, number of structural shocks, linearities vs nonlinearities, and so on.

The most important computational aspect of the new algorithm is that samples from the restricted $\mathbf{\Lambda}$ matrices are always accepted, making it very efficient in high-dimensions. This feature is in contrast with a large class of accept/reject algorithms used especially for sign restrictions in VARs; see for example the Bayesian algorithms of \cite{RubioRamirezetal2010} and \cite{BaumeisterHamilton2015} as well as the algorithm of \cite{OuliarisPagan2017}. In \cite{RubioRamirezetal2010}, for example, one has to first obtain posterior samples from the VAR covariance matrix $\mathbf{\Omega}$ and then rotate its Cholesky factor $\mathbf{P}$ using randomly generated orthogonal matrices $\mathbf{Q}$. If the random rotation $\mathbf{H} = \mathbf{P}\mathbf{Q}$ satisfies the required sign restrictions then $\mathbf{H}$ is a draw from the desired matrix of contemporaneous structural relationships. Inevitably, such an accept/reject scheme is deemed to fail in high-dimensions, when the desired restrictions may be so tight that no sample of $\mathbf{H}$ can be accepted by random chance, see also the discussion in Section 13.6.4 of \cite{kilianlutkepohl2017}. Many authors express the belief that when the accept/reject algorithm results to a high rejection rate, this is evidence that identification is sharp.\footnote{See for example ``principle 7'' in \cite{Uhlig2017} and the corresponding discussion.} However, this premise is not testable in a statistical sense, and can be misleading especially in high dimensions: a high rejection rate could either be because of the researcher imposing too many restrictions, or because of imposing restrictions that simply do not comply with the evidence in the data.

\section{Simulation studies}
\subsection{Numerical evaluation of the new algorithm}
In this section, the properties of the new algorithm are explored using artificially generated data. The core exercise involves generating multivariate time series from a data generating process (DGP) that fully matches equation \eqref{VAR_one_eq}, and estimating parameters and impulse response functions based on time series generated from this DGP. I first implement this experiment assuming that a correctly specified model is estimated using the artificial data. Subsequently, various cases of misspecification errors during the estimation process are considered -- that is, I estimate models that do not perfectly match the correct DGP. 

The DGP is of the form
\begin{eqnarray}
\mathbf{y}_{t} & = & \widehat{\mathbf{\Phi}} \mathbf{x}_{t} +  \widehat{\mathbf{\Lambda}} \mathbf{f}_{t} + \mathbf{v}_{t}, \text{ for } t=1,...,\widehat{T}, \label{DGP1}\\
\mathbf{v}_{t} & \sim & N \left( \mathbf{0}, \widehat{\mathbf{\Sigma}} \right), \text{ \ } \mathbf{f}_{t}  \sim  N\left( \mathbf{0}, \mathbf{I} \right), \label{DGP2}\\
\mathbf{y}_{(-p+1):0} & = & \mathbf{0}, \text{ \ } p=12, \text{ \ } r=3. \label{DGP3}
\end{eqnarray}
The DGP parameters $\widehat{\mathbf{\Phi}}$, $\widehat{\mathbf{\Lambda}}$, $\widehat{\mathbf{\Sigma}}$ are based on estimates of a VAR on real data. First, monthly data on 14 monthly macroeconomic variables are collected\footnote{The variables are: 1) real GDP, 2) GDP deflator, 3) federal funds rate, 4) commodity price index, 5) total reserves, 6) nonborrowed reserves, 7) S\&P 500, 8) M1 , 9) unemployment rate , 10) industrial production , 11) employment , 12) CPI , 13) core CPI , 14) core PCE. More details on these variables is provided in the online Appendix.} for the US over the period 1965M1 - 2007M12, providing $\widehat{T}=516$ observations.\footnote{In practice, I generate $\widehat{T} + 1000$ observations and discard the first 1000 observations in order to remove dependence to the initial values of the generated time series process.} At a second step, an estimate $\widehat{\mathbf{\Phi}}$ is obtained by applying OLS to an unrestricted VAR($12$) estimated with these 14 observed US variables. The third step is to obtain the first $r=3$ principal components of these OLS residuals, and store the estimate $\widehat{\mathbf{\Lambda}}$ using OLS in a regression between the VAR residuals and their principal components. Finally, the residuals from this latter regression provide the elements of the diagonal matrix $\widehat{\mathbf{\Sigma}}$, by means of equation-by-equation application of the usual least squares formula for the variance.

While it is not possible, or even interesting, to display all estimates $\widehat{\mathbf{\Phi}}$ used as input in the DGP, it is instead interesting to look at the estimates $\widehat{\mathbf{\Lambda}}$ obtained using the procedure described above. This is because both the signs and the magnitudes of the implied IRFs in the true DGP will be affected by those estimates. Panel (A) of \autoref{table:lambda_dgp} shows the OLS estimates, where the diagonal is normalized to be one, by dividing each element in the $m^{th}$ column of $\widehat{\mathbf{\Lambda}}$ with the original value of its $m^{th}$ element, $m=1,2,3$. While this matrix is the outcome of using real data and applying simple principal components followed by OLS estimates (which carry no structural restrictions), looking at the signs of the loadings of the first three variables (output, inflation, interest rate) allows for the classification of the three pseudo-shocks as aggregate supply, aggregate demand, and monetary policy, respectively. The estimated magnitudes, of course, are not necessarily economically meaningful. For example, in the first column a shock of $1\%$ in GDP increases inflation by $1.49\%$, which is probably not a representative magnitude for a true aggregate supply shock. Nevertheless, this is an exercise where the main aim is to check the numerical precision of the new algorithm, so the estimates in panel (A) of \autoref{table:lambda_dgp} are perfectly valid inputs for a DGP. Finally, panel (B) of \autoref{table:lambda_dgp} shows the sign restrictions imposed on $\mathbf{\Lambda}$, that is the matrix $\mathbf{S}$ introduced in equation \eqref{lam_prior}. These comply with the signs imposed in the DGP, and in 11 instances no sign restrictions are imposed (these entries are denoted as \emph{NA}). The choice of which signs are known in $\mathbf{S}$ is random, and the next subsection looks at varying assumptions about how many sign restrictions are known to the researcher.

\vskip 1cm
\begin{table}[H]
\centering
\caption{\emph{OLS estimates $\widehat{\mathbf{\Lambda}}$ used in the DGP, and sign restrictions $\mathbf{S}$ used for estimation}}  \label{table:lambda_dgp}
\resizebox{0.9\textwidth}{!}{ 
\begin{tabular}{lccccccc} \hline
& \multicolumn{3}{c}{\textsc{(a) True parameter values}} &  & \multicolumn{3}{c}{\textsc{(b) Sign restrictions}} \\
 Variable &  1$^{st}$ shock	&	2$^{nd}$ shock	&	3$^{rd}$ shock & &	1$^{st}$ shock	&	2$^{nd}$ shock	&	3$^{rd}$ shock \\ \hline\hline
real GDP growth	&	1.00	&	-1.39	&	-0.87	&		&	+	&	--	&	--	\\
GDP deflator inflation 	&	1.42	&	1.00	&	-0.71	&		&	+	&	+	&	--	\\
Fed funds rate	&	0.49	&	-0.28	&	1.00	&		&	NA	&	NA	&	+	\\
Commodity prices	&	0.16	&	0.16	&	-0.45	&		&	NA	&	NA	&	--	\\
Total reserves	&	-0.61	&	0.22	&	-3.48	&		&	NA	&	NA	&	NA	\\
Nonborrowed reserves	&	-0.91	&	0.25	&	-3.37	&		&	NA	&	NA	&	--	\\
Stock prices	&	-0.25	&	-0.30	&	-0.82	&		&	NA	&	NA	&	--	\\
M1	&	-1.03	&	-0.48	&	-1.27	&		&	--	&	--	&	--	\\
Unemployment	&	-0.63	&	0.51	&	0.43	&		&	--	&   +	&	+	\\
Industrial production	&	1.12	&	-1.34	&	-0.87	&		&	+	&	--	&	--	\\
Employment	&	0.88	&	-1.00	&	-1.01	&		&	+	&	--	&	--	\\
CPI inflation (total)	&	1.44	&	1.01	&	-0.75	&		&	+	&	+	&	--	\\
CPI inflation (core)	&	1.05	&	0.49	&	-1.12	&		&	+	&	+	&	--	\\
PCE inflation (core)	&	1.05	&	0.57	&	-0.80	&		&	+	&	+	&	--	\\
 \hline
\par
\multicolumn{8}{p{\dimexpr 1.05\linewidth}}{{\small Notes: Panel (A) shows true parameter values used as input in the data generating process (DGP), while panel (B) shows the sign restrictions imposed during econometric estimation using each artificial dataset from the DGP. Entries in panel (B) show the restrictions imposed: + for positive sign; -- for negative sign; NA for no restriction.  }} 
\end{tabular}
}
\end{table}

For estimation purposes five different scenarios are assumed: one correctly specified case and four misspecified cases. These are denoted as C1-C5, and are defined as follows:
\begin{itemize}
\item[\textbf{C1}] Correctly specified model with $n=14$ dependent variables, $p=12$ lags, $r=3$ shocks.
\item[\textbf{C2}] Misspecified model with $n=8$, using the first eight variables in \autoref{table:data_mc} in the online Appendix. All other settings are correct, that is, $p=12$ lags, $r=3$ shocks.
\item[\textbf{C3}] Misspecified model with $p=2$ lags. All other settings are correct, that is, $n=14$ and $r=3$.
\item[\textbf{C4}] Misspecified model with $r=2$ shocks, using only the restrictions on the first two shocks in panel (B) of \autoref{table:lambda_dgp}. All other settings are correct, that is, $n=14$ and $p=12$.
\item[\textbf{C5}] Misspecified model with $r=4$ shocks, using an additional shock.\footnote{This fourth shock is identified using the randomly  selected vector of restrictions $s = [+,+,+,-,+,NA,NA,-,+,+,+,+,+,+]$.} All other settings are correct, that is, $n=14$ and $p=12$.
\end{itemize}
500 datasets of size $T=516$ are generated and posterior mean estimates of all parameters, IRFs and DICs from all five cases above are obtained. Results presented next are based on the distribution of the posterior means over these 500 generated datasets.

Before evaluating precision of estimates over the Monte Carlo iterations, it is important to first evaluate general model fit using the DIC. \autoref{table:dic_mc} shows the value of the deviance information criterion attained by the estimates of the VAR model in each of the five specification cases. Because case C2 refers to a VAR with $n=8$, it is impossible to directly compare it with the other four cases that assume $n=14$.\footnote{Information criteria can only be used to compare models with the same dependent variable $\mathbf{y}$.} For that reason I present two DIC metrics, a full one based on all $n=14$ variables (with no value available for C2) and a reduced DIC which is the same formula evaluated only in the first eight VAR equations (which are common to all five cases). These are labeled in \autoref{table:dic_mc} as $DIC_{14}$ and $DIC_{8}$, respectively. According to both subsets of criteria, the correctly specified estimated model, case C1, is the best one as it attains the lowest DIC value. Interestingly, the case where an additional fourth shock is incorrectly estimated (C5), doesn't seem to harm estimation accuracy; at least not as much as the case of estimating one less shock (C4). This result makes sense because shocks in the proposed VAR are equivalent to factors. By far the worst type of misspecification seems to be the one related to the lag-length. This is a characteristic of the VAR model rather than a ``problem'' with the specific algorithm or prior. As long as the true DGP has $p=12$ important lags, estimating the VAR with $p=2$ provides a huge loss of structurally meaningful information. In contrast, reducing the VAR from $n=14$ (which is the truth in the DGP) to $n=8$ as in case C2 is less harmful for the general fit of the model. This is probably because the missing six variables are additional measures of output and prices, that do not offer more information compared to the real GDP and GDP deflator variables that are included in both the eight- and fourteen-variable VARs.

\vskip 0.5cm
\begin{table}[H]
\centering
\caption{\emph{DIC values attained by the correctly specified and misspecified models estimated on artificial data}} \label{table:dic_mc}
\resizebox{0.8\textwidth}{!}{
\begin{tabular}{lccccc} \hline
  & C1 & C2 & C3 & C4 & C5 \\ \hline\hline
$DIC_{14}$ value & 7393.14 & n/a & 27241.13 & 11859.45 & 9037.44 \\
$DIC_{8}$ value &  15300.63 & 17094.28 & 28630.21 & 27981.48 & 19269.55 \\ 
\hline
\par
\multicolumn{6}{p{\dimexpr 0.8\linewidth}}{{\small Notes: $DIC_{14}$ is the deviance information criterion applied jointly to all 14 VAR equations. $DIC_{8}$ is the same criterion applied jointly only to the first 8 VAR equations. Case C2 does not have a $DIC_{14}$ value because it assumes that the VAR has $n=8$ variables.  }} 
\end{tabular}
}
\end{table}

\begin{figure}[H]
\centering
\includegraphics[trim= 2cm 1cm 1cm 0cm, width=\textwidth]{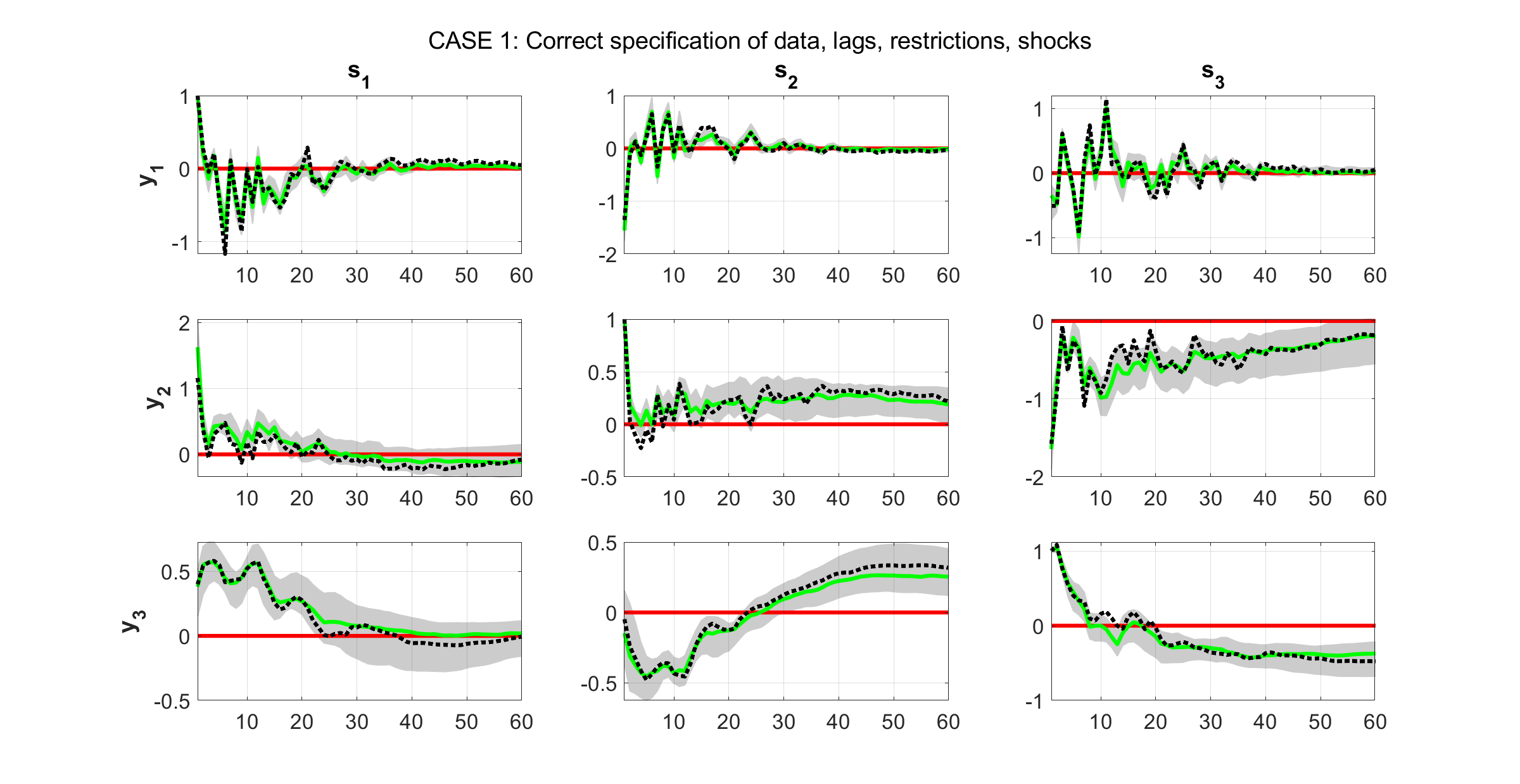}
\caption{\emph{Impulse response functions of the first three artificially generated variables (denoted as $y_{1},y_{2},y_{3}$) in response to the three identified shocks (denoted as $s_{1},s_{2},s_{3}$) in model C1 (correctly specified model). The green solid lines show the posterior median IRFs over the 500 Monte Carlo iterations, and the gray shaded areas their associated 90\% bands. The true IRFs based on the DGP are shown using the black dashed lines.}} \label{fig:irfc1_mc}
\end{figure}

Next, estimation accuracy of the proposed algorithm has to be evaluated. Since the main focus of sign restrictions algorithms is on impulse response analysis, I compare precision of the estimated impulse response functions using all generated datasets. IRFs are combinations of all VAR parameters $\mathbf{\Phi},\mathbf{\Lambda},\mathbf{f},\mathbf{\Sigma}$, therefore comparing their precision provides a convenient summary of overall estimation precision in a VAR model. \autoref{fig:irfc1_mc} shows the responses of the first three variables in the VAR to the three identified pseudo-shocks, in the correctly specified case (C1). Green solid lines are medians over the posterior IRFs in the 500 estimated VARs using an equal number of artificial datasets. Shaded areas show the 90\% probability bands of these IRFs. Finally, black dashed lines show the true IRFs implied by the parameters that are fed into the DGP. The 90\% bands always include the true IRF, which suggests that estimation precision is satisfactory. The online Appendix shows identical graphs for the four misspecified cases C2-C5. These graphs become a visual confirmation of the numerical results in \autoref{table:dic_mc}, that is, case C5 quite precisely captures the path of the true IRFs, while case C3 results in the largest estimation errors.
\vskip 0.5cm

\subsection{How fast is the new algorithm?} \label{sec:how_fast}
The next Section makes clear that in the context of the empirical application in \cite{Furlanettoetal2017}, the new algorithm is multiple times faster than the algorithm of \cite{RubioRamirezetal2010} in a six-variable VAR with five identified shocks. Nevertheless, it would be interesting to use artificial data in order to provide more thorough evidence on how fast the factor sign restrictions algorithm is, and how large a VAR it can scale to. For that reason artificial data are generated from the same DGP described in equations \eqref{DGP1}-\eqref{DGP2} for various values of the key parameters that affect the dimensionality of the VAR, namely $T$, $n$ and $r$. Due to the fact that this exercise pushes the VAR dimension $n$ to very large values, I fix $p=1$ in order to be able to ensure that the VAR process in the DGP is always stationary, and generation of explosive data is excluded. For the purposes of this exercise I set $\mathbf{\Phi} = 0.9\mathbf{I}_{n}$, $\mathbf{\Lambda}_{ij} \sim U(-1,1)$ and $\Sigma_{i} \sim U(0,1)$, for all $i=1,...,n$ and $j=1,...,r$. During estimation $nk$ sign restrictions are imposed, simply by obtaining the signs of the randomly generated matrix $\mathbf{\Lambda}$.\footnote{The purpose of this exercise is not to estimate meaningful restrictions, rather just to measure times. In this case, I impose the maximum number of restrictions possible on $\mathbf{\Lambda}$ in order to test the new algorithm in a worst-case scenario where all $nk$ of its elements are restricted and have to be generated from a truncated Normal posterior.}

\begin{table}[H]
\centering
\caption{\emph{Computer time in minutes (defined as $(seconds/60)$, rounded to the nearest integer) for obtaining $10,000$ post-burn-in draws ($12,000$ in total) using various VAR sizes. Here $T$ is the number of observations, $n$ the number of endogenous variables, and $r$ the number of shocks. All VARs have $p=1$ lag.}} \label{table:times_mc}
\resizebox{0.9\textwidth}{!}{ 
\begin{tabular}{lccccccc} \hline
	&		&	$T=200$	&		&		&		&	$T=500$	&		\\
	&	$n=15$	&	$n=50$	&	$n=100$	&		&	$n=15$	&	$n=50$	&	$n=100$	\\ \hline \hline
$r=3$	&	1	&	2	&	6	&	\hspace{1cm}	&	4	&	11	&	23	\\
$r=10$	&	1	&	4	&	8	&	\hspace{1cm}	&	4	&	12	&	23	\\
$r=20$	&	NA	&	6	&	11	&	\hspace{1cm}	&	NA	&	15	&	25	\\ \hline
\end{tabular}
}
\end{table}

\autoref{table:times_mc} shows the average, over 10 Monte Carlo iterations, machine time in minutes (defined as the total estimation time in seconds divided by 60 and then rounded to the nearest integer) needed to obtain $10,000$ draws from the posterior of all parameters after discarding $2,000$ draws (hence, $12,000$ draws in total). These results show that in a huge-dimensional VAR with $n=100$ series, $T=500$ observations, and $r=20$ shocks, it only takes 25 minutes to obtain $10,000$ draws from all parameter matrices, including the $1000$ sign-restricted elements in $\mathbf{\Lambda}$. For the smaller model with $n=15$ -- which is already much larger than the vast majority of models considered in the sign restrictions literature -- it only takes less than five minutes to obtain the same number of draws when $T=500$, and only one minute when $T=200$. These fantastic timings justify the choice to focus on carefully developing a Gibbs sampler that is computationally efficient.\footnote{The Gibbs sampler typically loses efficiency when there is high correlation in the samples from the posterior. In the online Appendix I show that, in order to draw $\mathbf{\Lambda}_{ij}$ from univariate (instead of the intractable multivariate) truncated Normal conditionals, we need to condition on $\mathbf{\Lambda}_{-ij}$, i.e. the set of all elements of $\mathbf{\Lambda}$ excluding the $ij^{th}$. This conditioning increases correlation relative to sampling directly the full matrix $\mathbf{\Lambda}$. However, inefficiency factors for the Gibbs sampler in the linear factor-VAR specification are still quite low (MCMC diagnostic results are available upon request). Additionally, given the ability of the algorithm to obtain quickly tens of thousands of draws from the posterior, concerns about possible correlation of draws can be alleviated by doing ``thinning'' -- i.e. the procedure of storing only every $\rho^{th}$ sample from the posterior, where is $\rho$ is the order of the highest significant autocorrelation in the chain.}

The results above are based on code written in MATLAB2019b and run in a personal computer with Intel Core i7 8700K, tuned at 4.9Ghz, and 32GB of RAM. Note that the Gibbs sampler algorithm iterates over each VAR equation independently and, thus, significant speed improvements can be achieved by taking advantage of parallel processing abilities of modern computers and high-performance clusters (HPCs). In MATLAB this is as simple as replacing \emph{for loops} with \emph{parfor loops}. Therefore, the algorithm indeed allows the estimation of arbitrarily large VAR models, as it is claimed in the Introduction.

In practical situations, the only issue that might inhibit the performance of the algorithm (and any Monte Carlo-based algorithm, to that effect) is the fact that in very large dimensions we may be sampling parameters $\mathbf{\Phi}$ in a region of the posterior that implies nonstationarity of the VAR. In order to make sense out of impulse response functions, forecast error variance decompositions, historical decompositions etc, we need to make sure we maintain only samples from the posterior which are stationary. For that reason it is important to stress that, throughout my experiments, the horseshoe prior does a great job (especially relative to a subjectively chosen Minnesota prior) in shrinking the coefficients $\mathbf{\Phi}$ towards a more numerically stable region of their posterior, where the VAR model is stationary.

\section{Empirics: Financial factors in economic fluctuations}

In this section I revisit the empirical exercise in \cite{Furlanettoetal2017}, who aim to measure various financial shocks to the US economy.\footnote{The online Appendix provides the results of an additional numerical exercise (\emph{measuring optimism shocks}) that builds on \cite{Ariasetal2018}.} Given computational restrictions, due to their use of the \cite{RubioRamirezetal2010} accept/reject algorithm, \cite{Furlanettoetal2017} end up estimating a series of smaller VARs in order to sequentially measure and label interesting financial shocks, such as uncertainty, credit and housing. Before illustrating how to use the new algorithm to collectively measure all these shocks in one high-dimensional data setting, I first replicate their benchmark results using a smaller VAR. That way, the new algorithm can be contrasted against the output of the \cite{RubioRamirezetal2010} algorithm, something that is not computationally feasible in the large VAR case.

\subsection{Financial shocks using a baseline VAR specification} \label{sec:baseline_fur}
Among all VAR specifications they use in their work, \cite{Furlanettoetal2017} specify a \emph{baseline} VAR specification with $p=5$ lags, using data on real GDP, consumer prices, interest rate, investment-to-output ratio, stock prices, and the external finance premium.\footnote{The external finance premium is defined as the spread between yields on Baa rated bonds and the federal funds rate. Notice that the three variables that are not already expressed as rate, ratio, or spread (i.e. GDP, consumer prices, and stock prices), are transformed only using logarithms of the levels and not growth rates. Also note that these authors use a noninformative (uniform) prior, while I use the shrinkage horseshoe prior described in Section 2.} All data are for the 1985Q1 - 2013Q2 period. The online Appendix provides exact details of all series and transformations used, which in this case they are identical to those reported in Table 11 of \cite{Furlanettoetal2017}. Five shocks in total are identified by the authors using the six-variable baseline VAR. The names of the shocks and the associated sign restrictions adopted are shown in \autoref{table:sign_baseline}. The first four shocks are standard macro-related shocks, and of interest is the fifth shock which is a generic financial sector shock.
\begin{table}[H]
\centering
\caption{\emph{Identified shocks and sign restrictions imposed on the matrix $\mathbf{\Lambda}$ in the baseline, six-variable, five-shock VAR model.}} \label{table:sign_baseline}
\resizebox{0.7\textwidth}{!}{ 
\begin{tabular}{rccccc} \hline
 & \multicolumn{5}{c}{\textsc{Identified Shocks}} \\
  	&	Supply	&	Demand	&	Monetary	&	Investment	&	Financial \\ \hline\hline
GDP	&	+	&	+	&	+	&	+	&	+	\\
prices	&	--	&	+	&	+	&	+	&	+	\\
interest rate	&	NA	&	+	&	--	&	+	&	+	\\
investment/output	&	NA	&	--	&	NA	&	+	&	+	\\
stock prices	&	+	&	NA	&	NA	&	--	&	+	\\
spread	&	NA	&	NA	&	NA	&	NA	&	NA	\\ \hline
\par
\multicolumn{6}{l}{{\emph{Notes: Entries in this table show the restrictions imposed: + positive sign;}}} \\ \multicolumn{6}{l}{{\small \emph{-- negative sign; NA no restriction.}  }}
\end{tabular}
}
\end{table}

\begin{figure}[H]
    \centering
    \begin{subfigure}[t]{0.45\textwidth}
        \centering
        \includegraphics[trim= 3cm 0cm 3cm 0cm,width=\linewidth]{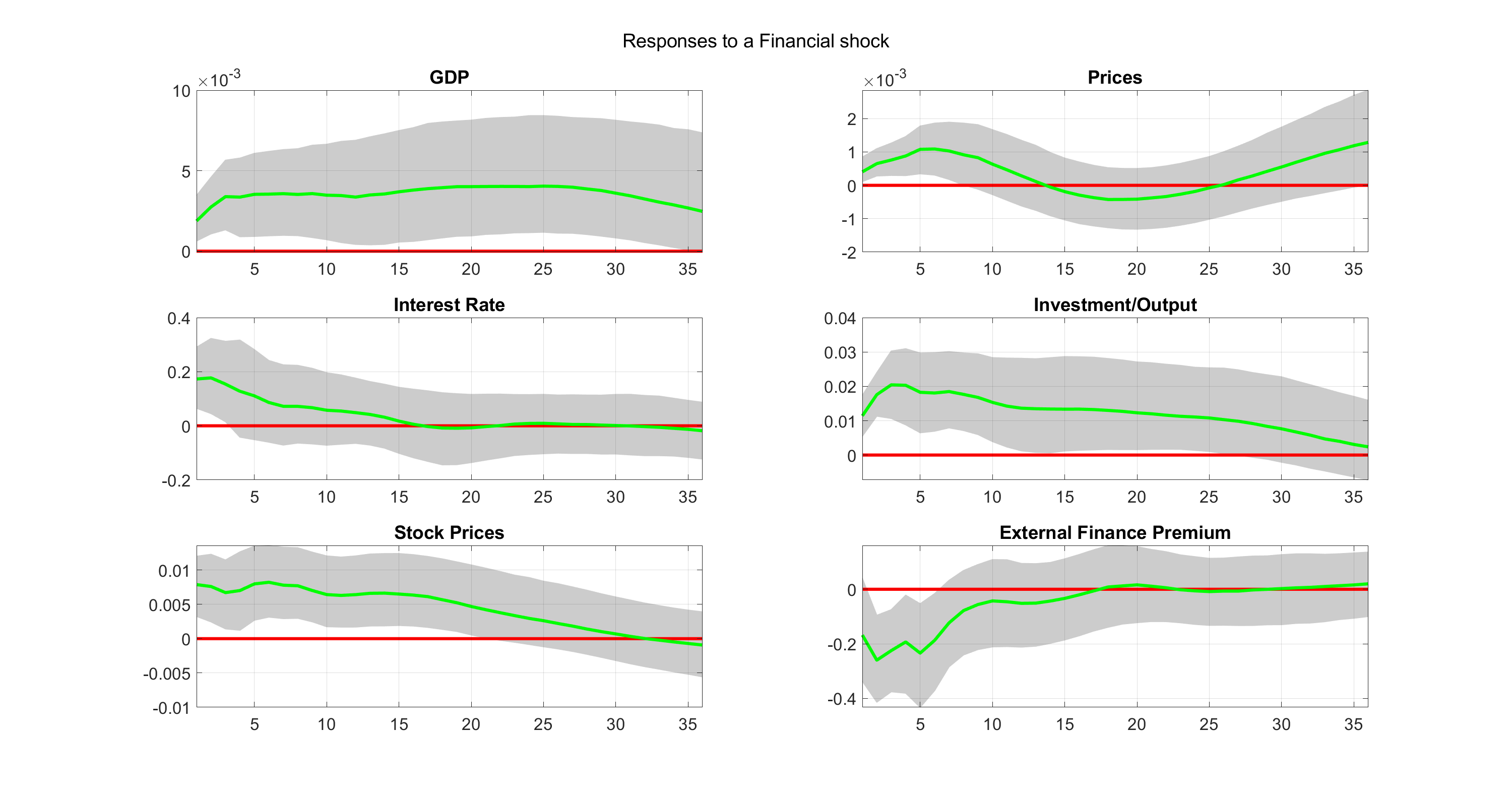}
        \caption{\cite{RubioRamirezetal2010} algorithm}
    \end{subfigure}
    \hfill
    \begin{subfigure}[t]{0.45\textwidth}
        \centering
        \includegraphics[trim= 3cm 0cm 3cm 0cm,width=\linewidth]{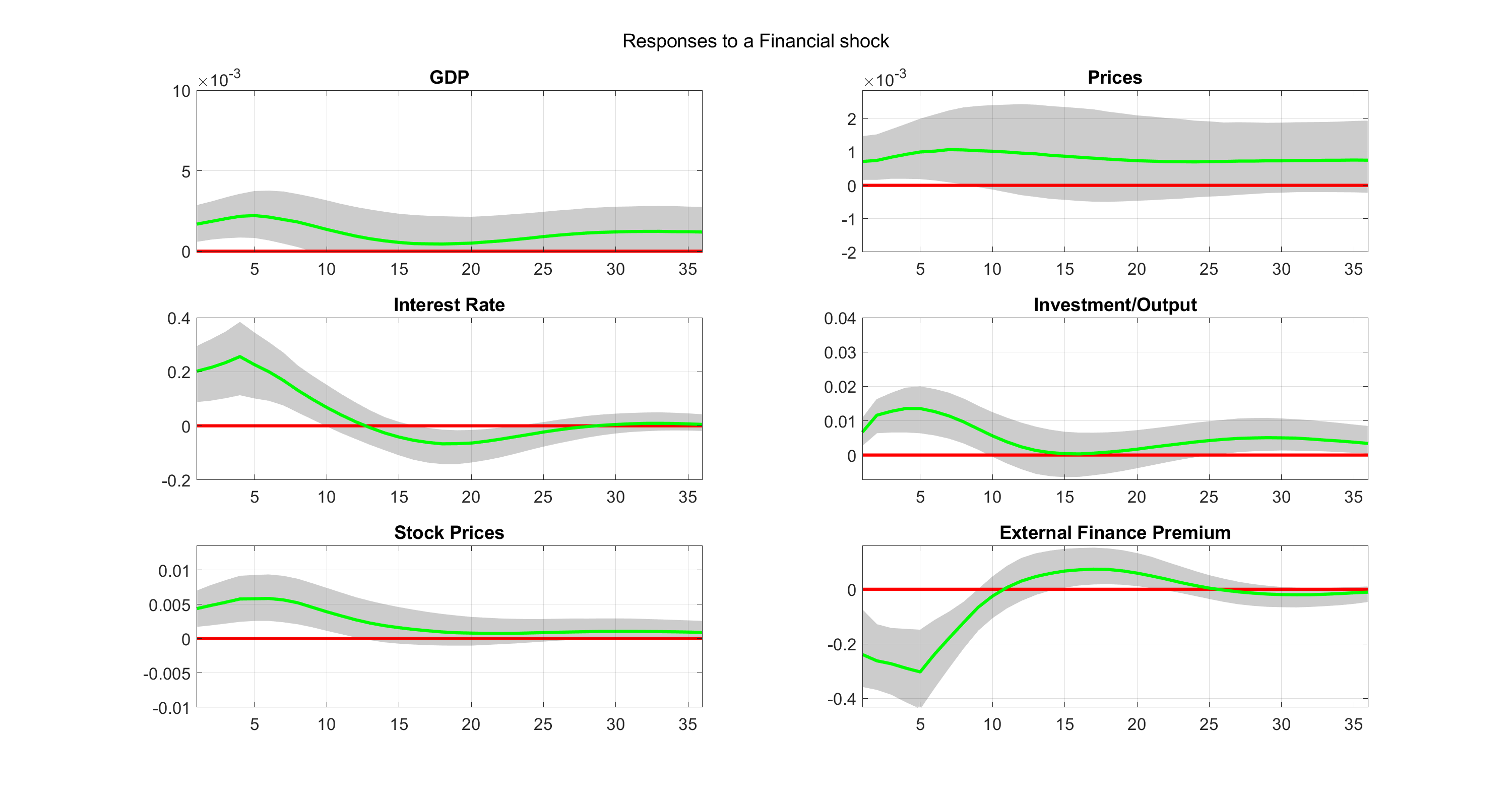}  
        \caption{Factor-based sign restrictions algorithm}
    \end{subfigure}
\caption{\emph{This figure replicates the impulse response functions (IRFs) to a financial shock using the baseline specification of \cite{Furlanettoetal2017}. Panel (a) shows results based on the exact configuration of \citet[see Figure 1]{Furlanettoetal2017}, using the algorithm of \cite{RubioRamirezetal2010}. Panel (b) replicates the same financial shock using the new sign restrictions algorithm. Solid lines are the posterior medians of the IRFs, and shaded areas the $68\%$ posterior bands.}} \label{fig:financial_benchmark}
\end{figure}

\autoref{fig:financial_benchmark} shows the effects of a financial shock identified as a shock that causes GDP, consumer prices, stock prices, interest rate and the investment/output ratio to react positively contemporaneously. The sign of the spread is left unrestricted. Panel (a) replicates the impulse responses also shown in Figure 1 of \cite{Furlanettoetal2017}, produced using the algorithm of \cite{RubioRamirezetal2010}. Panel (b) shows the same responses produced by application of the new algorithm for sign restrictions. The responses on impact in both panels are of almost identical magnitude, showing that the new algorithm produces sensible results. Any observed differences in the propagation of the impulse responses in subsequent periods, especially for GDP, prices and investment/output ratio, is due to the different estimates of the autoregressive coefficients (\cite{Furlanettoetal2017} use noninformative priors).\footnote{The online Appendix replicates \autoref{fig:financial_benchmark} by exchanging the horseshoe prior for a diffusing (flat) prior. This can be done by simply droping the local-global hyperparameters of the horseshoe and, instead, setting $\mathbf{V}_{i} = c \times \mathbf{I}$ for $c \rightarrow \infty$. In this case, the normal prior becomes locally uniform on the parameter support. \autoref{baseline_noninformative} in the online Appendix reveals that in the case of this noninformative prior the shapes of the shocks between the two algorithms become identical. However, the error bands in the new algorithm are still sharper. This is because \cite{Furlanettoetal2017} sample the VAR covariance matrix from the standard inverse Wishart posterior \citep{KoopKorobilis2010}, while the new algorithm samples the covariance matrix from the more parsimonious factor model.} Following up on the discussion in the previous section, it takes roughly four hours to obtain 2000 draws from the \cite{Furlanettoetal2017} using their MATLAB code and exact numerical settings based on the \cite{RubioRamirezetal2010} algorithm. In contrast, using the same PC\footnote{Specifications of the PC are reported in \autoref{sec:how_fast}.} it takes 20 minutes to obtain $600,000$ draws from the proposed Gibbs sampler (where out of these $600,000$ draws I discard $100,000$ and then save every $100^{th}$ draw, leading to $5,000$ draws from the posterior of VAR parameters and impulse response functions). Similar conclusions can be made for all other shocks in the system (supply, demand, monetary, investment), where impulse responses are qualitatively similar. Plots for these shocks are provided in the online Appendix.

The new algorithm relies on joint estimation of parameters and identification restrictions. Therefore, it could be argued that the qualitatively similar results in \autoref{fig:financial_benchmark} are an artifact as they can be very sensitive to the structural identification restrictions imposed. However, this is not the case and the algorithm works well in various different scenarios. I consider the following identification restrictions in $\mathbf{\Lambda}$
\begin{enumerate}
\item Benchmark case with $r=5$ shocks, identified as in \autoref{table:sign_baseline}.
\item Case with $n=r=6$ shocks. The first five shocks are identical to the benchmark case, and a sixth shock/factor is added with no restrictions. This is equivalent to adding to the restrictions in \autoref{table:sign_baseline} a column with six ``NA'' entries.
\item Case with $r=5$ shocks, but only a financial shock is identified. The first four shocks have no restrictions, that is, entries in the first four columns of \autoref{table:sign_baseline} are replaced with NA. The financial shock identified using the restrictions in the last column \autoref{table:sign_baseline}.
\item Case with $r=1$ identified shocks. This is only the financial shock identified using the restrictions in the last column of \autoref{table:sign_baseline}, but no macro shocks are identified or estimated.
\end{enumerate}
The first case is the one also plotted in panel (b) of \autoref{fig:financial_benchmark}. The second case is used as a means of showing that the algorithm is able to incorporate the case with as many factors as variables, and is not sensitive to the motivating assumption that only a few shocks drive the VAR. This motivating assumption seems reasonable in larger systems (see next subsection), but what if a researcher is interested in smaller systems with as many shocks as variables? The third case allows to find out to what extend identification of the financial shock is affected by the restrictions in the remaining four shocks. Since shocks are identical factors, the aim is to find how estimates of the fifth factor are affected by assumptions in the first four factors. Finally, case four simply removes any information in the first four shocks and simply estimates a model with one shock. Identifying a single shock of interest is a very popular practice in empirical papers that rely on the \cite{RubioRamirezetal2010} algorithm, as it allows for faster inference (less accept/reject algorithmic steps) and results remain quantitatively unchanged. In contrast, the current algorithm is affected by the assumptions on the number of restrictions. In the fourth case using one shock means that only one factor is used for estimation, which in turn means that estimates of the VAR parameters (covariance matrix, and coefficients of lagged variables) will be affected.

\begin{figure}[H]
    \centering
    \begin{subfigure}[t]{0.45\textwidth}
        \centering
        \includegraphics[trim= 3cm 0cm 3cm 0cm,width=\linewidth]{FACTOR5.png}
        \caption{\emph{Case 1: benchmark}}
    \end{subfigure}
    \hfill
    \begin{subfigure}[t]{0.45\textwidth}
        \centering
        \includegraphics[trim= 3cm 0cm 3cm 0cm,width=\linewidth]{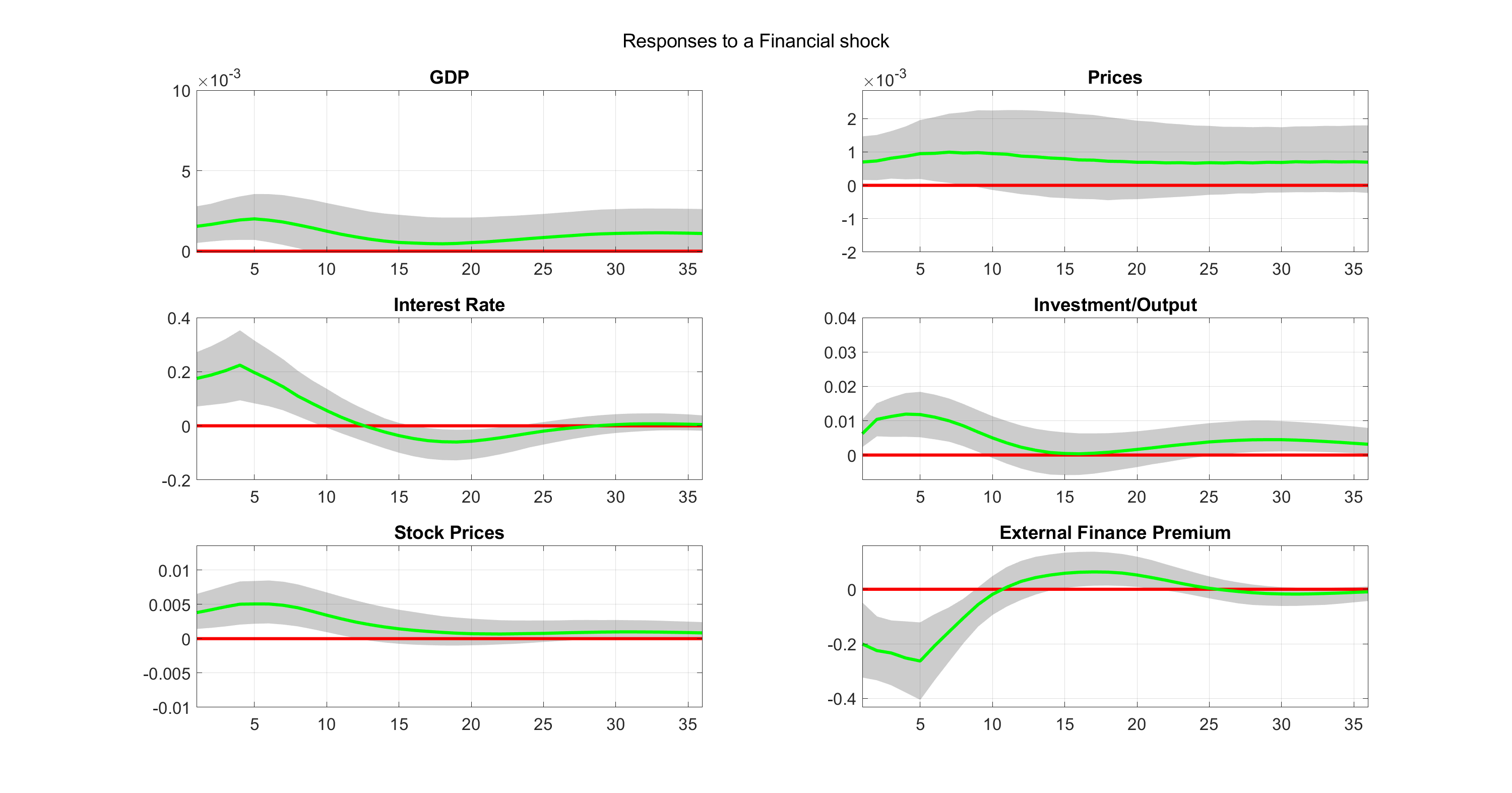}  
        \caption{\emph{Case 2: $n=r=6$ shocks}}
    \end{subfigure} \\
      \centering
    \begin{subfigure}[t]{0.45\textwidth}
        \centering
        \includegraphics[trim= 3cm 0cm 3cm 0cm,width=\linewidth]{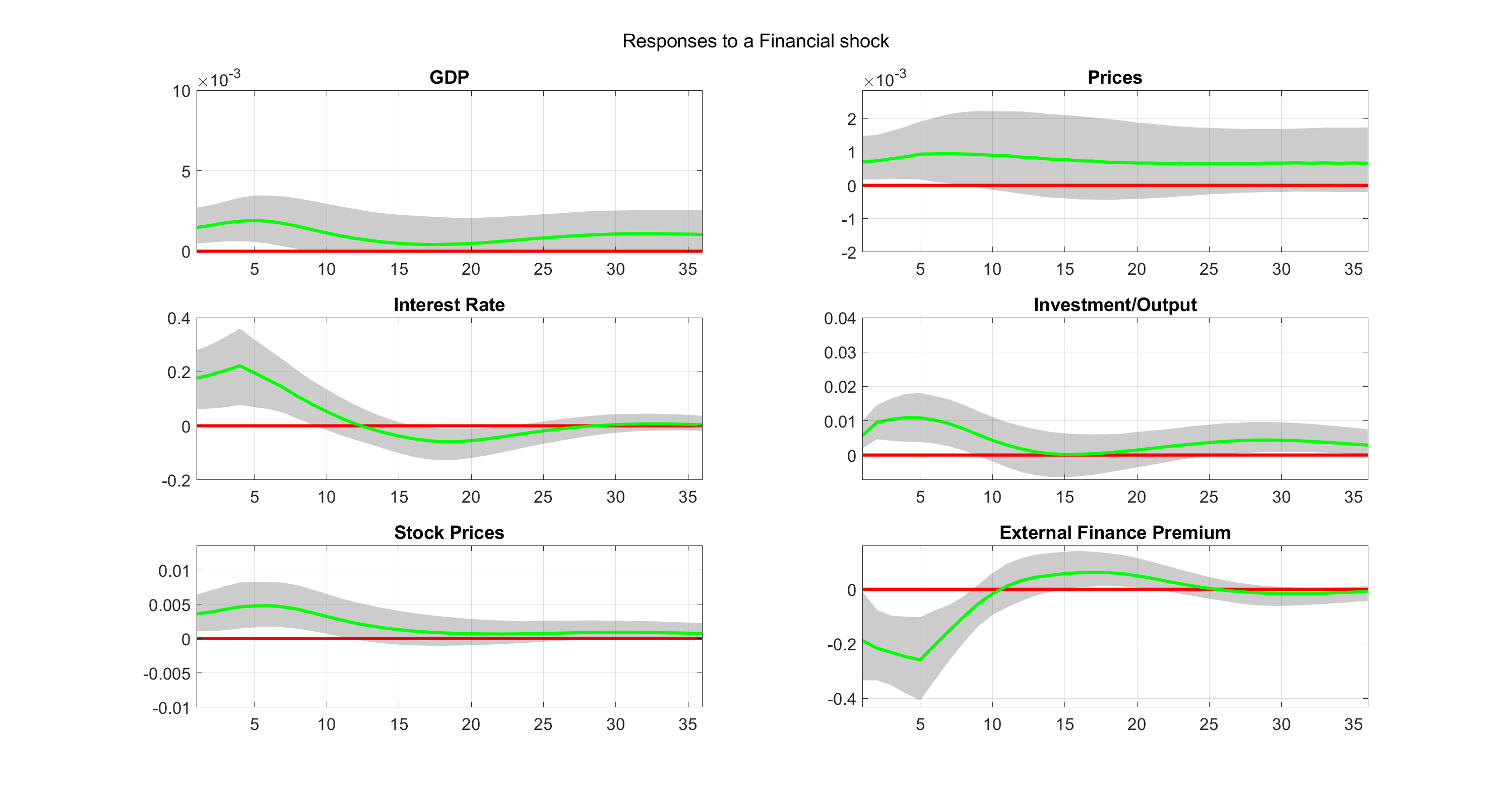}
        \caption{\emph{Case 3: no restrictions on macro shocks}}
    \end{subfigure}
    \hfill
    \begin{subfigure}[t]{0.45\textwidth}
        \centering
        \includegraphics[trim= 3cm 0cm 3cm 0cm,width=\linewidth]{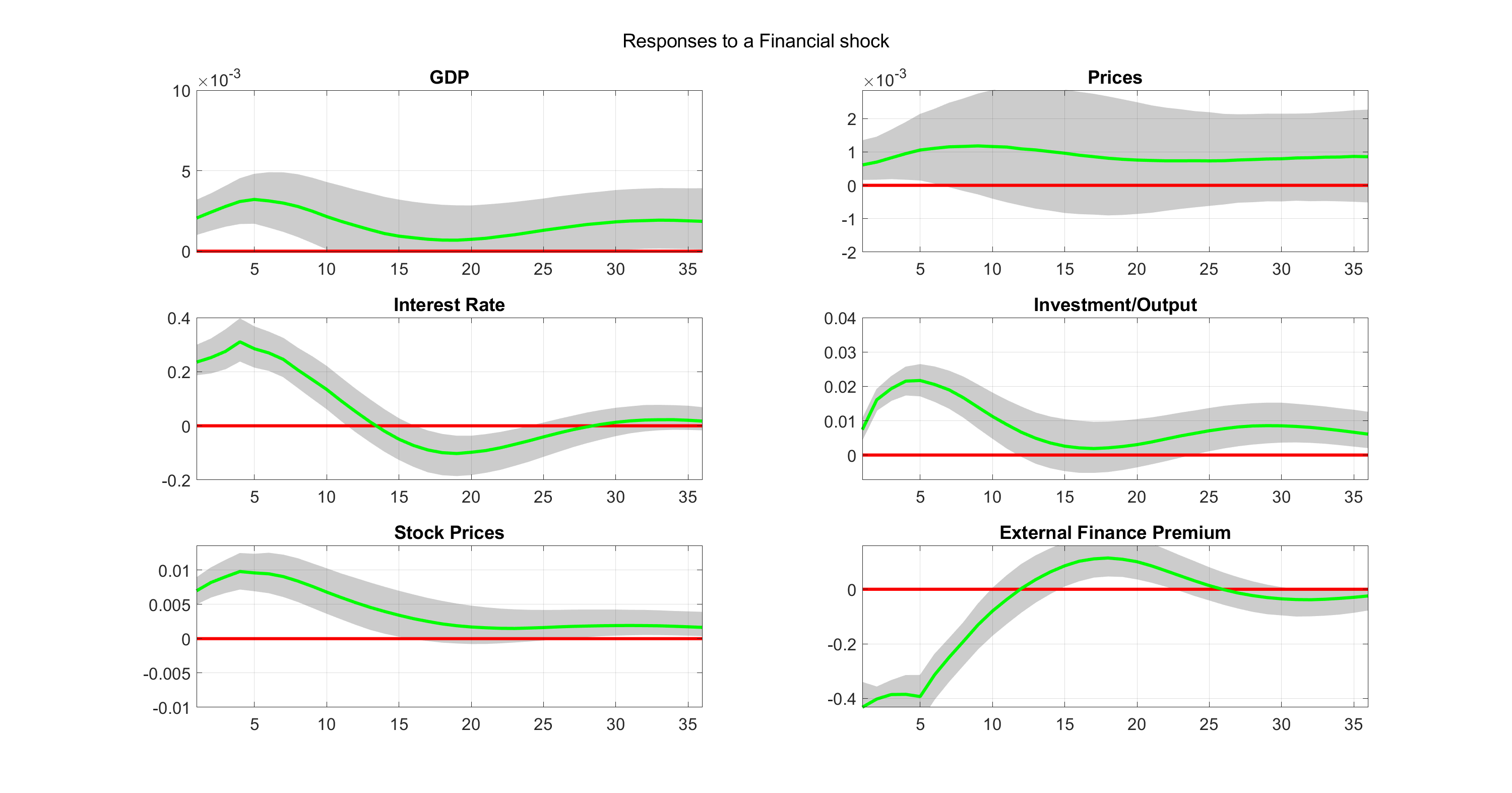}  
        \caption{\emph{Case 4: $r=1$, no macro shocks present}}
    \end{subfigure}  
\caption{\emph{This figure shows the robustness of the benchmark IRF results produced with the new algorithm, under different identification assumptions. See main text for description of each of the four cases presented in panels (a)-(d). Solid lines are the posterior medians of the IRFs, and shaded areas the $68\%$ posterior bands.}} \label{fig:financial_cases}        
\end{figure}

\autoref{fig:financial_cases} presents the impulse responses from all four cases. They are qualitatively and quantitatively identical. There are only a couple of differences when moving from the models with five shocks (factors) to the model with only one factor (panel (d) of the figure). In the latter case, the $68\%$ bands of the IRFs of prices are a bit wider, and the median impact response of the spread is twice as large (in absolute value) as the other three cases. Additionally, the curvature of the IRFs of GDP, interest rate, investment/output ratio, and stock prices, over the first 10 periods following the shock, is more pronounced in panel (d) relative to the other three panels. Therefore, \autoref{fig:financial_cases} suggests that the number of shocks is more important than identification assumptions made in shocks other than the shock of interest. 

In the previous Section, using artificially generated data, it was suggested that it can be hurtful to estimate less shocks compared to the true number of shocks in the DGP. In this case, when estimating $r=1$ shock does not distort the IRFs substantially, as from a statistical point of view one factor can be sufficient for a small, six-variable VAR. In order to find out if this is truly the case, \autoref{table:DIC_baseline} shows the DIC values attained by each of the four cases. The worst case is the one where six shocks are identified from the six series -- this corresponds to an overparametrized and unnecessary (from a statistical point of view) factor decomposition. The case with $r=1$ has the second highest DIC value, meaning that the true number of factors (again in a statistical sense) is larger than one and smaller than six. Surprisingly, Case 3 which is builds on the benchmark Case 1 but lifts all sign restrictions in the first four shocks, is the one that fits the best from a statistical point of view.\footnote{Notice that in this case the IRFs of supply, demand, monetary, and investment will be flat around zero -- as a matter of fact these should be named as Shock 1, Shock 2, Shock 3, Shock 4 exactly because there are no sign restrictions or identifying assumptions. As it was argued in the previous Sections, even though these four shocks are not structurally identified, the common component $\mathbf{\Lambda}\mathbf{f}_{t}$ is identified and covariance matrix estimation is feasible. Despite the fact that these shocks are not identified, the fifth shock (financial) is identified based on its restrictions and it is not affected by the lack of identification of the first four shocks/factors.} This DIC value suggests that identifying supply, demand, monetary and investment shocks, using sign restrictions based on economic theory, is statistically inferior to estimating four generic shocks with no sign restrictions. However, the model with the best statistical fit is not necessarily the best model for structural analysis \citep{Bernankeetal2005}, therefore, in this case the benchmark Case 1 should be preferred as it has a good fit and at the same time allows for identification of important macroeconomic shocks.

\begin{table}
\centering
\caption{\emph{Values of the Deviance Information Criterion (DIC) under four different identification schemes (see main text for details).  Lower values of the DIC indicate that a given restriction is more plausible than alternatives.} } \label{table:DIC_baseline}
\begin{tabular}{rcccc}
     & Case 1 & Case 2 & Case 3 & Case 4 \\ \hline\hline
DIC  & -8838.02 &  -7730.67  & -9418.08  &  -8452.36  \\ \hline
\end{tabular}
\end{table}

\subsection{A (reasonably) large-scale VAR model for measuring financial shocks}
We next proceed to demonstrate how the new algorithm can estimate one, large-dimensional system in order to measure in one setting all the financial shocks that \cite{Furlanettoetal2017} identify. The larger VAR that these authors specify has seven variables and six shocks: aggregate supply, aggregate demand, investment, housing, uncertainty, and credit. These authors do not identify a monetary shock using this larger VAR, possibly due to computational concerns. Here we attempt to use all available variables in \cite{Furlanettoetal2017} to identify seven shocks, that is, the six shocks just listed plus a monetary shock. We also use additional measures of output, consumer prices, stock prices, interest rate, and credit spread, in order to enhance identification. We end up with a 15-variable VAR with $p=5$ on the following variables: 1) real GDP; 2) prices (GDP deflator); 3) interest rate (3-month Tbill); 4) investment to output ratio; 5) stock prices (real S\&P500 prices); 6) credit spread (Baa minus Fed funds rate); 7) credit to real estate value ratio; 8) excess bond premium (EBP); 9) EBP to VIX ratio; 10) mortgage rate (30-year rates); 11) employment; 12) Federal funds rate; 13) core CPI; 14) stock prices 2 (real DJIA prices); and 15) credit spread 2 (``GZ'' spread). The online Appendix has detailed definitions of these variables, transformations used, and sources.

\autoref{table:sign_large} shows the signs imposed on each of the 15 variables in order to identify each of the seven structural shocks. This is a large matrix of restrictions, but the new algorithm can handle computationally the task of drawing $600,000$ samples from the posterior of all parameters (including the structural matrix of contemporaneous shocks) in a matter of minutes. As it was the case with the baseline VAR above, out of these $600,000$ draws $100,000$ are discarded and every $100^{th}$ sample is stored, resulting in $5,000$ samples used to produce numerical results from this large model. The horseshoe prior also has a crucial role in the estimation of this model, as we have $1140$ parameters in $\mathbf{\Phi}$ and only $114$ observations for each of the $15$ endogenous variables.

\begin{figure}[H]
\centering
\includegraphics[trim= 3cm 0cm 3cm 0cm,width=\linewidth]{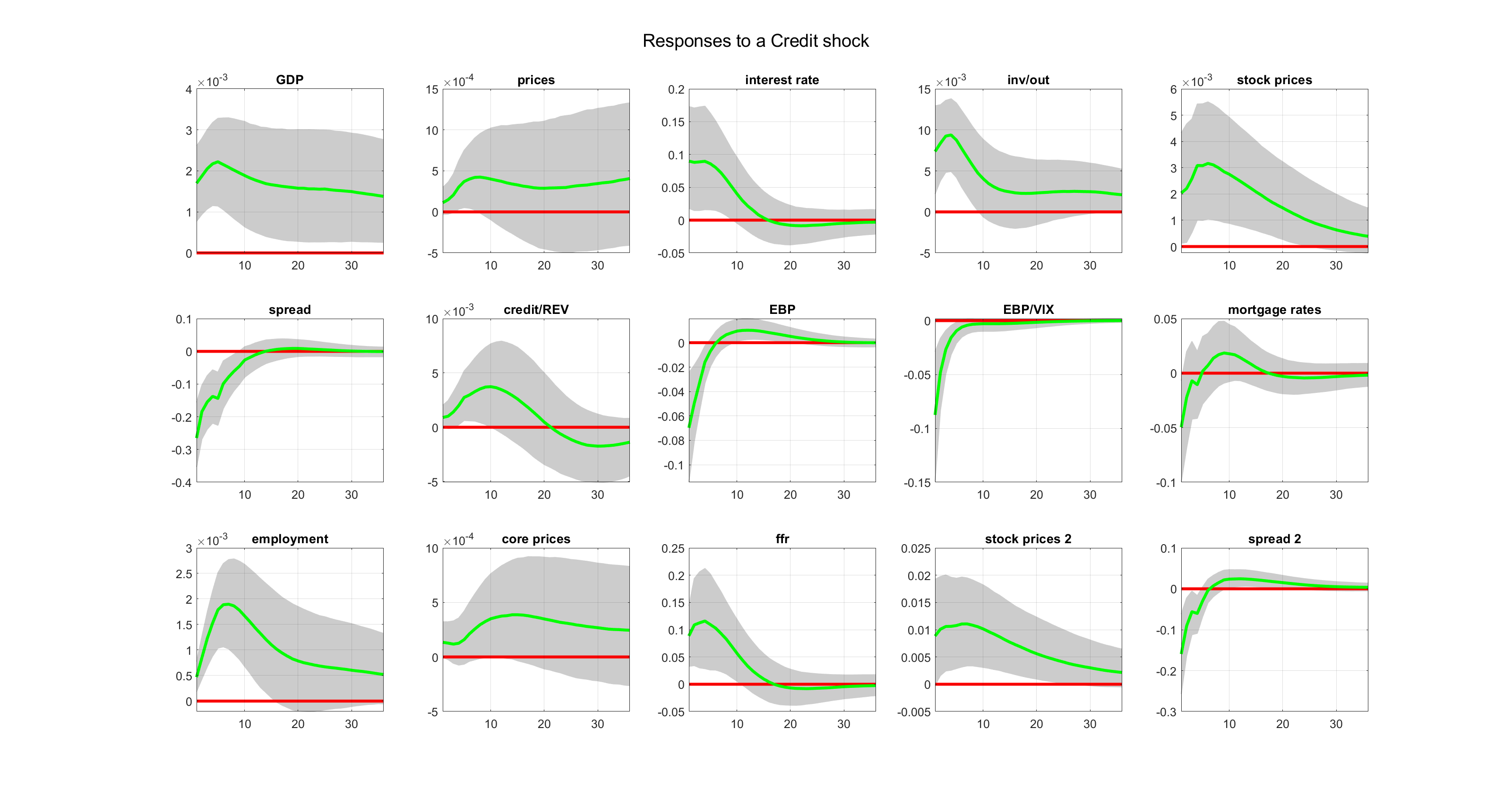}  
\caption{\emph{Impulse response functions to a credit shock in the large, 15-variable VAR with seven shocks identified in total.}} \label{fig:credit_large}
\end{figure}

\begin{table}[H]
\centering
\caption{\emph{Identified shocks and sign restrictions imposed on the matrix $\mathbf{\Lambda}$ in the large, 15-variable VAR model. These are the restrictions imposed by \cite{Furlanettoetal2017} using smaller VARs, accumulated into one integrated VAR setting with 15 variables and seven shocks.}} \label{table:sign_large}
\resizebox{0.9\textwidth}{!}{ 
\begin{tabular}{rccccccc} \hline
 & \multicolumn{7}{c}{\textsc{Identified Shocks}} \\
  	&	Supply	&	Demand	&	Monetary	&	Investment	&	Housing	&	Uncertainty	&	Credit	\\ \hline\hline
\multicolumn{8}{c}{\underline{\textsc{Original variables in \cite{Furlanettoetal2017}:}}}	\\
GDP	&	+	&	+	&	+	&	+	&	+	&	+	&	+	\\
prices	&	--	&	+	&	+	&	+	&	+	&	+	&	+	\\
interest rate	&	NA	&	+	&	--	&	+	&	+	&	+	&	+	\\
investment/output	&	NA	&	--	&	NA	&	+	&	+	&	+	&	+	\\
stock prices	&	+	&	NA	&	NA	&	--	&	+	&	+	&	+	\\
spread	&	NA	&	NA	&	NA	&	NA	&	NA	&	NA	&	NA	\\
credit/real estate value	&	NA	&	NA	&	NA	&	NA	&	--	&	+	&	+	\\
EBP	&	NA	&	NA	&	NA	&	NA	&	NA	&	--	&	--	\\
EBP/VIX	&	NA	&	NA	&	NA	&	NA	&	NA	&	+	&	--	\\
mortgage rates	&	NA	&	NA	&	NA	&	NA	&	NA	&	NA	&	--	\\
\multicolumn{8}{c}{\underline{\textsc{Additional measures of output, prices etc.:}}} \\
employment	&	+	&	+	&	+	&	+	&	+	&	+	&	+	\\
Federal funds rate	&	NA	&	+	&	--	&	+	&	+	&	+	&	+	\\
core prices	&	--	&	+	&	+	&	+	&	+	&	+	&	+	\\
stock prices 2	&	+	&	NA	&	NA	&	--	&	+	&	+	&	+	\\
spread 2	&	NA	&	NA	&	NA	&	NA	&	NA	&	NA	&	NA	\\ \hline
\par
\multicolumn{8}{l}{{\small \emph{Notes: Entries in this table show the restrictions imposed: + positive sign; -- negative sign; NA no restriction.}  }}
\end{tabular}
}
\end{table}
\autoref{fig:credit_large} shows the impulse responses of the 15 endogenous variables to a credit shock. The green lines are posterior medians, and the shaded areas 68\% bands. The magnitudes and shapes of the IRFs are consistent with the ones reported in \citet[Figure 7]{Furlanettoetal2017}, despite the fact that in the case of variables such as GDP the IRFs are strongly different from zero. The most interesting feature of this figure is the effect of a credit shock on the two credit spread variables we used in the same VAR. \cite{Furlanettoetal2017} use these spreads (plus an additional third spread we haven't included here) one at a time in their VAR in order to assess robustness of their results. These authors do not impose sign restrictions on the credit spread and they find that in their baseline specification this tends to be negative. In the large VAR case, the first credit spread variable has a strong negative contemporaneous response before subsequently moving to positive territory, while the second credit variable does not have a contemporaneous response different from zero and in subsequent period reacts positively. Such results show the important avenues for identifying various structural shocks that the new algorithm opens up: by using large information sets we can have the ability to identify several structural shocks in one setting, thus, making comparisons and testing of structural hypotheses more transparent. The online Appendix provides further results for this 15-variable VAR.

\section{Conclusions}
This paper outlines a new algorithm based on a VAR methodology that fully utilizes the interpretability and parsimony of factor models. In particular, the novel element of the proposed approach is the formulation of reduced-form VAR disturbances using a common factor structure, and the derivation of an algorithm that allows for efficient sampling of sign-restricted decompositions of the VAR covariance matrix. The new algorithm can handle VARs with possibly 100 or more variables and it provides sensible numerical results compared to the algorithm of \cite{RubioRamirezetal2010} -- despite the fact that the two algorithms rely on different modeling assumptions and are not directly comparable.\footnote{Additional numerical results are provided in the online Appendix.} Therefore, the new algorithm can be seen as a useful tool in the toolbox of modern macroeconomists, that complements existing algorithms and at the same time opens up new avenues for empirical research using large-scale VAR models.

\newpage

\addcontentsline{toc}{section}{References}
\bibliographystyle{BibTex/ecta}
\bibliography{BibTex/BK1}

\begin{thebibliography}{39}
\newcommand{\enquote}[1]{``#1''}
\expandafter\ifx\csname natexlab\endcsname\relax\def\natexlab#1{#1}\fi

\bibitem[\protect\citeauthoryear{Ahmadi and Uhlig}{Ahmadi and
  Uhlig}{2015}]{AhmadiUhlig2016}
\textsc{Ahmadi, P.~A. and H.~Uhlig} (2015): \enquote{Sign Restrictions in
  Bayesian FaVARs with an Application to Monetary Policy Shocks,} Working Paper
  21738, National Bureau of Economic Research.

\bibitem[\protect\citeauthoryear{Amir-Ahmadi and Drautzburg}{Amir-Ahmadi and
  Drautzburg}{2021}]{https://doi.org/10.3982/QE1277}
\textsc{Amir-Ahmadi, P. and T.~Drautzburg} (2021): \enquote{Identification and
  inference with ranking restrictions,} \emph{Quantitative Economics}, 12,
  1--39.

\bibitem[\protect\citeauthoryear{Anderson and Rubin}{Anderson and
  Rubin}{1956}]{anderson1956}
\textsc{Anderson, T.~W. and H.~Rubin} (1956): \enquote{Statistical Inference in
  Factor Analysis,} in \emph{Proceedings of the Third Berkeley Symposium on
  Mathematical Statistics and Probability, Volume 5: Contributions to
  Econometrics, Industrial Research, and Psychometry}, Berkeley, Calif.:
  University of California Press, 111--150.

\bibitem[\protect\citeauthoryear{Arias, Caldara, and Rubio-Ramírez}{Arias
  et~al.}{2019}]{Ariasetal2019}
\textsc{Arias, J.~E., D.~Caldara, and J.~F. Rubio-Ramírez} (2019):
  \enquote{The Systematic Component of Monetary Policy in SVARs: An Agnostic
  Identification Procedure,} \emph{Journal of Monetary Economics}, 101, 1 --
  13.

\bibitem[\protect\citeauthoryear{Arias, Rubio-Ramírez, and Waggoner}{Arias
  et~al.}{2018}]{Ariasetal2018}
\textsc{Arias, J.~E., J.~F. Rubio-Ramírez, and D.~F. Waggoner} (2018):
  \enquote{Inference Based on Structural Vector Autoregressions Identified With
  Sign and Zero Restrictions: Theory and Applications,} \emph{Econometrica},
  86, 685--720.

\bibitem[\protect\citeauthoryear{Armagan, Dunson, Lee, Bajwa, and
  Strawn}{Armagan et~al.}{2013}]{Armaganetal2013}
\textsc{Armagan, A., D.~B. Dunson, J.~Lee, W.~U. Bajwa, and N.~Strawn} (2013):
  \enquote{Posterior Consistency in Linear Models Under Shrinkage Priors,}
  \emph{Biometrika}, 100, 1011--1018.

\bibitem[\protect\citeauthoryear{Bai}{Bai}{2003}]{Bai2003}
\textsc{Bai, J.} (2003): \enquote{Inferential Theory for Factor Models of Large
  Dimensions,} \emph{Econometrica}, 71, 135--171.

\bibitem[\protect\citeauthoryear{Baumeister and Hamilton}{Baumeister and
  Hamilton}{2015}]{BaumeisterHamilton2015}
\textsc{Baumeister, C. and J.~D. Hamilton} (2015): \enquote{Sign Restrictions,
  Structural Vector Autoregressions, and Useful Prior Information,}
  \emph{Econometrica}, 83, 1963--1999.

\bibitem[\protect\citeauthoryear{Beaudry, Nam, and Wang}{Beaudry
  et~al.}{2011}]{NBERw17651}
\textsc{Beaudry, P., D.~Nam, and J.~Wang} (2011): \enquote{Do Mood Swings Drive
  Business Cycles and is it Rational?} Working Paper 17651, National Bureau of
  Economic Research.

\bibitem[\protect\citeauthoryear{Bernanke and Blinder}{Bernanke and
  Blinder}{1992}]{BernankeMihov1992}
\textsc{Bernanke, B.~S. and A.~S. Blinder} (1992): \enquote{The Federal Funds
  Rate and the Channels of Monetary Transmission,} \emph{The American Economic
  Review}, 82, 901--921.

\bibitem[\protect\citeauthoryear{Bernanke, Boivin, and Eliasz}{Bernanke
  et~al.}{2005}]{Bernankeetal2005}
\textsc{Bernanke, B.~S., J.~Boivin, and P.~Eliasz} (2005): \enquote{{Measuring
  the Effects of Monetary Policy: A Factor-Augmented Vector Autoregressive
  (FAVAR) Approach*},} \emph{The Quarterly Journal of Economics}, 120,
  387--422.

\bibitem[\protect\citeauthoryear{Bhattacharya, Chakraborty, and
  Mallick}{Bhattacharya et~al.}{2016}]{Bhattacharya2016}
\textsc{Bhattacharya, A., A.~Chakraborty, and B.~K. Mallick} (2016):
  \enquote{{Fast Sampling with Gaussian Scale Mixture Priors in
  High-Dimensional Regression},} \emph{Biometrika}, 103, 985--991.

\bibitem[\protect\citeauthoryear{Bhattacharya and Dunson}{Bhattacharya and
  Dunson}{2011}]{BhattacharyaDunson2011}
\textsc{Bhattacharya, A. and D.~B. Dunson} (2011): \enquote{Sparse Bayesian
  infinite factor models,} \emph{Biometrika}, 98, 291--306.

\bibitem[\protect\citeauthoryear{Botev}{Botev}{2017}]{Botev2017}
\textsc{Botev, Z.~I.} (2017): \enquote{The Normal Law Under linear
  Restrictions: Simulation and Estimation via Minimax Tilting,} \emph{Journal
  of the Royal Statistical Society: Series B (Statistical Methodology)}, 79,
  125--148.

\bibitem[\protect\citeauthoryear{Bruns and Piffer}{Bruns and
  Piffer}{2019}]{RePEc:diw:diwwpp:dp1796}
\textsc{Bruns, M. and M.~Piffer} (2019): \enquote{{Bayesian Structural VAR
  Models: A New Approach for Prior Beliefs on Impulse Responses},} Tech. rep.

\bibitem[\protect\citeauthoryear{Caldara, Fuentes-Albero, Gilchrist, and
  Zakraj\u{s}ek}{Caldara et~al.}{2016}]{CALDARA2016185}
\textsc{Caldara, D., C.~Fuentes-Albero, S.~Gilchrist, and E.~Zakraj\u{s}ek}
  (2016): \enquote{The macroeconomic impact of financial and uncertainty
  shocks,} \emph{European Economic Review}, 88, 185--207, sI: The Post-Crisis
  Slump.

\bibitem[\protect\citeauthoryear{Canova and Paustian}{Canova and
  Paustian}{2011}]{CanovaPaustian2011}
\textsc{Canova, F. and M.~Paustian} (2011): \enquote{Business Cycle Measurement
  with some Theory,} \emph{Journal of Monetary Economics}, 58, 345 -- 361.

\bibitem[\protect\citeauthoryear{Carriero, Clark, and Marcellino}{Carriero
  et~al.}{2019}]{Carrieroetal}
\textsc{Carriero, A., T.~E. Clark, and M.~Marcellino} (2019): \enquote{Large
  Bayesian vector autoregressions with stochastic volatility and non-conjugate
  priors,} \emph{Journal of Econometrics}, 212, 137 -- 154, big Data in Dynamic
  Predictive Econometric Modeling.

\bibitem[\protect\citeauthoryear{Carvalho, Polson, and Scott}{Carvalho
  et~al.}{2010}]{Carvalhoetal2010}
\textsc{Carvalho, C.~M., N.~G. Polson, and J.~G. Scott} (2010): \enquote{{The
  Horseshoe Estimator for Sparse Signals},} \emph{Biometrika}, 97, 465--480.

\bibitem[\protect\citeauthoryear{Furlanetto, Ravazzolo, and
  Sarferaz}{Furlanetto et~al.}{2019}]{Furlanettoetal2017}
\textsc{Furlanetto, F., F.~Ravazzolo, and S.~Sarferaz} (2019):
  \enquote{{Identification of Financial Factors in Economic Fluctuations},}
  \emph{The Economic Journal}, 129, 311--337.

\bibitem[\protect\citeauthoryear{Geweke}{Geweke}{1996}]{Geweke1996}
\textsc{Geweke, J.~F.} (1996): \enquote{Bayesian Inference for Linear Models
  Subject to Linear Inequality Constraints,} in \emph{Modelling and Prediction
  Honoring Seymour Geisser}, ed. by J.~C. Lee, W.~O. Johnson, and A.~Zellner,
  New York, NY: Springer New York, 248--263.

\bibitem[\protect\citeauthoryear{Ghosh, Tang, Ghosh, and Chakrabarti}{Ghosh
  et~al.}{2016}]{Ghoshetal2016}
\textsc{Ghosh, P., X.~Tang, M.~Ghosh, and A.~Chakrabarti} (2016):
  \enquote{Asymptotic Properties of Bayes Risk of a General Class of Shrinkage
  Priors in Multiple Hypothesis Testing Under Sparsity,} \emph{Bayesian Anal.},
  11, 753--796.

\bibitem[\protect\citeauthoryear{Giannone, Lenza, and Primiceri}{Giannone
  et~al.}{2015}]{Giannoneetal2015}
\textsc{Giannone, D., M.~Lenza, and G.~E. Primiceri} (2015): \enquote{Prior
  Selection for Vector Autoregressions,} \emph{The Review of Economics and
  Statistics}, 97, 436--451.

\bibitem[\protect\citeauthoryear{Gorodnichenko}{Gorodnichenko}{2005}]{Gorodnichenko2005}
\textsc{Gorodnichenko, Y.} (2005): \enquote{Reduced-Rank Identification of
  Structural Shocks in VARs,} SSRN working paper 590906, University of
  California, Berkeley.

\bibitem[\protect\citeauthoryear{Kilian and L\"{u}tkepohl}{Kilian and
  L\"{u}tkepohl}{2017}]{kilianlutkepohl2017}
\textsc{Kilian, L. and H.~L\"{u}tkepohl} (2017): \emph{Structural Vector
  Autoregressive Analysis}, Themes in Modern Econometrics, Cambridge University
  Press.

\bibitem[\protect\citeauthoryear{Koop and Korobilis}{Koop and
  Korobilis}{2010}]{KoopKorobilis2010}
\textsc{Koop, G. and D.~Korobilis} (2010): \enquote{{Bayesian Multivariate Time
  Series Methods for Empirical Macroeconomics},} \emph{Foundations and
  Trends(R) in Econometrics}, 3, 267--358.

\bibitem[\protect\citeauthoryear{Kowal, Matteson, and Ruppert}{Kowal
  et~al.}{2019}]{Kowal2019}
\textsc{Kowal, D.~R., D.~S. Matteson, and D.~Ruppert} (2019): \enquote{Dynamic
  Shrinkage Processes,} \emph{Journal of the Royal Statistical Society: Series
  B (Statistical Methodology)}, 81, 781--804.

\bibitem[\protect\citeauthoryear{Lopes and West}{Lopes and
  West}{2004}]{LopesWest2004}
\textsc{Lopes, H.~F. and M.~West} (2004): \enquote{Bayesian Model Assessment in
  Factor Analysis,} \emph{Statistica Sinica}, 14, 41--67.

\bibitem[\protect\citeauthoryear{Matthes and Schwartzman}{Matthes and
  Schwartzman}{2019}]{MatthesSchwartzman2019}
\textsc{Matthes, C. and F.~Schwartzman} (2019): \enquote{{What Do Sectoral
  Dynamics Tell Us About the Origins of Business Cycles?}} Working Paper 19-9,
  Federal Reserve Bank of Richmond.

\bibitem[\protect\citeauthoryear{Mountford and Uhlig}{Mountford and
  Uhlig}{2009}]{MountfordUhlig2009}
\textsc{Mountford, A. and H.~Uhlig} (2009): \enquote{What Are the Effects of
  Fiscal Policy Shocks?} \emph{Journal of Applied Econometrics}, 24, 960--992.

\bibitem[\protect\citeauthoryear{Neal}{Neal}{2003}]{neal2003}
\textsc{Neal, R.~M.} (2003): \enquote{Slice sampling,} \emph{The Annals of
  Statistics}, 31, 705--767.

\bibitem[\protect\citeauthoryear{Ouliaris and Pagan}{Ouliaris and
  Pagan}{2016}]{OuliarisPagan2017}
\textsc{Ouliaris, S. and A.~Pagan} (2016): \enquote{A Method for Working with
  Sign Restrictions in Structural Equation Modelling,} \emph{Oxford Bulletin of
  Economics and Statistics}, 78, 605--622.

\bibitem[\protect\citeauthoryear{Ramey}{Ramey}{2016}]{Ramey2016}
\textsc{Ramey, V.} (2016): \enquote{Chapter 2 - Macroeconomic Shocks and Their
  Propagation,} in \emph{Handbook of Macroeconomics}, ed. by J.~B. Taylor and
  H.~Uhlig, Elsevier, vol.~2, 71 -- 162.

\bibitem[\protect\citeauthoryear{Rubio-Ram\'{i}rez, Waggoner, and
  Zha}{Rubio-Ram\'{i}rez et~al.}{2010}]{RubioRamirezetal2010}
\textsc{Rubio-Ram\'{i}rez, J.~F., D.~F. Waggoner, and T.~Zha} (2010):
  \enquote{Structural Vector Autoregressions: Theory of Identification and
  Algorithms for Inference,} \emph{The Review of Economic Studies}, 77,
  665--696.

\bibitem[\protect\citeauthoryear{Spiegelhalter, Best, Carlin, and Van
  Der~Linde}{Spiegelhalter et~al.}{2002}]{Spiegelhalteretal2002}
\textsc{Spiegelhalter, D.~J., N.~G. Best, B.~P. Carlin, and A.~Van Der~Linde}
  (2002): \enquote{Bayesian Measures of Model Complexity and Fit,}
  \emph{Journal of the Royal Statistical Society: Series B (Statistical
  Methodology)}, 64, 583--639.

\bibitem[\protect\citeauthoryear{Stock and Watson}{Stock and
  Watson}{2005}]{StockWatson2005}
\textsc{Stock, J.~H. and M.~W. Watson} (2005): \enquote{Implications of Dynamic
  Factor Models for VAR Analysis,} Working Paper 11467, National Bureau of
  Economic Research.

\bibitem[\protect\citeauthoryear{Uhlig}{Uhlig}{2005}]{Uhlig2005}
\textsc{Uhlig, H.} (2005): \enquote{What are the Effects of Monetary Policy on
  Output? Results from an Agnostic Identification Procedure,} \emph{Journal of
  Monetary Economics}, 52, 381 -- 419.

\bibitem[\protect\citeauthoryear{Uhlig}{Uhlig}{2017}]{Uhlig2017}
---\hspace{-.1pt}---\hspace{-.1pt}--- (2017): \enquote{Shocks, Sign
  Restrictions, and Identification,} in \emph{Advances in Economics and
  Econometrics: Eleventh World Congress}, ed. by B.~Honor\'{e}, A.~Pakes,
  M.~Piazzesi, and L.~Samuelson, Cambridge University Press, vol.~2 of
  \emph{Econometric Society Monographs}, 95--127.

\bibitem[\protect\citeauthoryear{van~der Pas, Kleijn, and van~der
  Vaart}{van~der Pas et~al.}{2014}]{vanderpas2014}
\textsc{van~der Pas, S.~L., B.~J.~K. Kleijn, and A.~W. van~der Vaart} (2014):
  \enquote{The horseshoe estimator: Posterior concentration around nearly black
  vectors,} \emph{Electronic Journal of Statistics}, 8, 2585--2618.

\end{thebibliography}

\clearpage

\onehalfspacing

\begin{center}
\LARGE{Online Appendix to ``A new algorithm for sign restrictions in vector autoregressions''} \\
\hfill \\
\large{Dimitris Korobilis \\ \emph{University of Glasgow}}
\end{center}
\vskip 1cm

\begin{appendix}

\setcounter{page}{1}
\renewcommand{\theequation}{A.\arabic{equation}} \setcounter{equation}{0} %
\renewcommand{\thetable}{A\arabic{table}} \setcounter{table}{0}
\setcounter{footnote}{0}
\renewcommand{\thefigure}{A\arabic{figure}} \setcounter{figure}{0}
\setcounter{footnote}{0}
\section{Technical Appendix}
\subsection{Full Gibbs sampler for VARs with sign restrictions}
The Gibbs sampler presented in the main paper is a quite accurate description of the core algorithmic steps required in order to estimate the VAR model with factor structure in the residuals. Nevertheless, in order to produce empirical and other results, we have relied on the hierarchical horseshoe prior of \cite{Carvalhoetal2010}, two fast algorithms from drawing from the Normal \citep{Bhattacharya2016} and truncated Normal \citep{Botev2017} distributions, respectively, and the slice sampler of \cite{neal2003} in order to update the horseshoe prior parameters. Therefore, it is important to rewrite the Gibbs sampling algorithm in full, and give further explanations about the three enhancements that guarantee a fast and reliable algorithm in high dimensions. 

I repeat the full prior specification, which now includes the hierarchical horseshoe prior on $\bm{\phi}_{i}$. The priors for the $i^{th}$ VAR equation, $i=1,...,n$ is:
\begin{eqnarray}
\bm{\phi}_{i} \vert \sigma^{2}_{i}, \tau_{i}^{2}, \mathbf{\Psi}_{i}^{2}  & \sim & N_{k} \left( \mathbf{0}, \sigma^{2}_{i} \tau_{i}^{2} \mathbf{\Psi}_{i}^{2} \right), \text{ \ } \mathbf{\Psi}_{i}^{2} = diag\left(\psi_{i,1}^{2},...,\psi_{i,k}^{2} \right), \\
\psi_{i,j} & \sim & Cauchy^{+} \left(0, 1 \right), \text{ \ } j=1,...,k, \\
\tau_{i}  & \sim & Cauchy^{+} \left(0, 1 \right), \\
\mathbf{\Lambda}_{ij}  & \sim & \left\lbrace \begin{array}{ll} N\left(0, \underline{h}_{ij} \right) I( \Lambda_{ij} > 0), & \text{if $S_{ij} = 1$}, \\
N\left(0,\underline{h}_{ij} \right) I( \Lambda_{ij} < 0),  & \text{if $S_{ij} = -1$}, \\
\delta_{0} \left( \mathbf{\Lambda}_{ij}\right),  & \text{if $S_{ij} = 0$}, \\
N\left(0,\underline{h}_{ij} \right),  & \text{otherwise},
\end{array} \right. \text{ \ } j=1,...,r, \\
\mathbf{f}_{t} & \sim & N_{r} \left( \mathbf{0}, \mathbf{I} \right), \\
\sigma^{2}_{i} & \sim & inv-Gamma \left(\underline{\rho}_{i},\underline{\kappa}_{i} \right),
\end{eqnarray}
where we set $\underline{h}_{ij}=4$, $\underline{\rho}_{i}=1$ and $\underline{\kappa}_{i}=0.01$.

Under these priors, the full factor sign restrictions algorithm takes the following form
\textbf{Factor sign restrictions (FSR) algorithm}
\begin{enumerate}
\item Sample $\bm{\phi}_{i}$ for $i=1,...,n$ from
\begin{equation}
\bm{\phi}_{i} \vert \mathbf{\Sigma},\mathbf{\Lambda}, \mathbf{f}, \mathbf{y} \sim N_{k} \left( \overline{\mathbf{V}}_{i} \left( \sum_{t=1}^{T} \sigma_{i}^{-2} \mathbf{x}_{t}^{\prime} \widetilde{\mathbf{y}}_{it} \right), \overline{\mathbf{V}}_{i} \right),
\end{equation}
where $\widetilde{\mathbf{y}}_{it} = \mathbf{y}_{it} - \mathbf{\Lambda}_{i} \mathbf{f}_{t}$ and $\overline{\mathbf{V}}_{i}^{-1} = \left( \underline{\mathbf{V}}_{i}^{-1} + \sum_{t=1}^{T} \sigma_{i}^{-2} \mathbf{x}_{t}^{\prime} \mathbf{x}_{t} \right)$.
We use the efficient sampler of \cite{Bhattacharya2016} in order to sample these elements.
\item Sample $\psi_{ij}$ using slice sampling \citep{neal2003}
\begin{enumerate}
\item[a.] Set $\eta_{ij} = 1/\psi_{ij}^{2}$ using the last available sample of $\psi_{ij}^{2}$.
\item[b.] Sample a random variable $u$ from
\begin{equation}
u \vert \eta_{ij} \sim Uniform\left( 0, \frac{1}{1+\eta_{ij}} \right).
\end{equation}
\item[c.] Sample $\eta_{ij}$ from
\begin{equation}
\eta_{ij} \sim e^{\frac{\phi_{ij}^{2}}{2\sigma_{i}^{2}}\eta_{ij}} I \left( \frac{u}{1-u} >\eta_{ij} \right)
\end{equation}
and set $\psi_{ij} = 1/\sqrt{\eta_{ij}}$.
\end{enumerate}
\item Sample $\tau_{i}$ using slice sampling \citep{neal2003}
\begin{enumerate}
\item[a.] Set $\xi_{i} = 1/\tau_{i}^{2}$ using the last available sample of $\tau_{i}^{2}$.
\item[b.] Sample a random variable $u$ from
\begin{equation}
v \vert \xi_{ij} \sim Uniform\left( 0, \frac{1}{1+\xi_{ij}} \right).
\end{equation}
\item[c.] Sample $\xi_{i}$ from
\begin{equation}
\xi_{i} \sim \gamma\left( (k+1)/2, v\frac{2\sigma^{2}}{\sum \left( \frac{\phi_{ij}}{\psi_{ij}}\right)^{2}} \right),
\end{equation}
where $\gamma \left( \bullet \right)$ is the lower incomplete gamma function, and set $\tau_{i} = 1/\sqrt{\xi_{i}}$
\end{enumerate}
\item Sample $\mathbf{\Lambda}_{ij}$ from univariate conditional posteriors \citep{Geweke1996} of the form
\begin{equation}
\mathbf{\Lambda}_{ij} \vert \mathbf{\Lambda}_{-ij} \mathbf{\Phi},\mathbf{\Sigma}, \mathbf{f}, \mathbf{y}  \sim TN_{(\mathbf{a}_{ij},\mathbf{b}_{ij})} \left( \overline{\lambda}_{ij} - \overline{h}_{ij}\sum_{l \neq j} \overline{h}_{il}^{-1} \left( \mathbf{\Lambda}_{il} - \overline{\lambda}_{il} \right)   ,\overline{h}_{ij} \right),
\end{equation}
where $\overline{\lambda}_{ij}$ and $\overline{h}_{ij}$ denote the $ij^{th}$ elements of the joint posterior mean and variance, respectively, of $\mathbf{\Lambda}_{i}$. The joint posterior variance is $\overline{\mathbf{H}}^{-1} = \left( \underline{\mathbf{H}}^{-1} + \sum_{t=1}^{T} \sigma_{i}^{-2} \mathbf{f}_{t}^{\prime}\mathbf{f}_{t} \right)$ and the joint posterior mean is $\overline{\mathbf{H}} \left(\sum_{t=1}^{T}  \sigma_{i}^{-2} \mathbf{f}_{t}^{\prime} \widehat{\mathbf{y}}_{it} \right)$ with $\widehat{\mathbf{y}}_{it} \equiv \bm{\varepsilon}_{it} = \mathbf{y}_{it} - \mathbf{\phi}_{i} \mathbf{x}_{t}$. Here $TN_{(\mathbf{a}_{ij},\mathbf{b}_{ij})} \left( \bullet \right)$ denotes the \textbf{univariate} truncated Normal distribution with bounds:
\begin{equation}
(\mathbf{a}_{ij},\mathbf{b}_{ij}) = \left \lbrace \begin{array}{ll}
(-\infty,0) & \text{if $S_{ij} = -1$}, \\
(0, \infty) & \text{if $S_{ij} = 1$}, \\
(0,0) & \text{if $S_{ij} = 0$}, \\
(- \infty, \infty) & \text{otherwise,} 
\end{array} \right. 
\end{equation}
We use the efficient univariate truncated Normal generator provided by \cite{Botev2017} in order to sample these elements.
\item Sample $\mathbf{f}_{t}$ for $t=1,...,T$ from
\begin{equation}
\mathbf{f}_{t} \vert \mathbf{\Lambda}, \mathbf{\Sigma}, \mathbf{\Phi}, \mathbf{y} \sim N\left(\overline{\mathbf{G}} \left( \mathbf{\Lambda} \mathbf{\Sigma}^{-1} \widehat{\mathbf{y}}_{t} \right), \overline{\mathbf{G}} \right),
\end{equation}
where $\overline{\mathbf{G}}^{-1} = \left( \mathbf{I}_{r} + \mathbf{\Lambda}^{\prime} \mathbf{\Sigma} \mathbf{\Lambda} \right)$. Post-process the draws of the $T \times r$ matrix $\mathbf{f} = \left(\mathbf{f}_{1},...,\mathbf{f}_{T} \right)^{\prime}$ such that its $r$ columns (corresponding to structural shocks) are uncorrelated and standardized to unit variance. This is done by applying first the Gram-Schmidt procedure and subsequently dividing each column of $\mathbf{f}$ with its standard deviation.
\item Sample $\sigma_{i}^{2} $ for $i=1,...,n$ from
{\footnotesize
\begin{equation}
\sigma_{i}^{2} \vert \mathbf{\Lambda}, \mathbf{f}, \mathbf{\Phi}, \mathbf{y} \sim inv-Gamma \left( \frac{T}{2} + \underline{\rho}_{i}, \left[ \underline{\kappa}_{i}^{-1} + \sum_{t=1}^{T} \left( \mathbf{y}_{it} - \mathbf{\phi}_{i} \mathbf{x}_{t} - \mathbf{\Lambda}_{i} \mathbf{f}_{t} \right)^{\prime} \left( \mathbf{y}_{it} - \mathbf{\phi}_{i} \mathbf{x}_{t} - \mathbf{\Lambda}_{i}\mathbf{f}_{t} \right) \right]^{-1} \right)
\end{equation}
}
\end{enumerate}

\subsection{Priors and efficient sampling} \label{section:horseshoe}
While the Gibbs sampler presented above is based on standard sampling steps for Bayesian VARs and static factor models \citep{LopesWest2004}, I discuss here how to build an even more reliable algorithm that can handle larger models with the same ease it can estimate smaller models. The main computational concerns with the core algorithm presented above stem from the high dimensions of $\mathbf{\Phi}$\footnote{For example, in a 50 variable VAR with 4 lags and an intercept this matrix has 10,000+ elements.} and the fact that $\mathbf{\Lambda}$ is both latent and has to be sampled from a restricted (truncated) Normal distribution.

First, overall the proposed factor-VAR had disturbances that have a diagonal covariance matrix. Therefore, the parameters $\mathbf{\Phi}$ can be sampled equation-by-equation (which is what \citealp{BaumeisterHamilton2015} also do in the SVAR setting) using \eqref{phi_post}. Instead of having to sample one large vector of $nk$ parameters, $n$ independent univariate autoregressions with $k$ parameters each can be estimated -- a step that is also trivial to parallelize using modern computers. This point has been established recently in \cite{Carrieroetal}. 

Second, I additionally follow ideas in \cite{Bhattacharya2016} and use an efficient sampler for the Normal distribution, which makes full use of the Woodbury identity in order to sample from equation \eqref{phi_post}efficiently. The idea is that during sampling from the Normal distribution one has to obtain the Cholesky of a large matrix ($\overline{\mathbf{V}}_{i}$) which requires $\mathcal{O} \left( k^3\right)$ algorithmic operations. The transformation of \cite{Bhattacharya2016} allows sampling with worst case algorithmic complexity of $\mathcal{O} \left( T^{2}k\right)$ operations. Therefore, gains from replacing this step with the \cite{Bhattacharya2016} approach can be very substantial as either the dimension $n$ of the VAR or the number of lags $p$ increase. In particular, as $k=np + 1$ becomes larger than $T$, a significantly smaller number of algorithmic operations will be required in order to sample from the posterior of $\mathbf{\Phi}$.

Third, as the dimension of $\mathbf{\Phi}$ increases polynomially in $n$ and $p$, computation is not the sole concern; regularization also becomes an important aspect of statistical inference. Traditionally, empirical Bayes priors such as the Minnesota prior, have been used both in reduced-form and structural VARs; see \cite{Giannoneetal2015} and \cite{BaumeisterHamilton2015}, respectively. However, little theory exists on the high-dimensional shrinkage properties and posterior consistency of such empirical Bayes rules. In contrast, the \emph{horseshoe prior for sparse signals} proposed by \cite{Carvalhoetal2010} has been shown to lead to Bayes estimates that are consistent a-posteriori, and that attain a risk equivalent to the (Bayes) oracle; see \cite{Armaganetal2013} and \cite{Ghoshetal2016} and references therein.\footnote{\cite{vanderpas2014} also show that the horseshoe has good frequentist properties and can attain minimax-adaptive risk up to a constant, for squared error ($l_{2}$) loss. Recently, \cite{Kowal2019} also establish the excellent shrinkage properties of the horseshoe in a dynamic linear model for time-series data.} Using equation \eqref{phi_prior}, I specify the prior covariance matrix to have the following hierarchical structure
\begin{eqnarray}
\bm{\phi}_{i} \vert  \sigma^{2}_{i} \tau_{i}^{2} \mathbf{\Psi}_{i} & \sim & N_{k} \left( \mathbf{0}, \underline{\mathbf{V}}_{i} \right), \\
\underline{\mathbf{V}}_{i,(jj)} & = & \sigma^{2}_{i} \tau_{i}^{2} \psi_{i,j}^{2}, \\
\psi_{i,j} & \sim & Cauchy^{+} \left(0, 1 \right),  \\
\tau_{i}  & \sim & Cauchy^{+} \left(0, 1 \right),
\end{eqnarray}
where $\underline{\mathbf{V}}_{i,(jj)}$ denotes the $j^{th}$ diagonal element of prior covariance matrix $\underline{\mathbf{V}}_{i}$, $\mathbf{\Psi}_{i} = diag\left(\psi_{i,1}^{2},..., \psi_{i,k}^{2} \right)$, and $Cauchy^{+}$ denotes the half-Cauchy distribution with support on the set of positive real numbers ${\rm I\!R}^{+}$.\footnote{Note that the prior covariance matrix of $\bm{\phi}_{i}$ is a function of the VAR variance $\sigma^{2}_{i}$. This is done in order to enhance numerical stability when the endogenous variables in the VAR are measured in different units, even though in practice it is trivial to specify this prior to be independent of $\sigma^{2}_{i}$ (which would be essential in case we want to specify $\sigma^{2}_{i}$ to be time-varying using, for instance, a stochastic volatility specification.} The most important aspect of the horseshoe prior is that it requires absolutely no input from the researcher, while retaining at the same time its excellent shrinkage properties. There are various reparametrizations of this prior and associated MCMC sampling schemes. Here I follow \cite{neal2003} and use a slice sampler (within the Gibbs sampler algorithm) that allows to update $\psi_{i,(jj)}$ and $\tau_{i}$ efficiently in high-dimensions.

Fourth, a major challenge in the proposed algorithm is that in Step 2, $\mathbf{\Lambda}_{ij}$ has to be sampled from a multivariate truncated Normal distribution which is notoriously hard to simulate from \citep{Geweke1996}. In order to deal with this significant computational challenge, I follow \cite{Geweke1996} and sample each $\mathbf{\Lambda}_{ij}$ conditional on all remaining $\mathbf{\Lambda}_{-ij}$ elements. Most importantly, I adopt the exact minimax tilting method for generating i.i.d. data from a truncated normal distribution that was recently proposed by \cite{Botev2017}. This method allows orders of magnitude computational improvements relative to the truncated normal sampler proposed by \cite{Geweke1996} and others; see \cite{Botev2017} for a review of this literature. Given absence of prior information on the variance of $\mathbf{\Lambda}$ (above and beyond the information we have on the sign restrictions) I set $\underline{h}_{ij} = 4$ in the remainder of the paper, which is a noninformative choice for the parameters of a loadings matrix.

All the considerations above ensure that the algorithm proposed in the previous subsection is not only fast, but is also numerically stable and can scale up to much larger VAR dimensions than ever considered before in the literature. Note that there are remaining prior settings and algorithmic steps (those involving parameters $\mathbf{f}_{t}$ and $\mathbf{\Sigma}$), but these steps are already computationally trivial and may not be made  more efficient. Additionally, the prior for $\mathbf{f}_{t}$ is fixed by the required identification restrictions, and the $\mathbf{\Sigma}$ are integrated out with fairly noninformative values (I set $\rho_{i}=1$ and $\kappa_{i}=0.01$ for all $i$, in all VARs estimated in this paper).

\subsection{Likelihood-based testing of identifying assumptions in VARs}

Our modeling assumption that the structural shocks $\mathbf{f}_{t}$ are latent parameters that have to be estimated from the likelihood, has an enormous implication: sign, or any other structural restrictions, become part of the model likelihood and can be explicitly tested the same way economists test other parametric restrictions in regression models (e.g. inequality constraints as in \citealp{Geweke1996}). This concept might not make sense for economic shocks that are indisputable such as the effects of an aggregate demand/supply shock, however, there are cases of shocks where the expected sign might not be known a-priori with certainty. Alternatively, a researcher might want to statistically test the plausibility of certain zero restrictions, or simply compare the performance of two different VAR models (e.g. with different number of lags and/or variables) given the same set of identifying restrictions. In the context of the proposed VAR model with factor structure, all these cases can be tested explicitly using marginal likelihoods or Bayes factors. Parametric (i.e. zero or sign) restrictions on $\mathbf{\Lambda}$ that agree with the information in the data will result in higher marginal data likelihood relative to restrictions that are not supported by the data. Put differently, given the decomposition in equation \eqref{factor_decomp}, plausible restrictions in $\mathbf{\Lambda}$ will result in estimation of a more precise unconditional VAR covariance matrix $\mathbf{\Omega}$.

Marginal likelihoods are not numerically stable in high-dimensional VARs \citep{Giannoneetal2015} and they can be demanding to compute even in smaller VARs with many layers of latent parameters (e.g. hierarchical priors like the horseshoe; stochastic variances; Markov-switching coefficients). In order to deal with this computational aspect, I instead propose to calculate the Deviance Information Criterion (DIC) of \cite{Spiegelhalteretal2002} as a default criterion for assessing model fit. For the matrix of VAR model parameters $\mathbf{\Theta} = \left( \mathbf{\Phi}, \mathbf{\Lambda}, \lbrace \mathbf{f}_{t} \rbrace_{t=1}^{T} , \mathbf{\Lambda} \right)$ the DIC is defined as the quantity
\begin{equation}
DIC  =  - 4 E_{ p \left( \mathbf{\Theta} \vert \mathbf{y} \right) } \left( \log f \left( \mathbf{y} \vert \mathbf{\Theta} \right) \right) + 2 \log f \left( \mathbf{y} \vert \widehat{\mathbf{\Theta}} \right),
\end{equation}
where $\log f \left( \mathbf{y} \vert \mathbf{\Theta} \right)$ is the log of the likelihood function implied by the regression in \eqref{VAR_one_eq}. The first term in the criterion above is the expectation of the likelihood w.r.t the parameter posterior, and can be obtained numerically using Monte Carlo integration by simply evaluating the likelihood at each MCMC draw from the posterior of the parameters $\mathbf{\Theta}$. The second term is the likelihood function evaluated at an estimate $\widehat{\mathbf{\Theta}}$ of high posterior density (typically posterior mean or mode). As with all other information criteria used in statistics/econometrics, lower values signify better fit. The DIC is not a first-order approximation to the marginal likelihood, in the same way that the Bayesian Information Criterion (BIC) is. The marginal likelihood, also known as \emph{prior predictive distribution}, addresses the issue of how well the data are predicted by the priors. In this sense, the DIC is a criterion that is closely related to measuring fit according to the \emph{posterior predictive distribution}, rather than marginal likelihoods. As a result, for the purpose of assessing the fit of a VAR that is intended to be used for out-of-sample projections (impulse responses, forecast error variance decompositions etc), the DIC can be considered as a more appropriate predictive measure of fit compared to marginal likelihoods or alternative in-sample measures of fit.

\clearpage
\renewcommand{\theequation}{B.\arabic{equation}} \setcounter{equation}{0} %
\renewcommand{\thetable}{B\arabic{table}} \setcounter{table}{0}
\renewcommand{\thefigure}{B\arabic{figure}} \setcounter{figure}{0}
\setcounter{footnote}{0}
\section{Data Appendix}
\subsection{Data for the simulation study}
The data used in the simulation exercise are an augmented version of the data used in \cite{Ariasetal2019}. The measure of commodity prices they provide comes from Global Financial Data, while all remaining variables come originally from St Louis' Federal Reserve Economic Data (FRED; \href{https://fred.stlouisfed.org/}{https://fred.stlouisfed.org/}). Monthly GDP and GDP deflator are constructed using interpolation, and the reader is referred to the data supplement of \cite{Ariasetal2019} for more information. The variables these authors use are augmented with stock prices, M1, unemployment rate, industrial production, employment, consumer prices (total), core consumer prices (total, less food and energy) and personal consumption deflator.

All variable mnemonics, short descriptions, and sources are visible in \autoref{table:data_mc}. The last column (Tcode) refers to the stationarity transformation used for each variable, namely 1: levels, and 5: first differences of logarithm (growth rates).

\begin{table}[H]
\caption{Data used in Monte Carlo exercise}  \label{table:data_mc}
\centering
\resizebox{\textwidth}{!}{ 
\begin{tabular}{lllc} \hline
Mnemonic	&	Description	&	Source	&	Tcode	\\ \hline\hline
GDPC1	&	Monthly real GDP	&	\cite{Ariasetal2019}	&	5	\\
GDPDEFL	&	Monthly GDP deflator	&	\cite{Ariasetal2019}	&	5	\\
FEDFUNDS	&	Fed funds rate	&	\cite{Ariasetal2019}	&	1	\\
CPRINDEX	&	Commodity price index	&	\cite{Ariasetal2019}	&	5	\\
TRARR	&	Total reserves	&	\cite{Ariasetal2019}	&	5	\\
BOGNONBR	&	Nonborrowed reserves	&	\cite{Ariasetal2019}	&	5	\\
$\hat{~}$GSPC	&	S\&P 500 prices	&	Yahoo! Finance	&	5	\\
M1REAL	&	Real M1 money stock	&	FRED	&	5	\\
UNRATE	&	Unemployment rate	&	FRED	&	1	\\
INDPRO	&	Industrial production index, all industries	&	FRED	&	5	\\
PAYEMS	&	Employment, total	&	FRED 	&	5	\\
CPIAUSL	&	Consumer price index, all items	&	FRED 	&	5	\\
CPILFESL	&	Core CPI	&	FRED 	&	5	\\
PCEPILFE	&	Core PCE deflator	&	FRED	&	5	\\ \hline
\end{tabular}
}
\end{table}

\subsection{Data used for the large-scale VAR model for the US}
The full list of variables is shown in \autoref{table:data_credit} and they pertain to the sample 1985Q1 - 2013Q2. The original mortgage rate used by \cite{Furlanettoetal2017} was only available after 1990Q1, and for that reason it has been replaced by the 30-year mortgage rate provided by FRED (contemporaneous correlation between the two series is 0.9967). All remaining series used in the empirical exercise are exactly those described in Table 11 of \cite{Furlanettoetal2017}, augmented with a few additional measures of output, consumer prices, and stock prices. The fifteen variables used in the large VAR can be seen in the first column of \autoref{table:sign_large}. In this list \textbf{stock prices} refers to S\&P500 while \textbf{stock prices 2} refers to Dow Jones Industrial Average. Similarly, \textbf{spread} refers to the Baa minus fed funds rate spread, while \textbf{spread 2} refers to the GZ credit spread.
\vskip 1cm
\begin{table}[H]
\caption{Data used in the empirical exercise}  \label{table:data_credit}
\centering
{\scriptsize
\begin{tabularx}{\textwidth}{lXX} \hline
Variable &  Description  & Source \\ \hline\hline
GDP	&	Log of real GNP/GDP	&	Federal Reserve Bank of Philadelphia	\\
GDP deflator	&	Log of price index for GNP/GDP	&	Federal Reserve Bank of Philadelphia	\\
Interest rate	&	3-month treasury bill	&	Federal Reserve Bank of St. Louis	\\
Investment	&	Log of real gross private domestic investment	&	Federal Reserve Bank of St. Louis	\\
Stock prices	&	Log of real S\&P 500	&	Yahoo! Finance	\\
Total credit	&	Log of loans to non-financial private sector	&	Board of Governors of the Federal Reserve System	\\
Mortgages	&	Log of home mortgages of households and non-profit organizations	&	Board of Governors of the Federal Reserve System	\\
Real estate value	&	Log of real estate at market value of households and non-profit organizations	&	Board of Governors of the Federal Reserve System	\\
Corporate bond yield	&	Moody’s baa corporate bond yield	&	Federal Reserve Bank of St. Louis	\\
Federal funds rate	&	Federal funds rate	&	Federal Reserve Bank of St. Louis	\\
GZ credit spread	&	Senior unsecured corporate bond spreads (non-financial firms)	&	Gilchrist and Zakraj{\v{s}}ek (2012)$^{a}$	\\
EBP	&	Excess bond premium	&	Gilchrist and Zakraj{\v{s}}ek (2012)$^{a}$	\\
VIX	&	Stock market volatility index	&	Bloom (2009)$^{b}$	\\
Mortgage rates	&	Home mortgages, fixed 30YR, Effective interest rate	&	Federal Reserve Bank of St. Louis	\\
Employment	&	Log of total nonfarm employment	&	Federal Reserve Bank of Philadelphia	\\
Core prices	&	Log of core consumer price index	&	Federal Reserve Bank of Philadelphia	\\
Stock prices 2	&	Log of real DJIA	&	Yahoo! Finance	\\ \hline
\par
\multicolumn{3}{p{\dimexpr \textwidth}}{$^{a}$ Gilchrist, Simon, and Egon Zakraj{\v{s}}ek (2012), Credit Spreads and Business Cycle Fluctuations. American Economic Review, 102 (4): 1692-1720. \hspace{8cm} \linebreak $^{b}$  Bloom, Nicholas (2009), The Impact of Uncertainty Shocks. Econometrica, 77: 623-685. \hfill } 
\end{tabularx}
}
\end{table}

\clearpage
\renewcommand{\theequation}{C.\arabic{equation}} \setcounter{equation}{0} %
\renewcommand{\thetable}{C\arabic{table}} \setcounter{table}{0}
\renewcommand{\thefigure}{C\arabic{figure}} \setcounter{figure}{0}
\setcounter{footnote}{0}
\section{Additional results}

\subsection{Additional results for the first Monte Carlo exercise}

\begin{figure}[H]
\centering
\includegraphics[trim= 2cm 1cm 1cm 1cm, width=\textwidth]{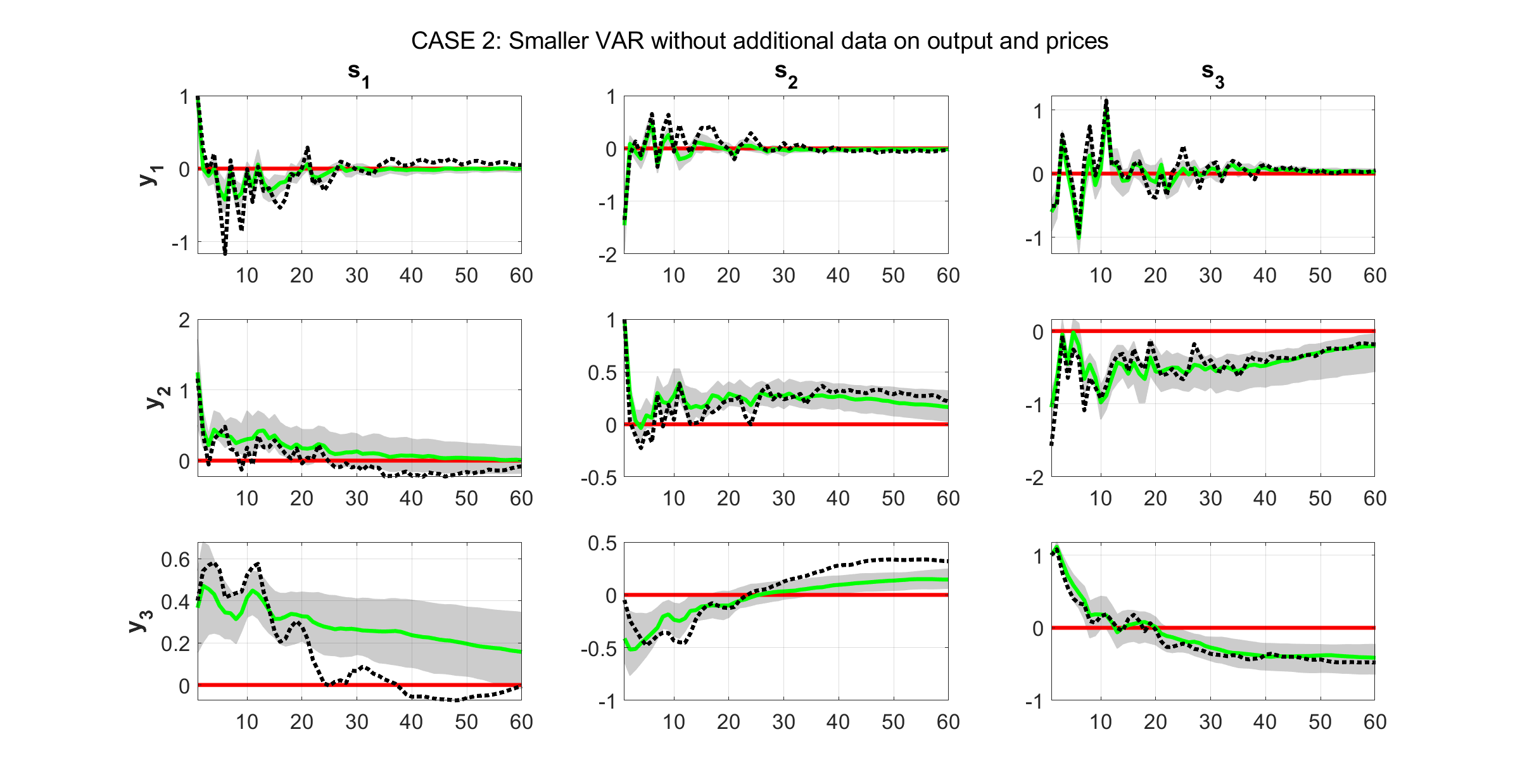}
\caption{\emph{Impulse response functions of the first three artificially generated variables (denoted as $y_{1},y_{2},y_{3}$) in response to the three identified shocks (denoted as $s_{1},s_{2},s_{3}$) in model C2 (misspecified VAR dimension). The green solid lines show the posterior median IRFs over the 500 Monte Carlo iterations, and the gray shaded areas their associated 90\% bands. The true IRFs based on the DGP are shown using the black dashed lines.}}
\end{figure}

\begin{figure}[H]
\centering
\includegraphics[trim= 2cm 1cm 1cm 1cm, width=\textwidth]{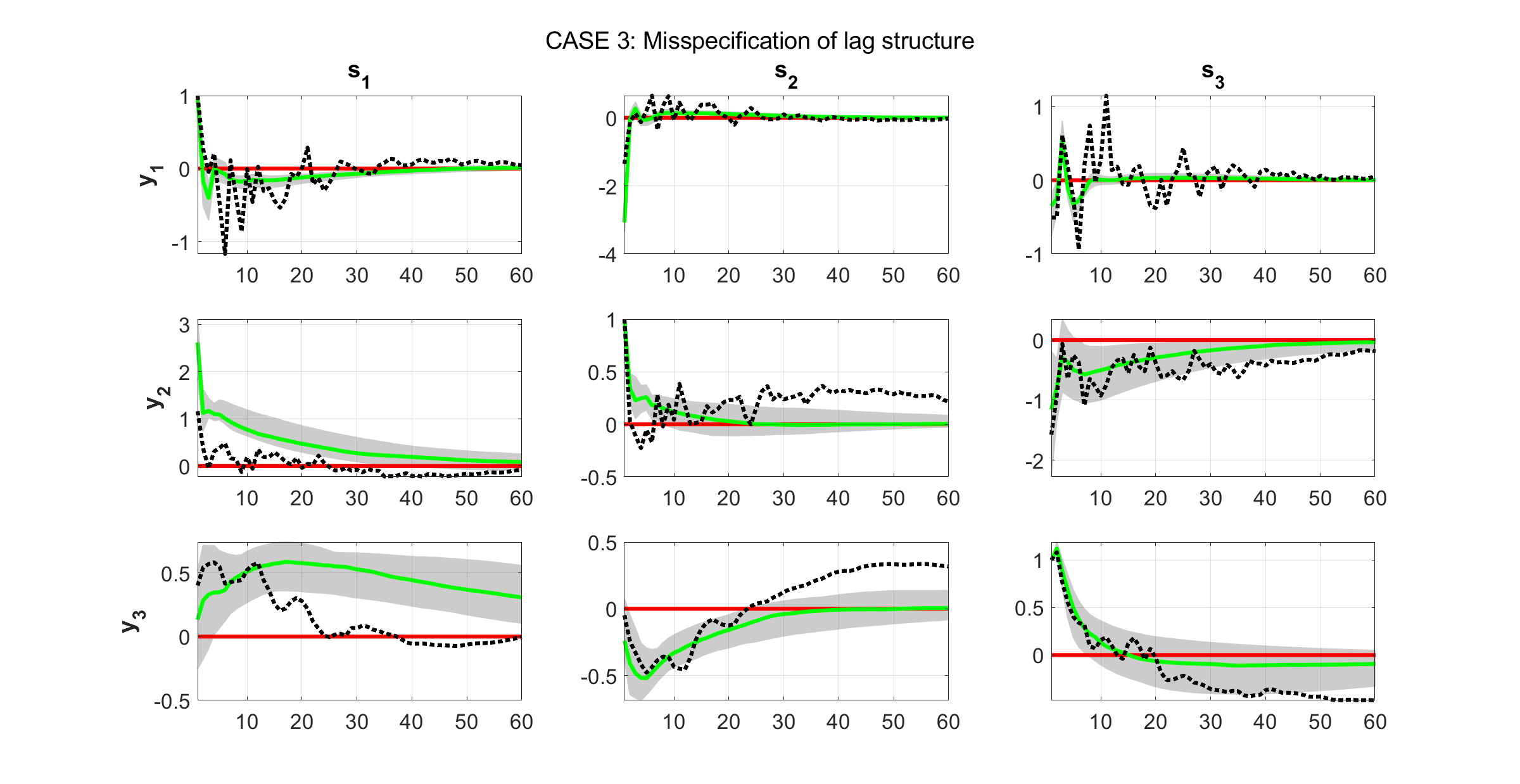}
\caption{\emph{Impulse response functions of the first three artificially generated variables (denoted as $y_{1},y_{2},y_{3}$) in response to the three identified shocks (denoted as $s_{1},s_{2},s_{3}$) in model C3 (misspecified number of lags). The green solid lines show the posterior median IRFs over the 500 Monte Carlo iterations, and the gray shaded areas their associated 90\% bands. The true IRFs based on the DGP are shown using the black dashed lines.}}
\end{figure}

\begin{figure}[H]
\centering
\includegraphics[trim= 2cm 1cm 1cm 1cm, width=\textwidth]{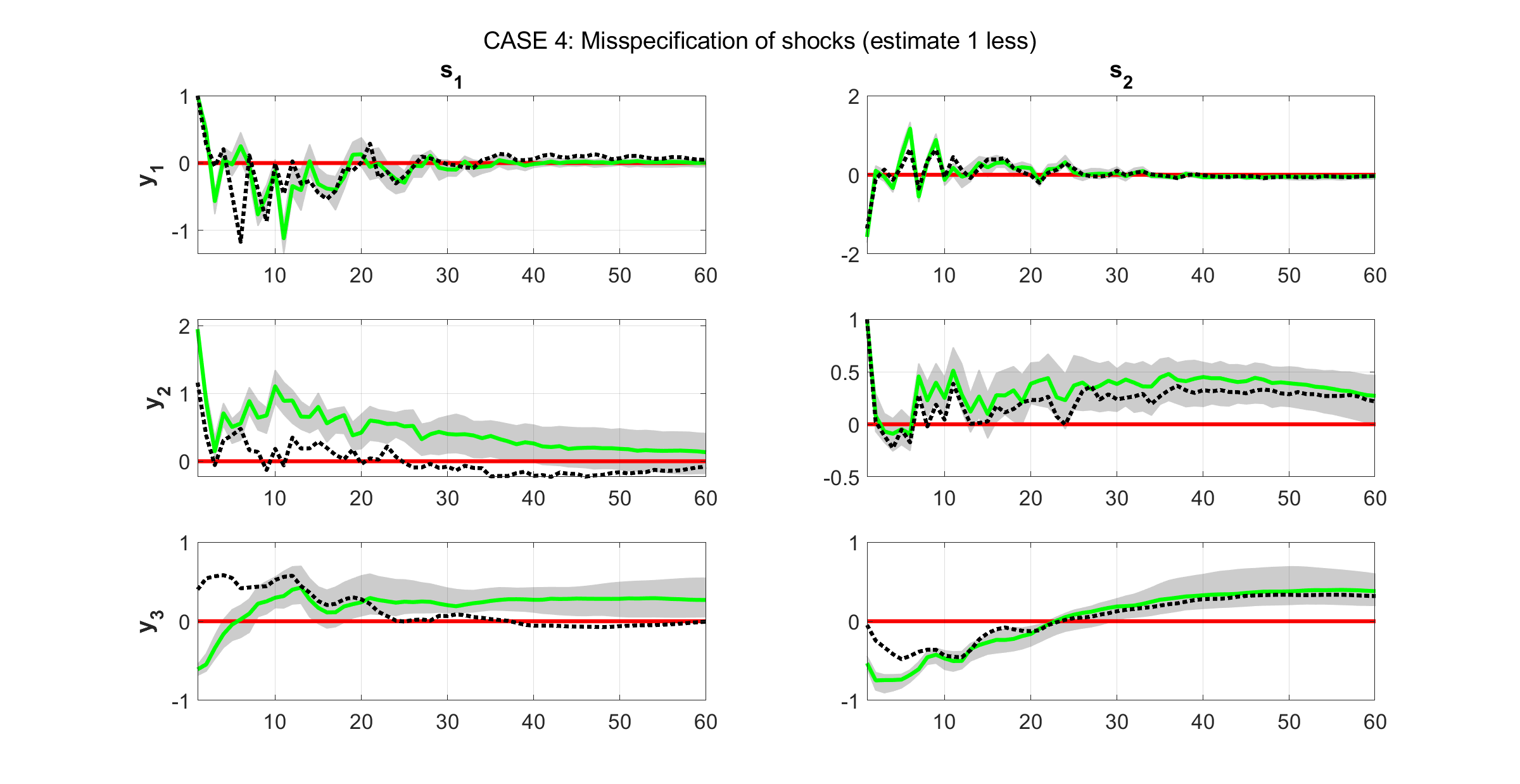}
\caption{\emph{Impulse response functions of the first three artificially generated variables (denoted as $y_{1},y_{2},y_{3}$) in response to the three identified shocks (denoted as $s_{1},s_{2},s_{3}$) in model C4 (misspecified number of shocks -- one less). The green solid lines show the posterior median IRFs over the 500 Monte Carlo iterations, and the gray shaded areas their associated 90\% bands. The true IRFs based on the DGP are shown using the black dashed lines.}}
\end{figure}

\begin{figure}[H]
\centering
\includegraphics[trim= 2cm 1cm 1cm 1cm, width=\textwidth]{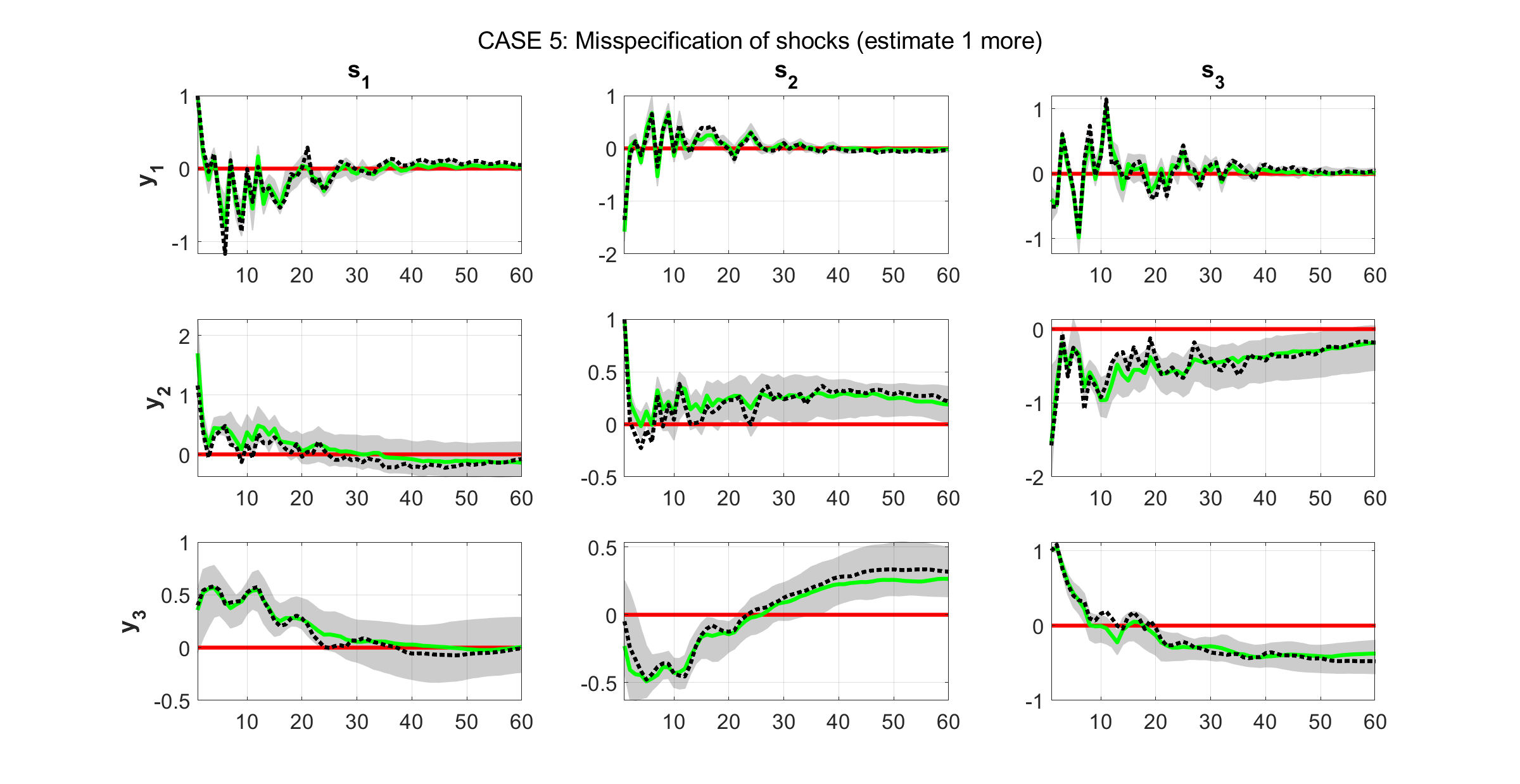}
\caption{\emph{Impulse response functions of the first three artificially generated variables (denoted as $y_{1},y_{2},y_{3}$) in response to the three identified shocks (denoted as $s_{1},s_{2},s_{3}$) in model C5 (misspecified number of shocks -- one more). The green solid lines show the posterior median IRFs over the 500 Monte Carlo iterations, and the gray shaded areas their associated 90\% bands. The true IRFs based on the DGP are shown using the black dashed lines.}}
\end{figure}

\subsection{Additional empirical exercise: Measuring optimism shocks}
\begin{figure}[H]
    \centering
    \begin{subfigure}[t]{0.48\textwidth}
        \centering
        \includegraphics[trim= 3cm 1cm 3cm 1cm,width=\linewidth]{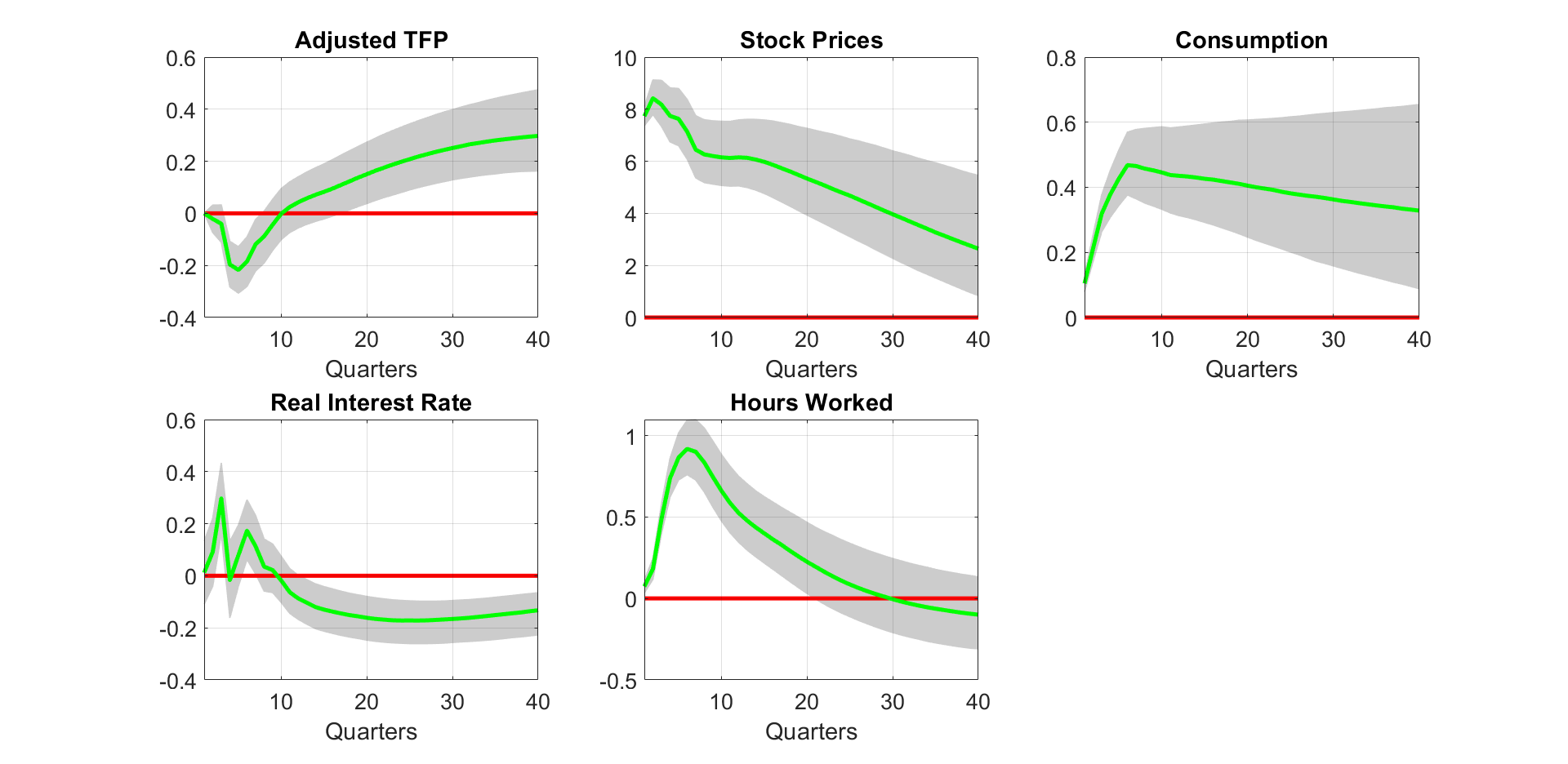}
        \caption{\cite{MountfordUhlig2009} algorithm}
    \end{subfigure}
    \hfill
    \begin{subfigure}[t]{0.48\textwidth}
        \centering
        \includegraphics[trim= 3cm 1cm 3cm 1cm,width=\linewidth]{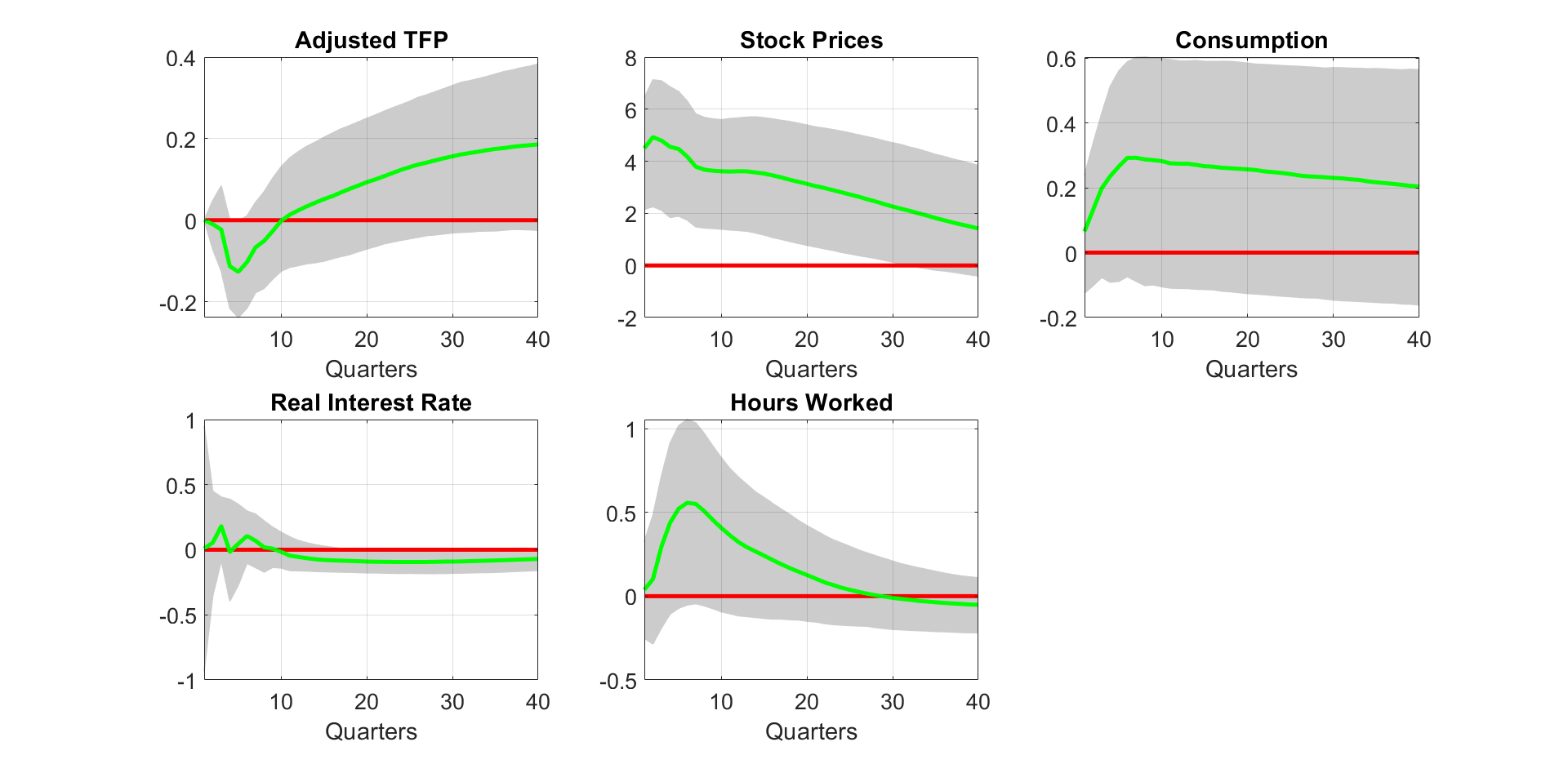}  
        \caption{\cite{Ariasetal2018} algorithm}
    \end{subfigure}
    \vskip 1cm
    \begin{subfigure}[t]{0.48\textwidth}
        \centering
        \includegraphics[trim= 3cm 1cm 3cm 0.5cm,width=\linewidth]{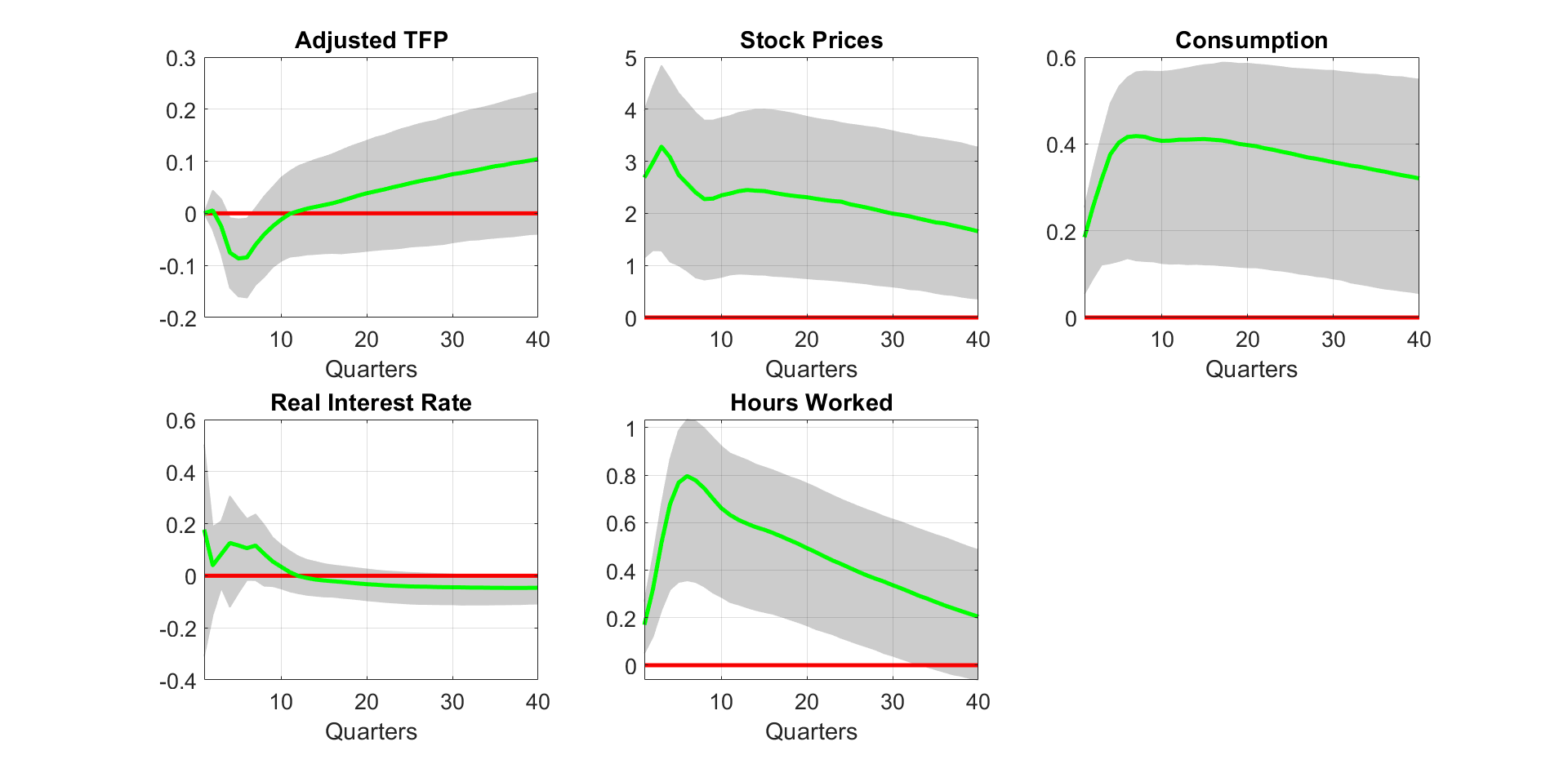} 
        \caption{Factor sign restrictions algorithm}
    \end{subfigure}
    \hfill
    \begin{subfigure}[t]{0.48\textwidth}
        \centering
        \includegraphics[trim= 3cm 1cm 3cm 0.5cm,width=\linewidth]{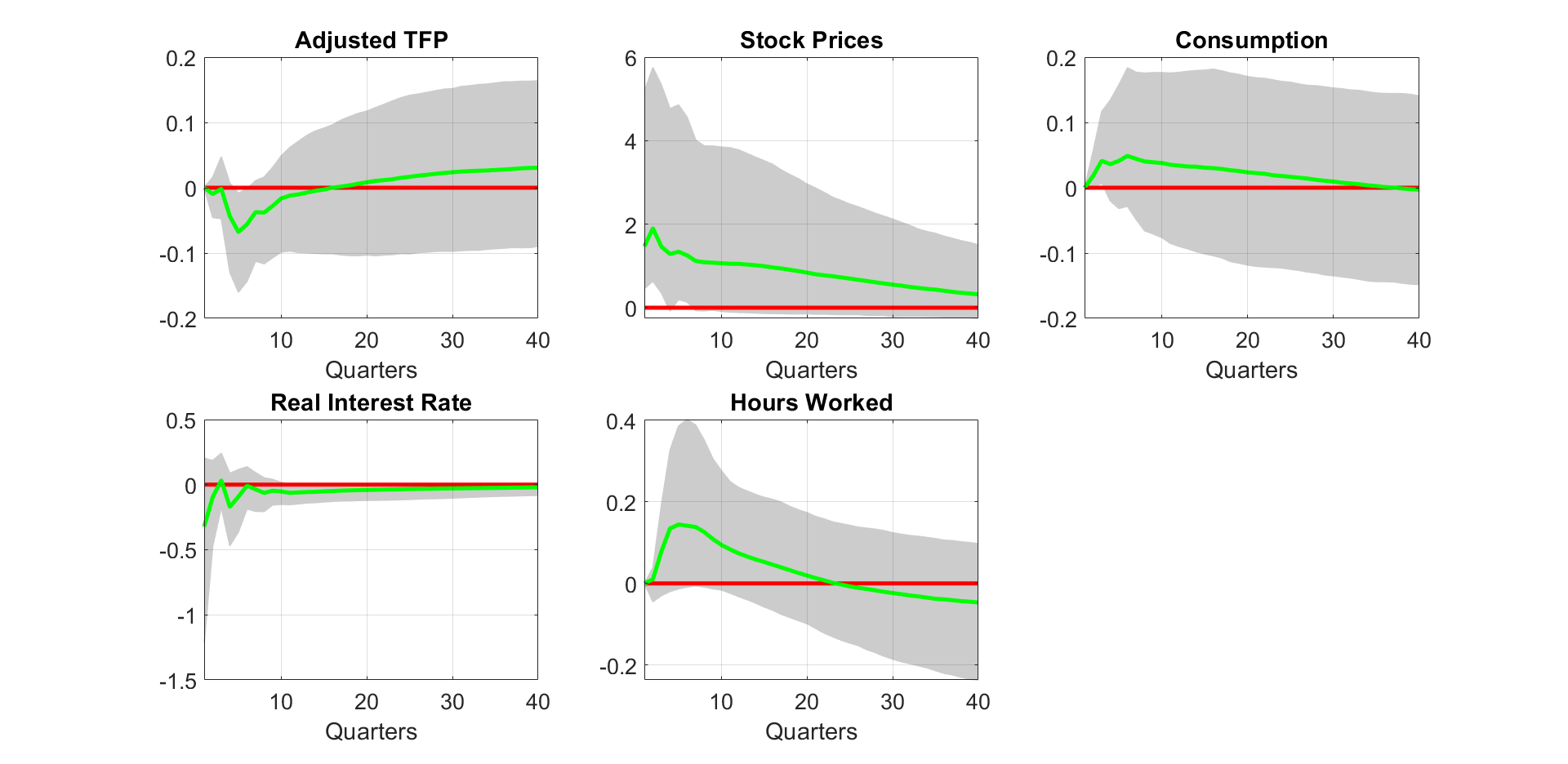} 
        \caption{Factor sign restrictions algorithm, with additional zero restrictions} 
    \end{subfigure}
\caption{\emph{This figure replicates the results in \cite{NBERw17651}, using a five variable VAR for assessing the effects of an optimism shock. The sign-restricted impulse responses of the five variables are estimated using (a) the PFA algorithm of \cite{MountfordUhlig2009}; (b) the importance sampling approach of \cite{Ariasetal2018}; (c) the Gibbs sampler algorithm proposed in this paper; and (d) the algorithm of this paper, with additional zero restrictions in consumption and hours.}} \label{fig:optimism_graph}
\end{figure}

Here I undertake an additional empirical exercise that will help shed more light on the performance of the new algorithm in real data. This empirical exercise is based on Section 6 of \cite{Ariasetal2018}. These authors use the example in \cite{NBERw17651} in order to compare their novel importance sampling algorithm to the penalty function approach (PFA) of \cite{MountfordUhlig2009}. For the sake of comparability, we maintain their empirical setting, and for that reason we estimate VAR($4$) models using the following five dependent variables: adjusted TFP, stock prices, consumption, the real interest rate, and hours worked. We identify a single optimism shock by restricting contemporaneously TFP to have a zero response, and stock prices to react positively. The signs in the remaining three variables are not restricted. Panels (a) and (b) in \autoref{fig:optimism_graph} show the estimated responses using the PFA and importance sampling algorithms, respectively. These two panels are identical to panels (a) and (b) in Figure 1 of \cite{Ariasetal2018}. The main point these authors make in their study is that the PFA algorithm ends up distorting the responses of stock prices, consumption and hours. Once their proposed importance sampling algorithm is considered, the significant responses found in \cite{NBERw17651} disappear.

Panel (c) of \autoref{fig:optimism_graph} shows the results from the factor sign restrictions algorithm using the same zero restriction on TFP and positive sign restriction on stock prices. The response of stock prices is not as pronounced as in \cite{NBERw17651}, and in general the responses for TFP, stock prices and real interest rate are equivalent to \cite{Ariasetal2018}. However, the responses of consumption and hours are still strongly different from zero, even though they have error bands and shapes that look much closer to those produced by the algorithm of \cite{Ariasetal2018}. Nevertheless, one aspect of the factor sign restrictions is that we can explicitly derive the implied fit to the VAR of imposing various restrictions. Therefore, we can explicitly test the premise of \cite{Ariasetal2018} that consumption and hours are not affected by optimism shocks. Panel (d) in \autoref{fig:optimism_graph} repeats estimation of the VAR using factor sign restrictions algorithm with additional zero restrictions in consumption and hours. The IRFs now look quantitatively and qualitatively closer to those in panel (b). Most importantly, we are able to test whether the model in panel (c) or (d) is supported by the data, that is, test whether the zero restrictions in consumption and hours. The DIC for the model without these restrictions is -13109.49 while the DIC for the model with the two zero restrictions is -15267.64. Thus, data evidence (which is conditional, of course, on the specific parametric likelihood specification and prior) suggests that the premise of \cite{Ariasetal2018} -- that optimism shocks do not affect consumption and hours -- is correct.

\subsection{Additional results for the baseline VAR of \cite{Furlanettoetal2017}}

\begin{figure}[H]
    \centering
    \begin{subfigure}[t]{0.49\textwidth}
        \centering
        \includegraphics[trim= 3cm 1cm 3cm 1cm,width=\linewidth]{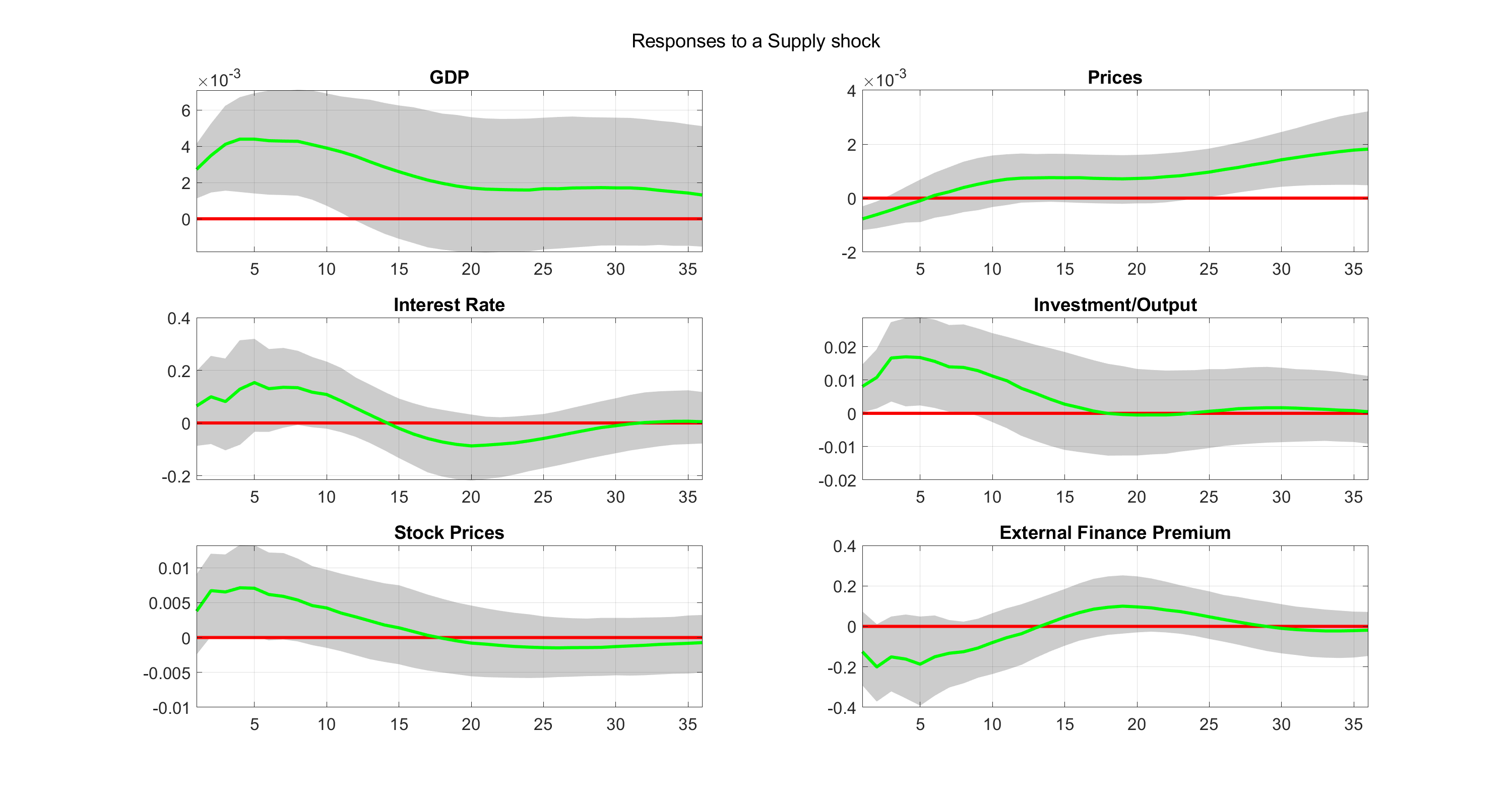}
        \caption{\cite{Furlanettoetal2017} algorithm}
    \end{subfigure}
    \hfill
    \begin{subfigure}[t]{0.49\textwidth}
        \centering
        \includegraphics[trim= 3cm 1cm 3cm 1cm,width=\linewidth]{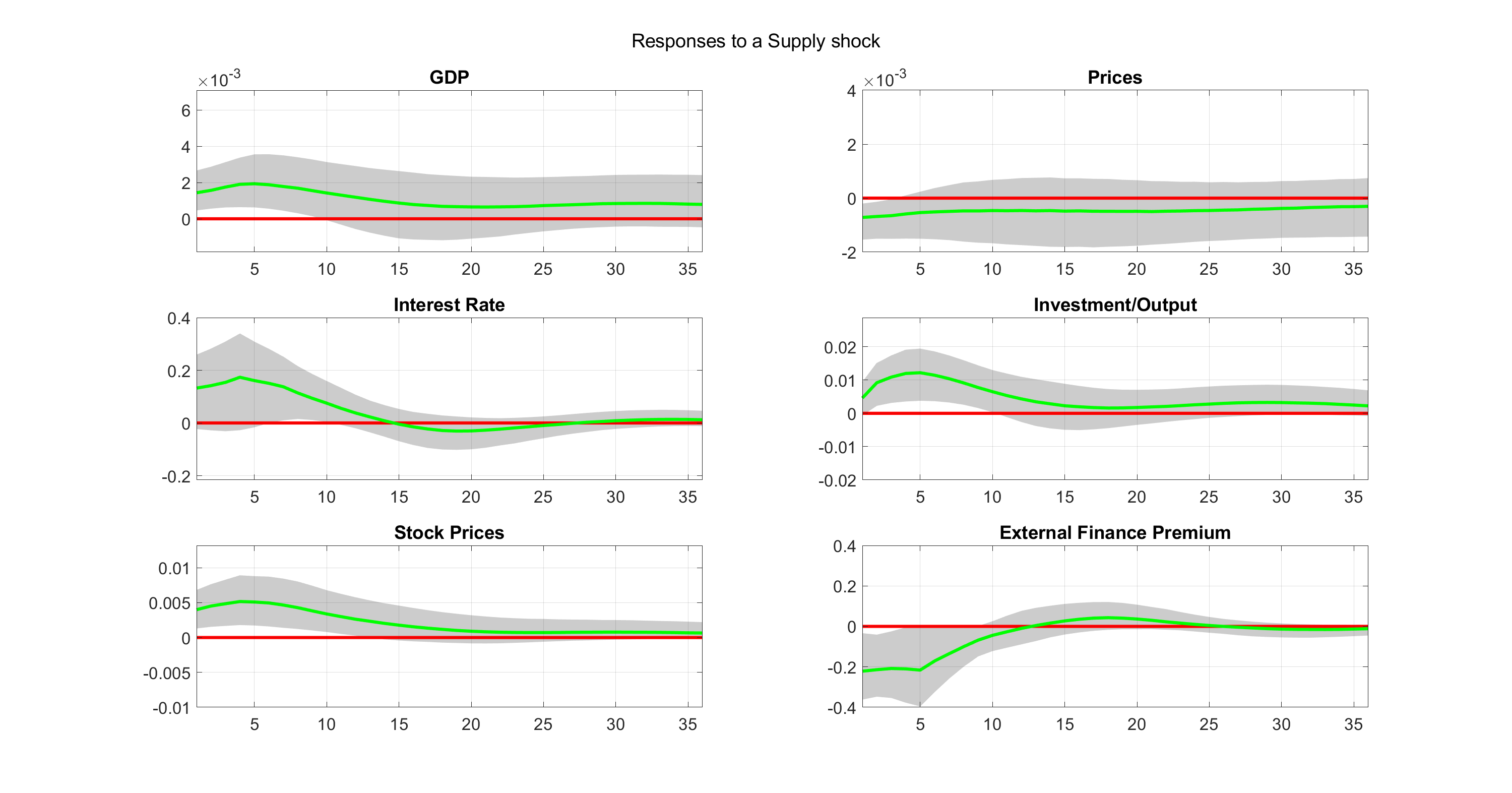}  
        \caption{Factor sign restrictions algorithm}
    \end{subfigure}
\caption{\emph{This figure replicates the impulse response functions to an aggregate supply shock using the baseline specification of \cite{Furlanettoetal2017}. Panel (a) shows results based on the exact configuration of \citet[see Figure 1]{Furlanettoetal2017}, using the algorithm of \cite{RubioRamirezetal2010}. Panel (b) replicates the same financial shock using the new sign restrictions algorithm.}} \label{baseline_supply}
\end{figure}

\begin{figure}[H]
    \centering
    \begin{subfigure}[t]{0.49\textwidth}
        \centering
        \includegraphics[trim= 3cm 1cm 3cm 1cm,width=\linewidth]{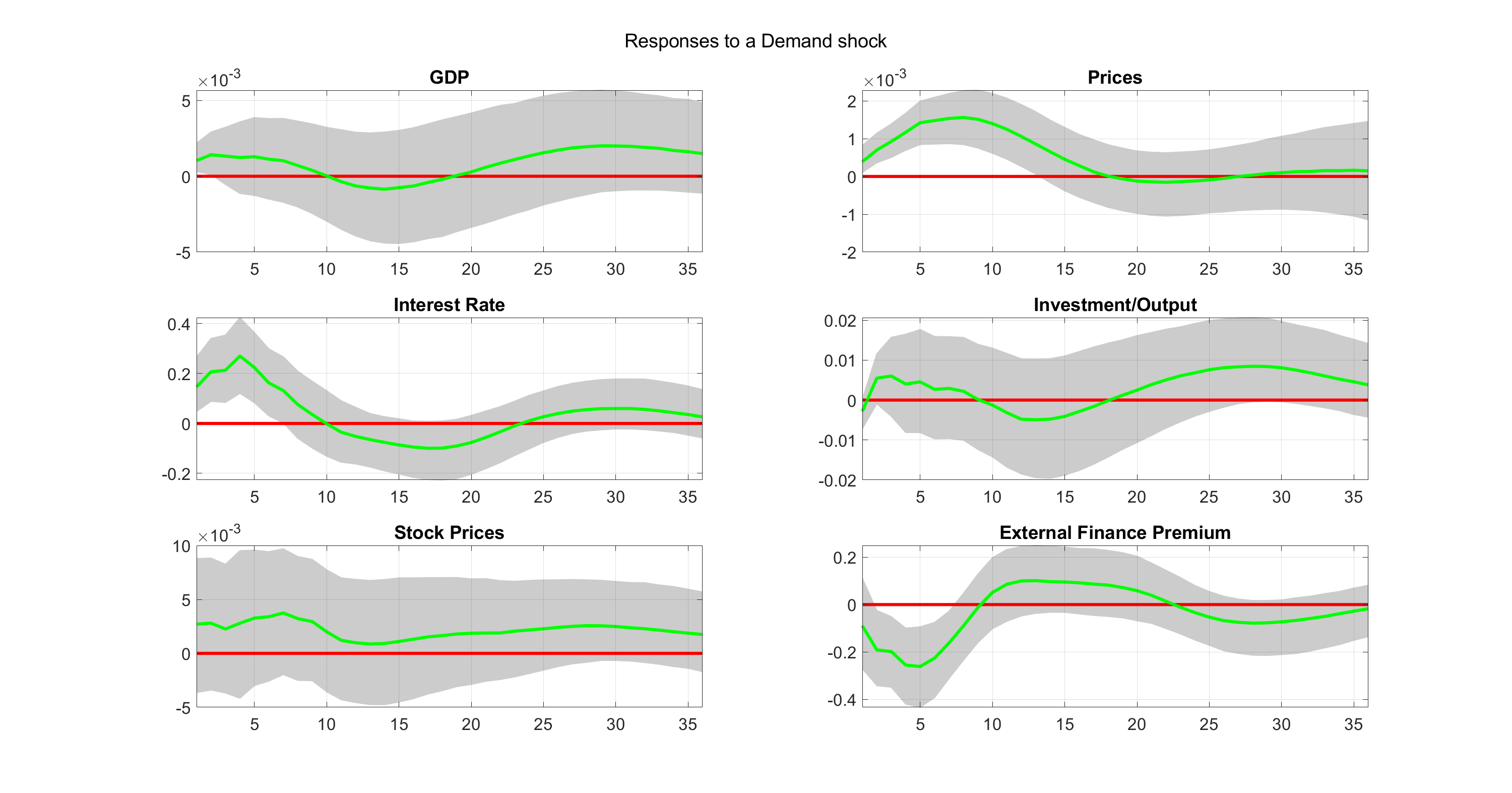}
        \caption{\cite{Furlanettoetal2017} algorithm}
    \end{subfigure}
    \hfill
    \begin{subfigure}[t]{0.49\textwidth}
        \centering
        \includegraphics[trim= 3cm 1cm 3cm 1cm,width=\linewidth]{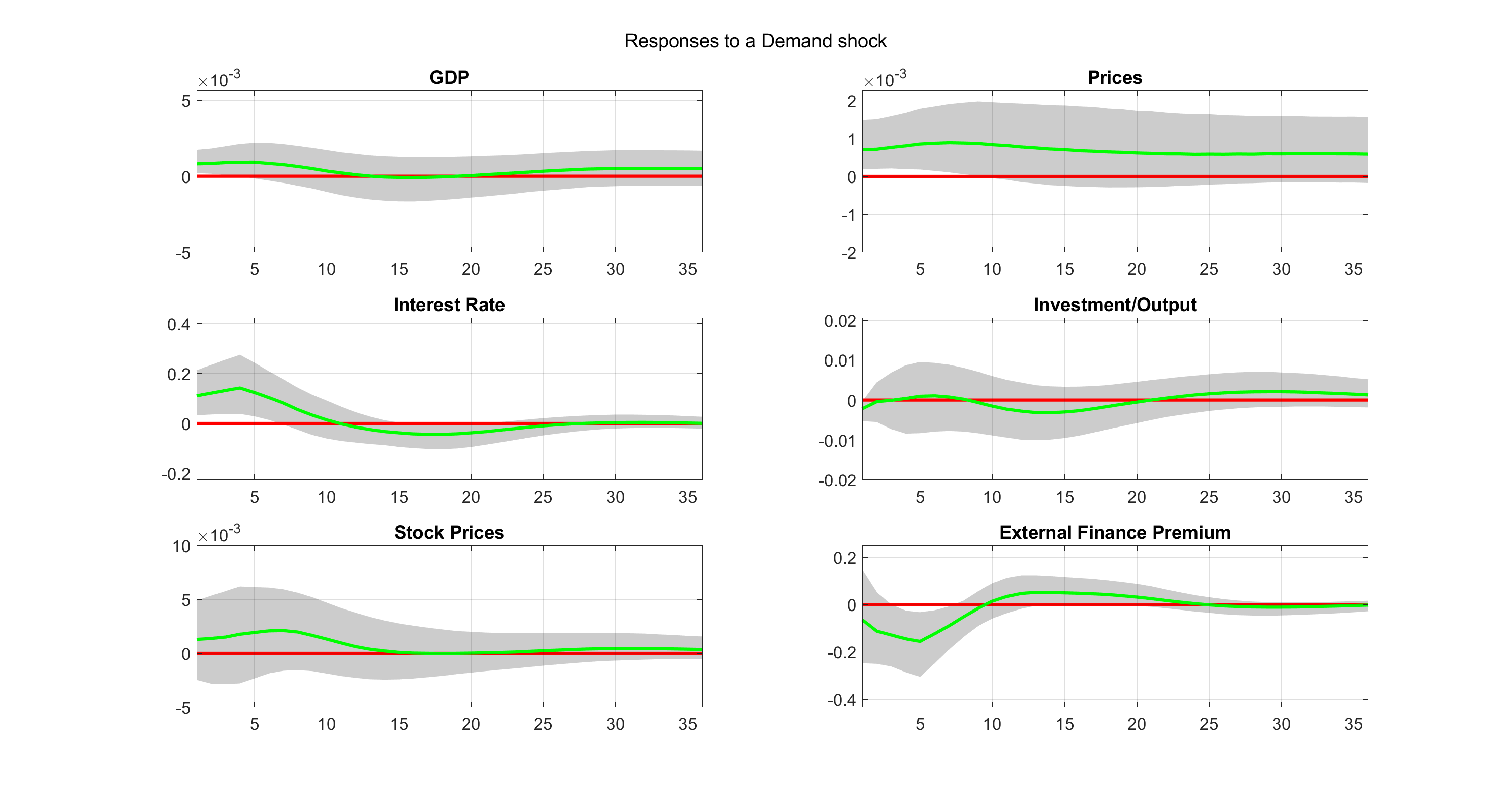}  
        \caption{Factor sign restrictions algorithm}
    \end{subfigure}
\caption{\emph{This figure replicates the impulse response functions to an aggregate demand shock using the baseline specification of \cite{Furlanettoetal2017}. Panel (a) shows results based on the exact configuration of \citet[see Figure 1]{Furlanettoetal2017}, using the algorithm of \cite{RubioRamirezetal2010}. Panel (b) replicates the same financial shock using the new sign restrictions algorithm.}} \label{baseline_demand}
\end{figure}

\begin{figure}[H]
    \centering
    \begin{subfigure}[t]{0.49\textwidth}
        \centering
        \includegraphics[trim= 3cm 1cm 3cm 1cm,width=\linewidth]{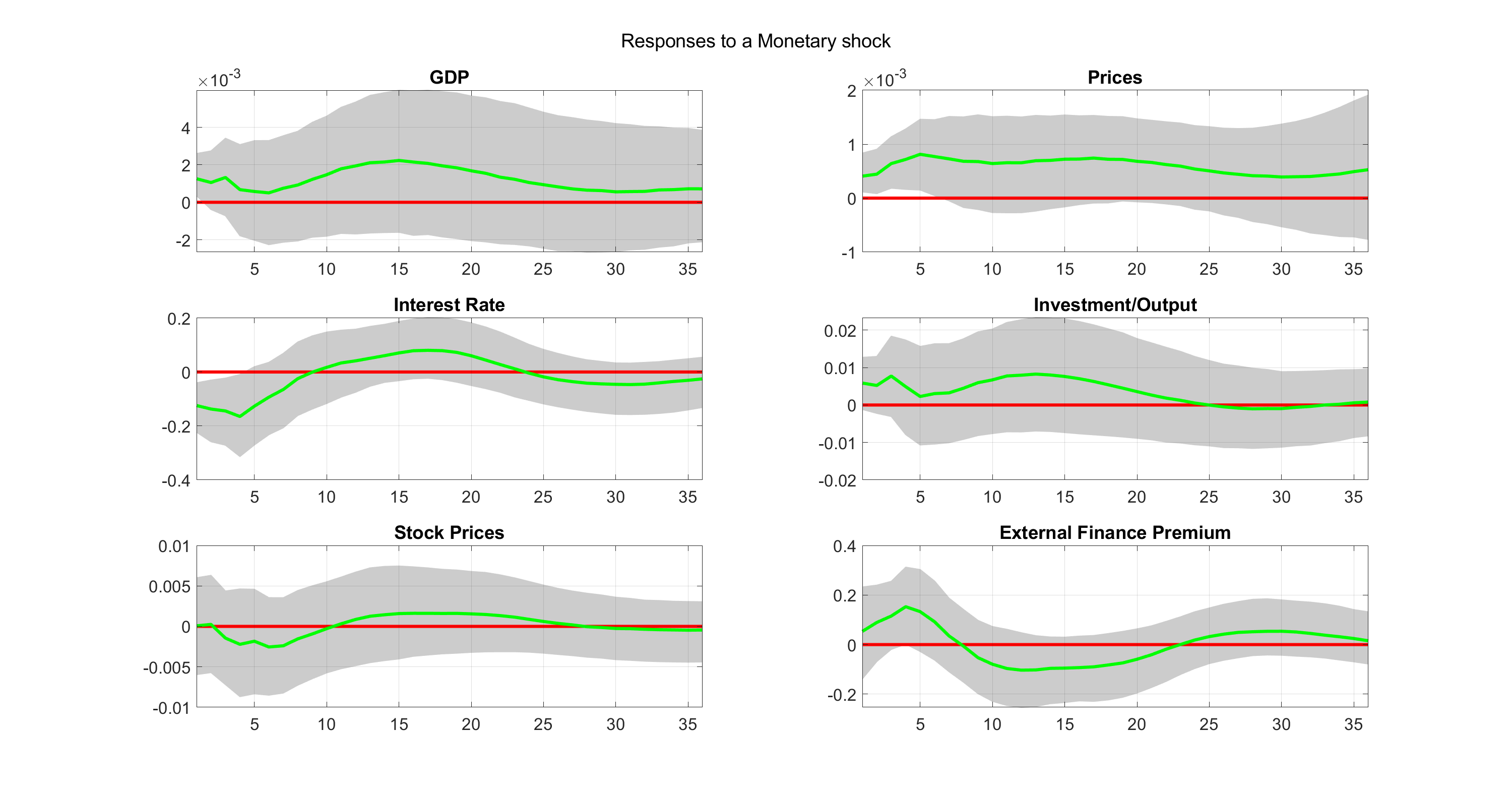}
        \caption{\cite{Furlanettoetal2017} algorithm}
    \end{subfigure}
    \hfill
    \begin{subfigure}[t]{0.49\textwidth}
        \centering
        \includegraphics[trim= 3cm 1cm 3cm 1cm,width=\linewidth]{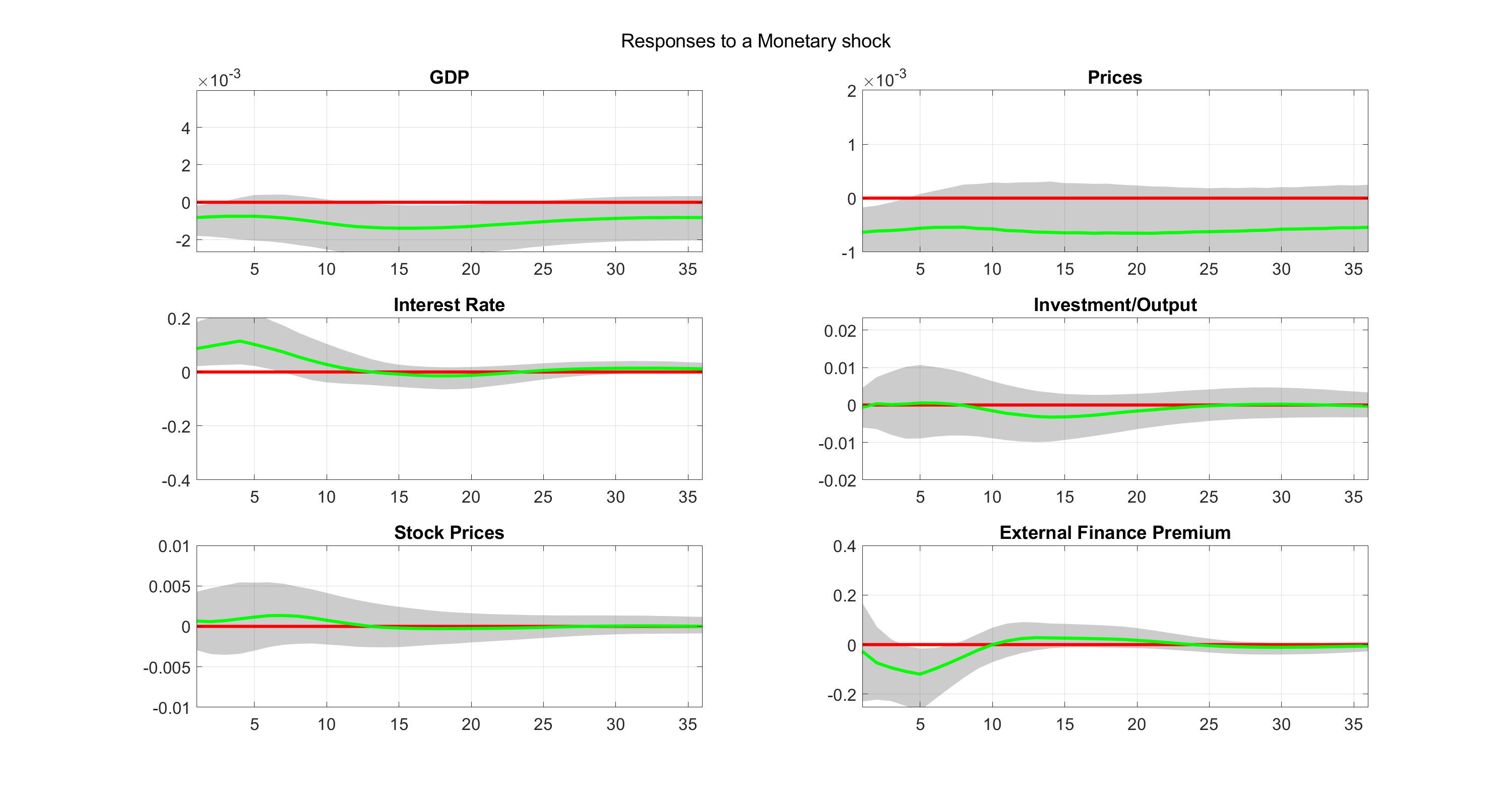}  
        \caption{Factor sign restrictions algorithm}
    \end{subfigure}
\caption{\emph{This figure replicates the impulse response functions to a monetary policy shock using the baseline specification of \cite{Furlanettoetal2017}. Panel (a) shows results based on the exact configuration of \citet[see Figure 1]{Furlanettoetal2017}, using the algorithm of \cite{RubioRamirezetal2010}. Panel (b) replicates the same financial shock using the new sign restrictions algorithm.}} \label{baseline_monetary}
\end{figure}

\begin{figure}[H]
    \centering
    \begin{subfigure}[t]{0.49\textwidth}
        \centering
        \includegraphics[trim= 3cm 1cm 3cm 1cm,width=\linewidth]{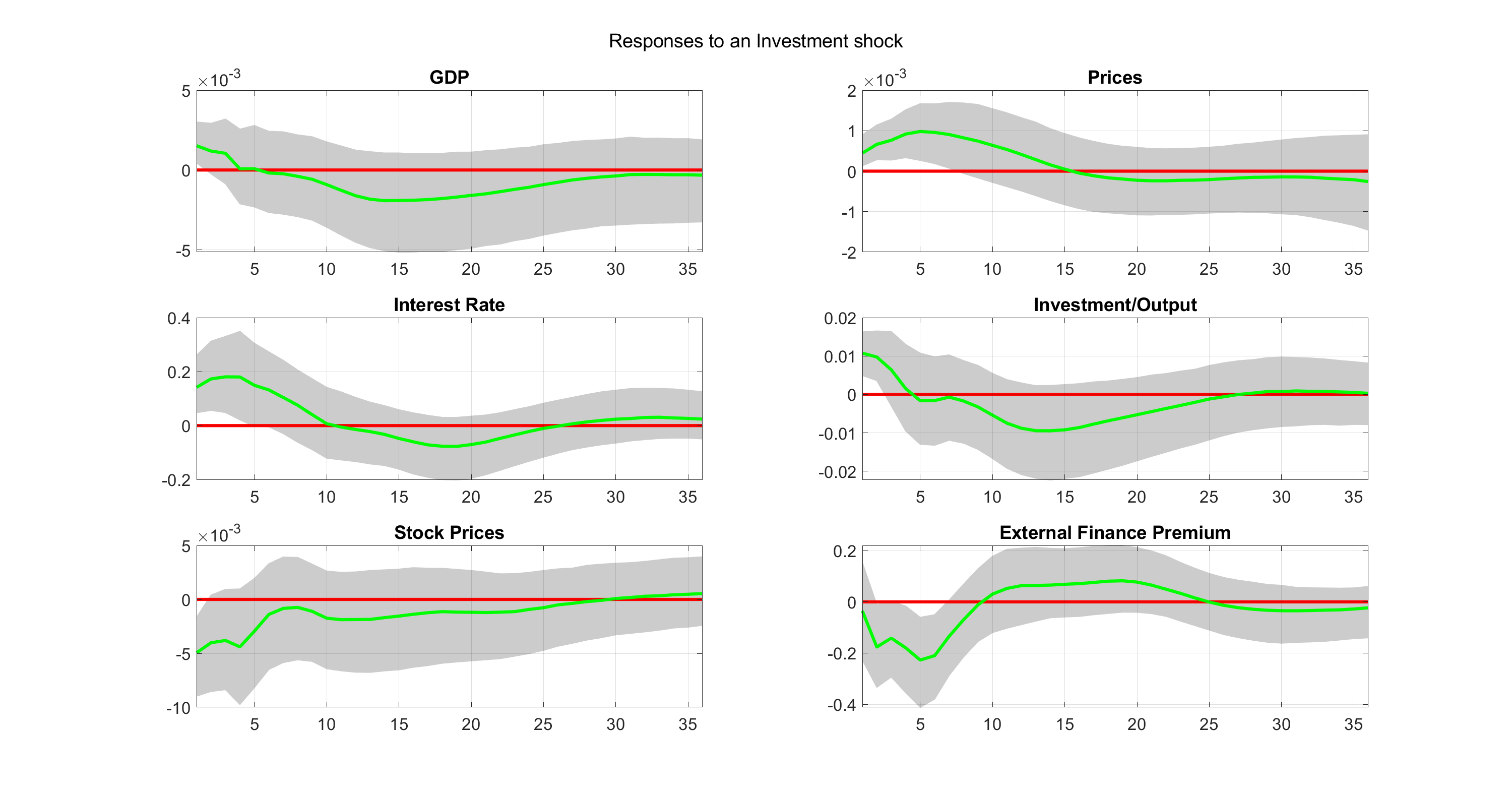}
        \caption{\cite{Furlanettoetal2017} algorithm}
    \end{subfigure}
    \hfill
    \begin{subfigure}[t]{0.49\textwidth}
        \centering
        \includegraphics[trim= 3cm 1cm 3cm 1cm,width=\linewidth]{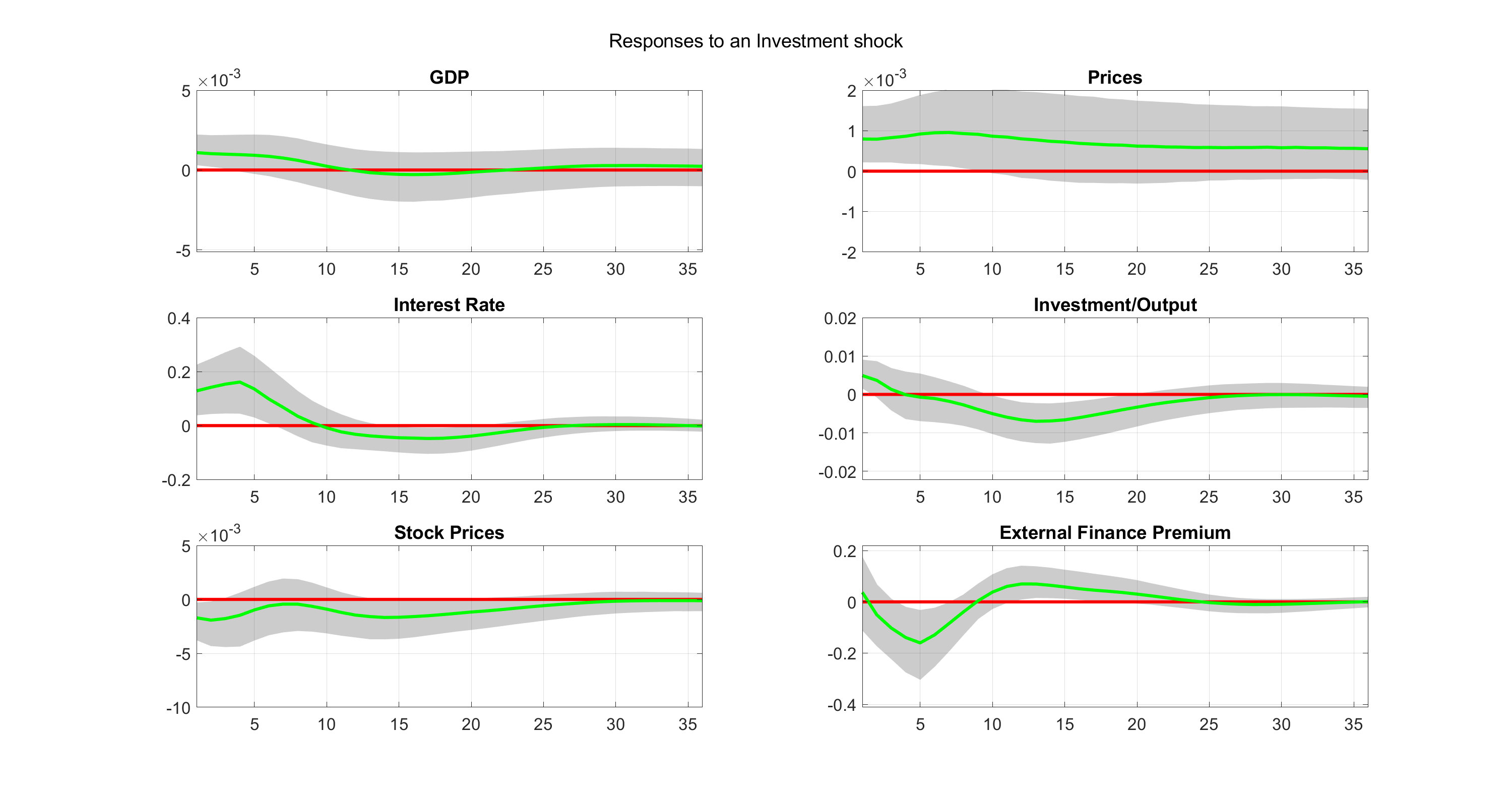}  
        \caption{Factor sign restrictions algorithm}
    \end{subfigure}
\caption{\emph{This figure replicates the impulse response functions to an investment shock using the baseline specification of \cite{Furlanettoetal2017}. Panel (a) shows results based on the exact configuration of \citet[see Figure 1]{Furlanettoetal2017}, using the algorithm of \cite{RubioRamirezetal2010}. Panel (b) replicates the same financial shock using the new sign restrictions algorithm.}} \label{baseline_investment}
\end{figure}

The main paper presents only the impulse responses to a financial shock using the baseline VAR specification of \cite{Furlanettoetal2017}. Figures \ref{baseline_supply} - \ref{baseline_investment} illustrate that when looking at the four other macro shocks, results of the new algorithm and the \cite{RubioRamirezetal2010} algorithm are qualitatively identical and quantitatively very similar.

\vskip 1cm

\subsubsection{Financial shock using a noninformative prior}
\begin{figure}[H]
    \centering
    \begin{subfigure}[t]{0.49\textwidth}
        \centering
        \includegraphics[trim= 3cm 1cm 3cm 1cm,width=\linewidth]{FRS5.png}
        \caption{\cite{Furlanettoetal2017} algorithm}
    \end{subfigure}
    \hfill
    \begin{subfigure}[t]{0.49\textwidth}
        \centering
        \includegraphics[trim= 3cm 1cm 3cm 1cm,width=\linewidth]{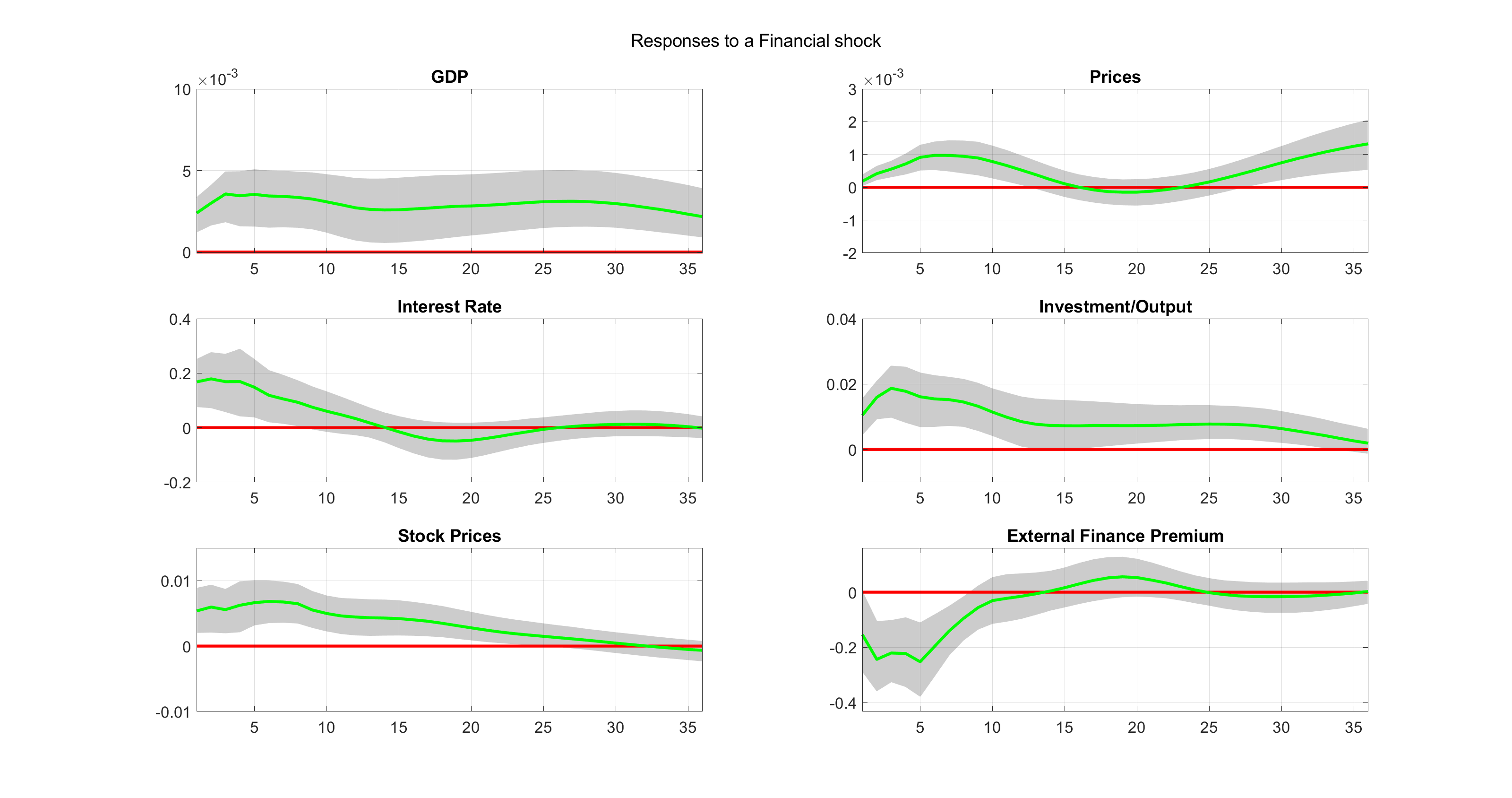}  
        \caption{Factor sign restrictions algorithm}
    \end{subfigure}
\caption{\emph{This figure replicates the impulse response functions to an financial shock using the baseline specification of \cite{Furlanettoetal2017}. This is identical to \autoref{fig:financial_benchmark}}} \label{baseline_noninformative}
\end{figure}

Just for this specific exercisxe, it would be interesting to exchange the informative (shrinkage) horseshoe prior for a noninformative prior in order to bring the new methodology closer to the settings adopted by \cite{Furlanettoetal2017}. These authors estimate the baseline VAR using diffuse prior distributions.  \autoref{baseline_noninformative} plots these results and compared to \autoref{fig:financial_benchmark}, the shapes of the IRFs are now identical. The IRFs in panel (a) still have wider bands, but this is to be expected since the covariance matrix in panel (b) is sampled from a parsimonious factor representation (and not from an inverse Wishart posterior).

\subsection{Additional impulse responses from the large-scale, 15-variable VAR}

\begin{landscape}
\begin{figure}[H]
\centering
\includegraphics[trim= 3cm 0cm 3cm 0cm,width=\linewidth]{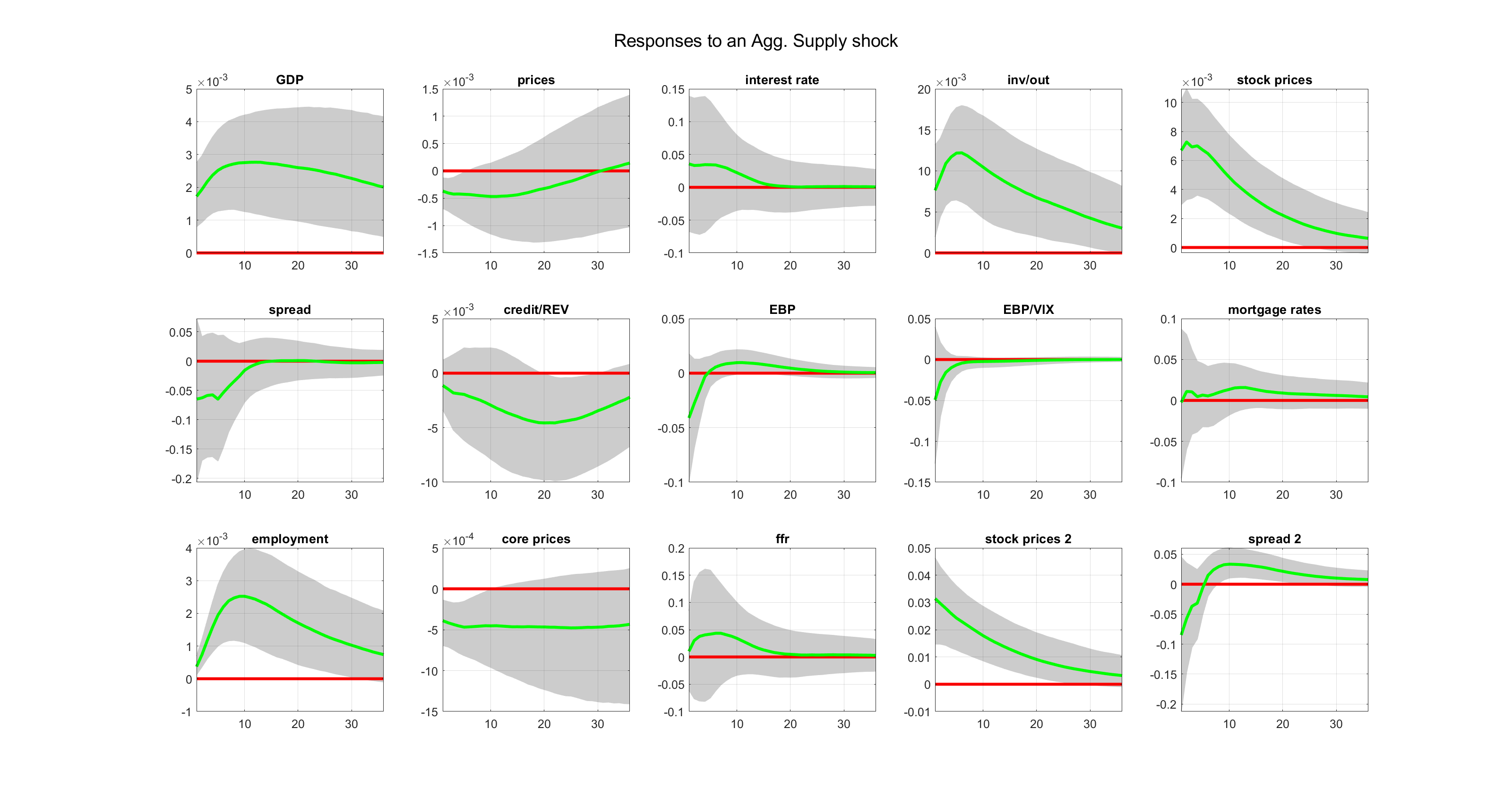}  
\caption{\emph{Impulses response functions to an aggregate supply shock in the large, 15-variable VAR with seven shocks identified in total.}} \label{fig:aggsup_large}
\end{figure}
\end{landscape}

\begin{landscape}
\begin{figure}[H]
\centering
\includegraphics[trim= 3cm 0cm 3cm 0cm,width=\linewidth]{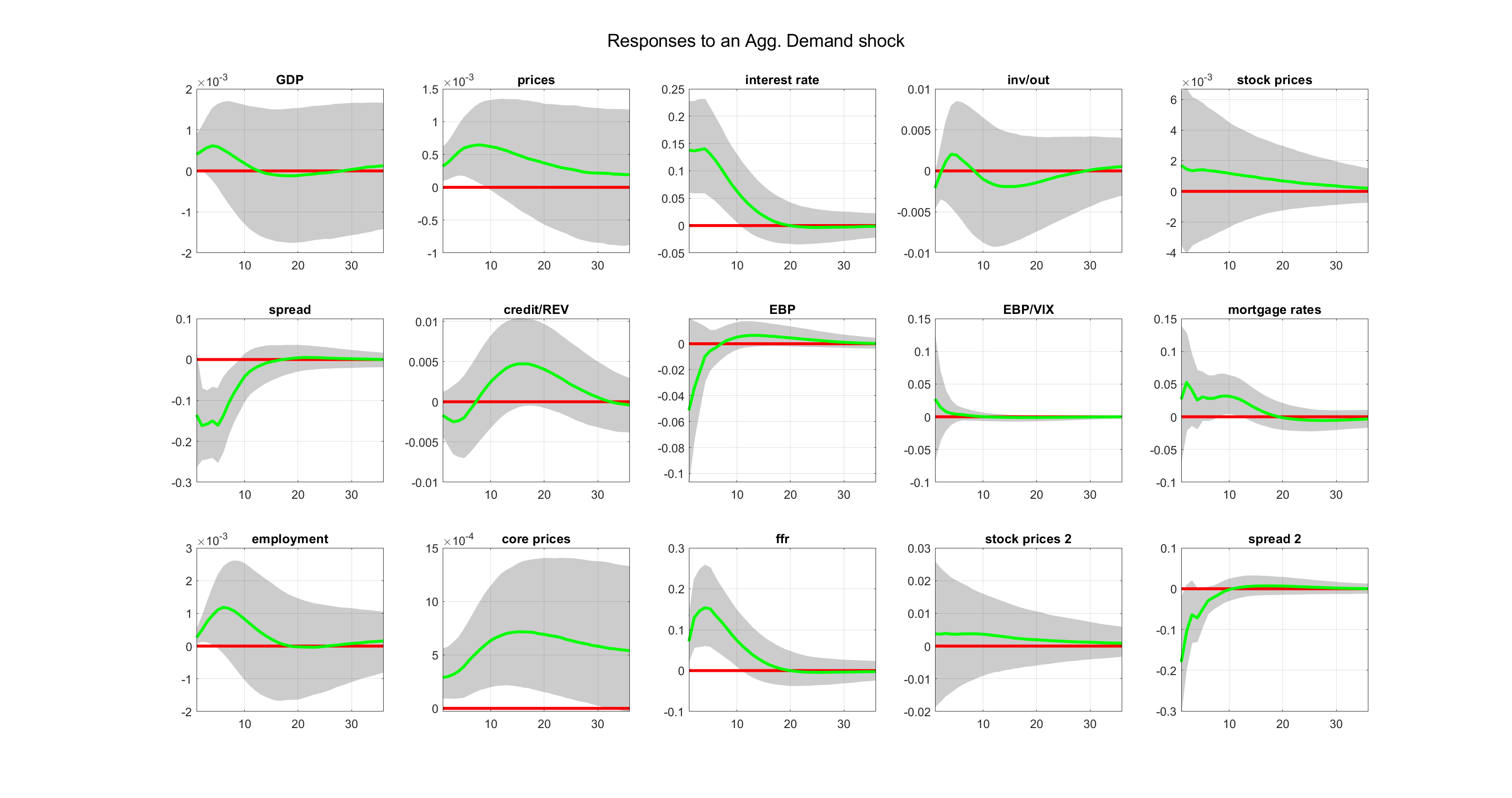}  
\caption{\emph{Impulses response functions to an aggregate demand shock in the large, 15-variable VAR with seven shocks identified in total.}} \label{fig:aggdem_large}
\end{figure}
\end{landscape}

\begin{landscape}
\begin{figure}[H]
\centering
\includegraphics[trim= 3cm 0cm 3cm 0cm,width=\linewidth]{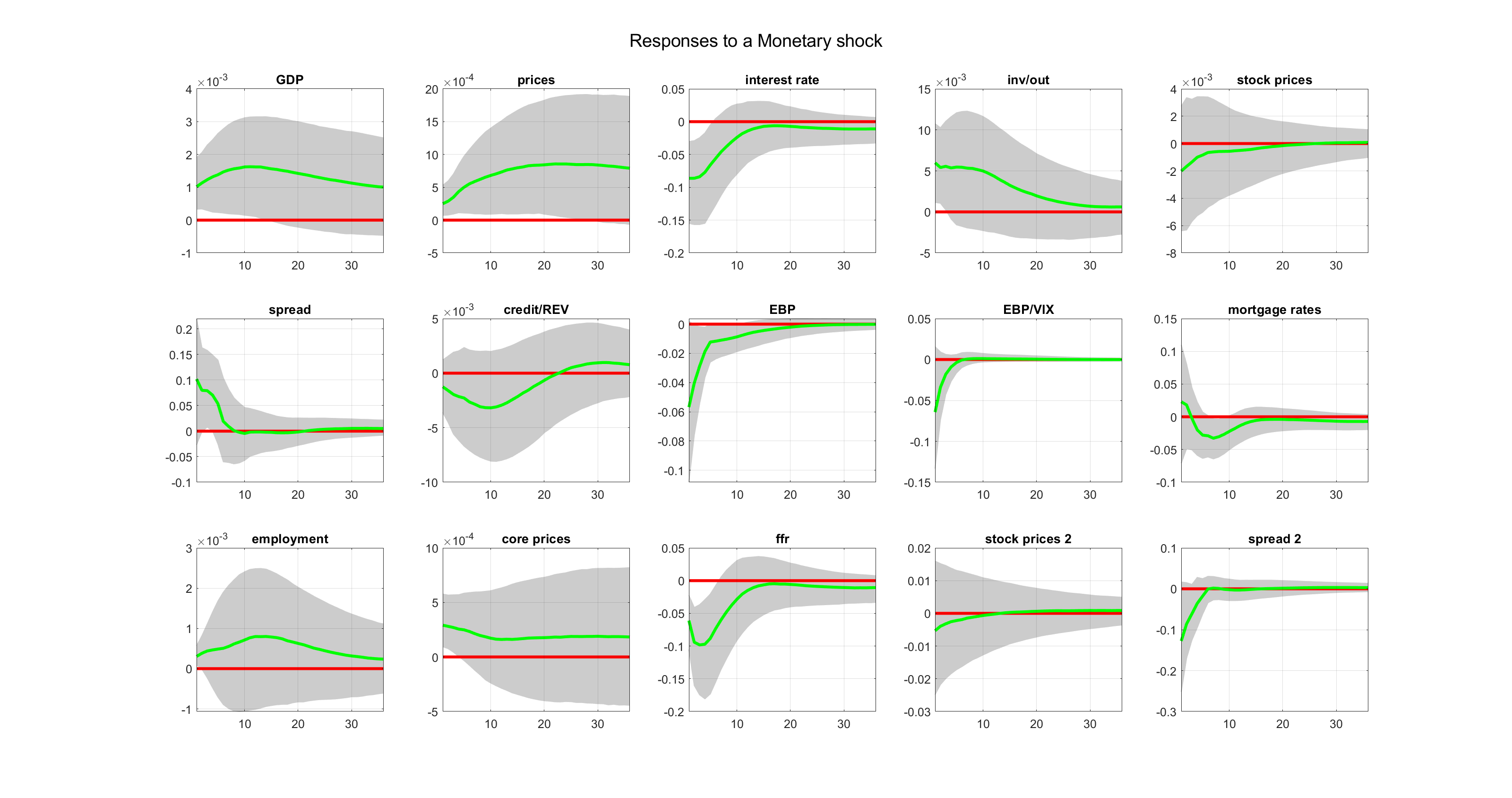}  
\caption{\emph{Impulses response functions to a monetary policy shock in the large, 15-variable VAR with seven shocks identified in total.}} \label{fig:monetary_large}
\end{figure}
\end{landscape}

\begin{landscape}
\begin{figure}[H]
\centering
\includegraphics[trim= 3cm 0cm 3cm 0cm,width=\linewidth]{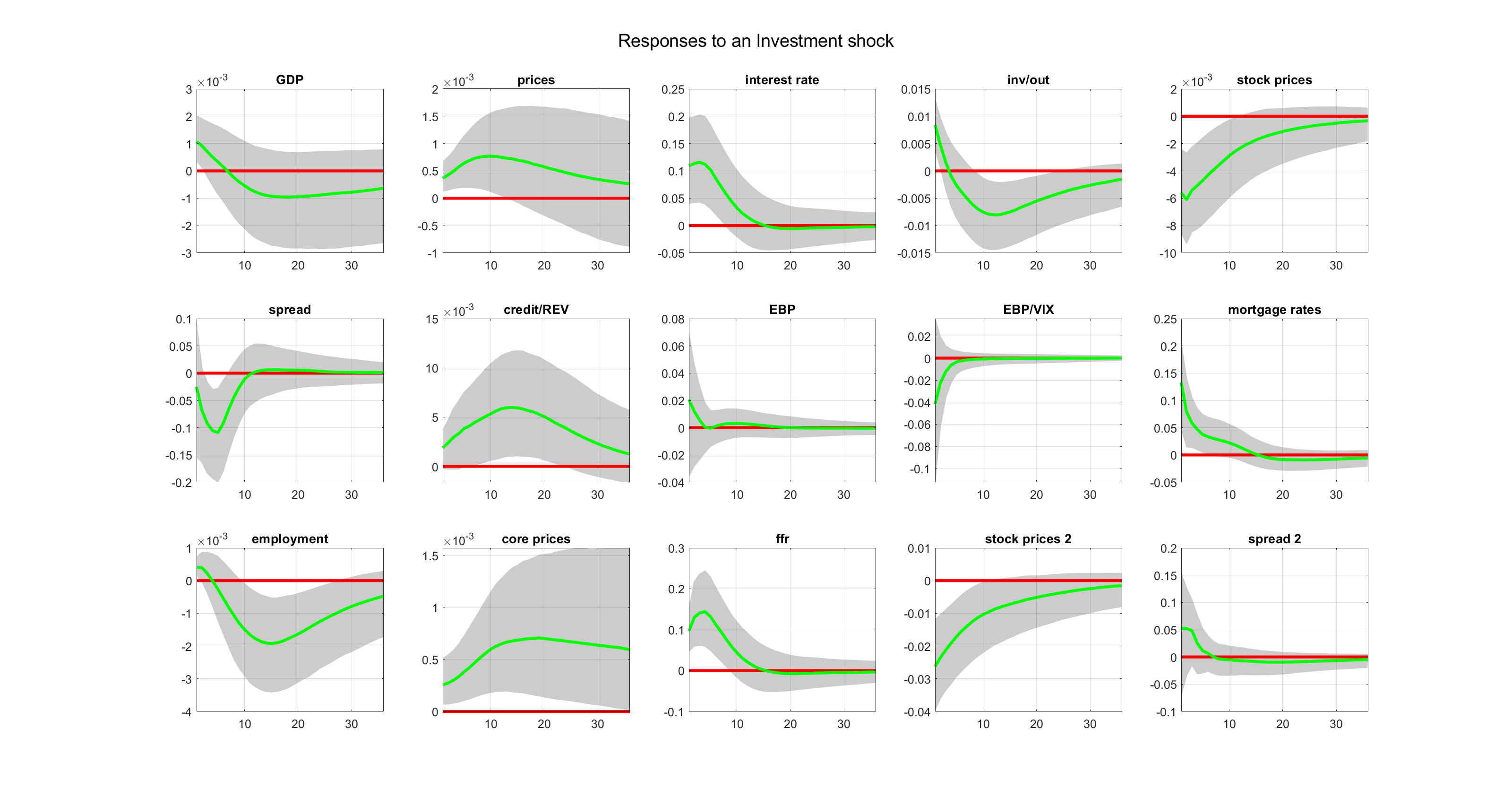}  
\caption{\emph{Impulses response functions to an investment shock in the large, 15-variable VAR with seven shocks identified in total.}} \label{fig:investment_large}
\end{figure}
\end{landscape}

\begin{landscape}
\begin{figure}[H]
\centering
\includegraphics[trim= 3cm 0cm 3cm 0cm,width=\linewidth]{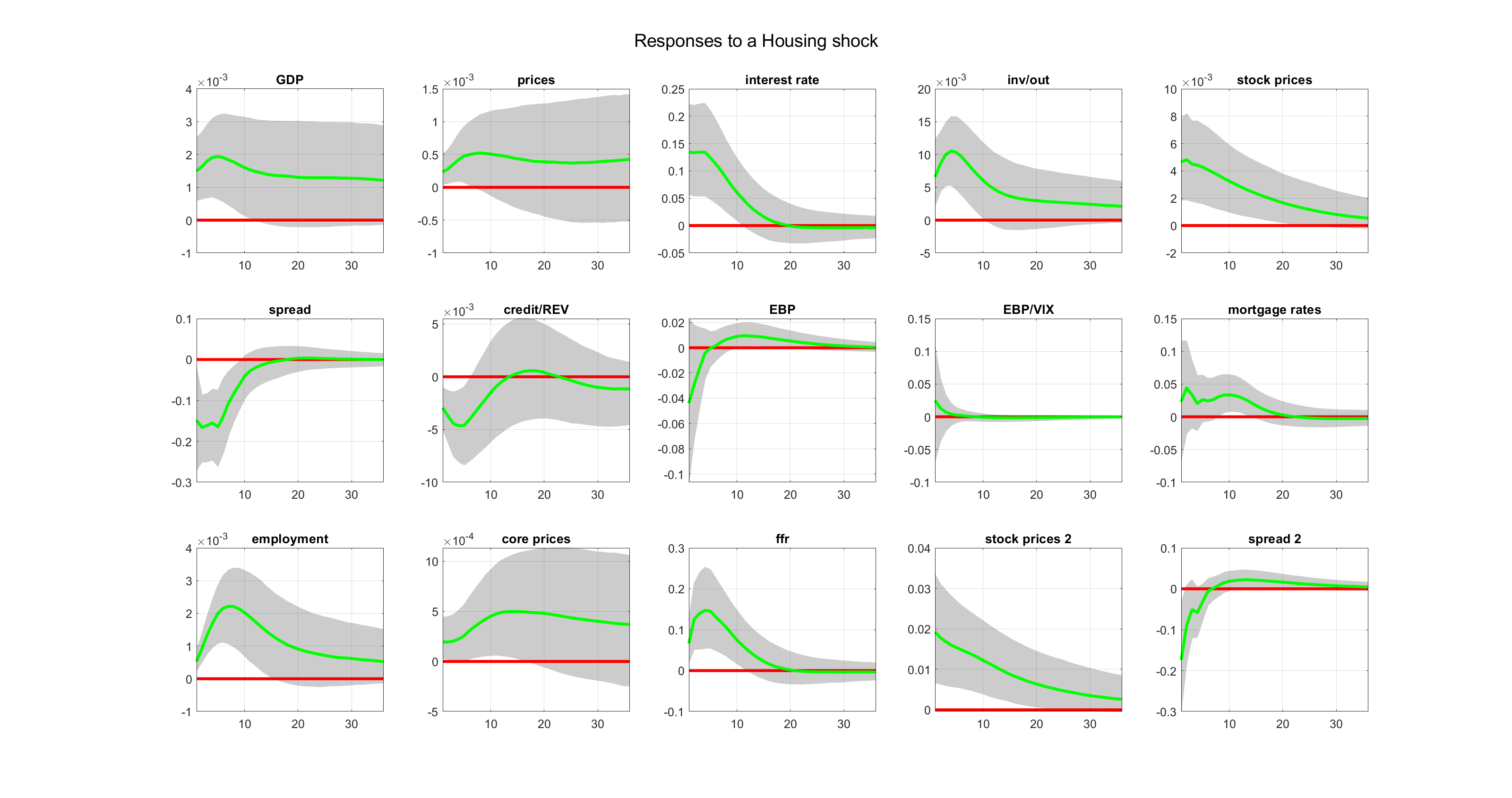}  
\caption{\emph{Impulses response functions to a housing shock in the large, 15-variable VAR with seven shocks identified in total.}} \label{fig:housing_large}
\end{figure}
\end{landscape}

\begin{landscape}
\begin{figure}[H]
\centering
\includegraphics[trim= 3cm 0cm 3cm 0cm,width=\linewidth]{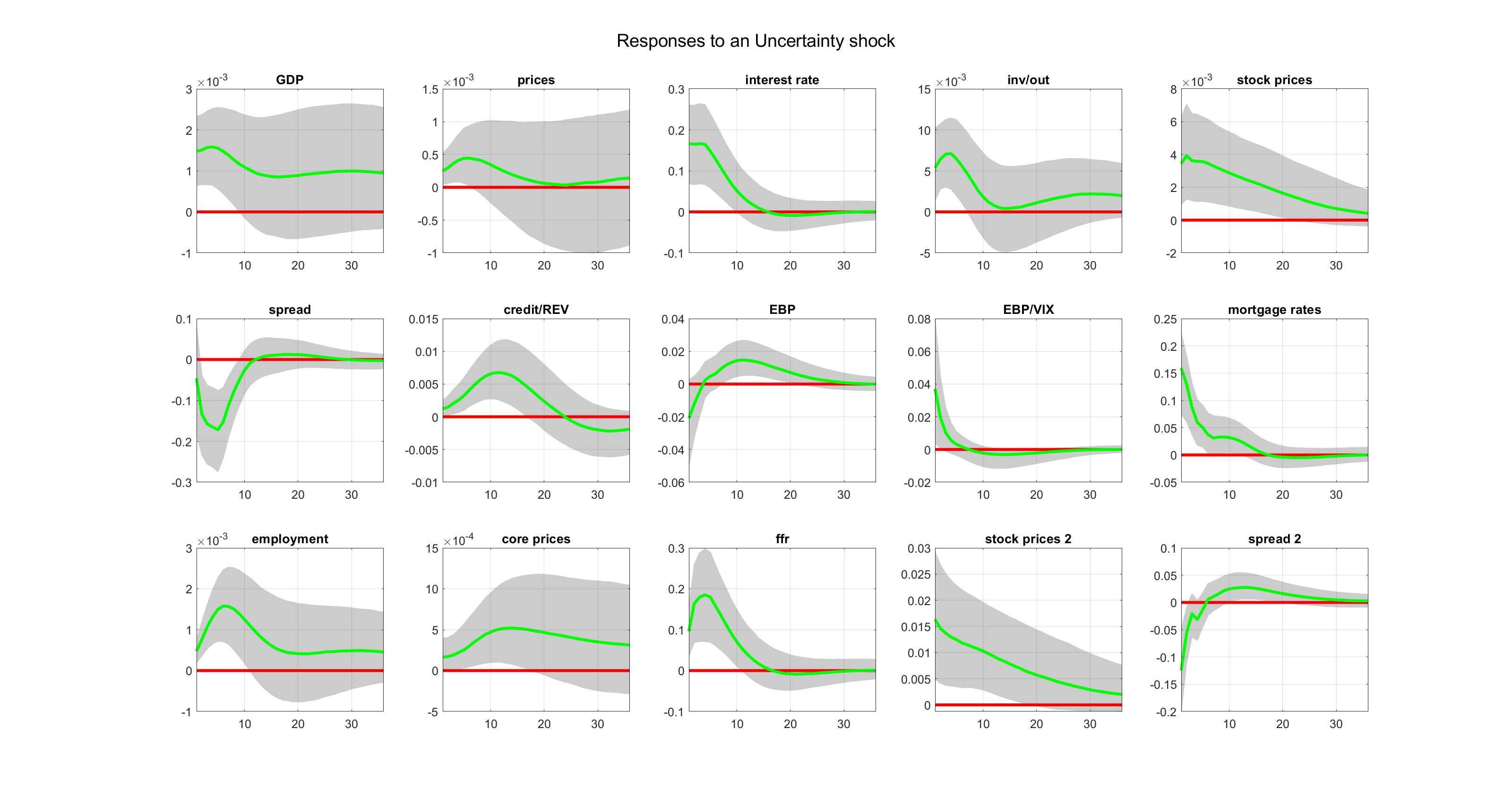}  
\caption{\emph{Impulses response functions to an uncertainty shock in the large, 15-variable VAR with seven shocks identified in total.}} \label{fig:uncertainty_large}
\end{figure}
\end{landscape}

\end{appendix}
\end{document}